\documentclass[twocolumns]{aa}
\usepackage{graphicx}
\usepackage{epsfig}
\usepackage{textcomp}
\usepackage{lscape}
\usepackage{longtable}

\def\kms  {km~s$^{-1}$}

\begin{document}

\title{Absolute positions of 
6.7-GHz methanol masers}

\subtitle{}

\author{Y. Xu\inst{1,2}, M. A. Voronkov\inst{3,4}, J. D.
Pandian\inst{1}, J. J. Li\inst{5}, A. M. Sobolev\inst{6}, A.
Brunthaler\inst{1}, B. Ritter\inst{1}, K. M. Menten\inst{1}}

   \offprints{Y. Xu}
   \institute{
   Max-Planck-Institute f$\rm\ddot{u}$r Radioastronomie,
Auf dem H$\rm\ddot{u}$gel 69, 53121 Bonn, Germany\\
         \email{xuye@pmo.ac.cn}
         \and
   Purple Mountain Observatory, Chinese Academy of Sciences, Nanjing 210008, China
         \and
   Australia Telescope National Facility CSIRO, PO Box 76, Epping, NSW 1710, Australia
         \and
   Astro Space Centre, Profsouznaya st. 84/32, 117997 Moscow, Russia
         \and
   Key Laboratory for Research in Galaxies and Cosmology,\\
   Shanghai Astronomical Observatory, Chinese Academy of Sciences, Shanghai, 20030, China
           \and
   Ural State University, Ekaterinburg, 620083, Russia}

\date{Received date; accepted date}
 \authorrunning{Y. Xu et al.}
 \titlerunning{Methanol maser positions}

\abstract{The ATCA, MERLIN and VLA interferometers were used to
 measure the absolute positions of 35 6.7-GHz methanol masers to subarcsecond
 or higher accuracy. Our measurements represent essential preparatory data for Very Long Baseline
 Interferometry, which can provide accurate parallax
 and proper motion determinations of the star-forming regions harboring the masers.
 Our data also allow associations to be established
 with infrared sources at different wavelengths. Our findings support the view
 that the 6.7~GHz masers are
 associated with the earliest phases of high-mass star formation.

}

\maketitle

\section{Introduction}
The $5_1-6_0$ A$^+$ transition of methanol at 6.7-GHz produces the
brightest methanol masers (Menten 1991). The masers are widespread
in the Galaxy and more than 550 sources have been detected to date,
including the compilations of Xu et al. (2003), Malyshev \& Sobolev
(2003), and Pestalozzi et al. (2005), and the searches of Caswell et
al. (1995a), Caswell (1996a, 1996b), MacLeod et al. (1998), Szymczak
et al. (2000), Pandian et al. (2007), Ellingsen (2007), and Xu et
al. (2008).

It has been shown that 12.2-GHz methanol masers are excellent tools
for determining the distances to massive star-forming regions  by
measuring their trigonometric parallax using Very Long Baseline
Interferometry (VLBI) (e.g., Xu et al. 2006a). The $2_0-3_{-1}$ E
transition at 12.2-GHz is the second brightest methanol maser
transition and the locations of the 6.7-GHz and 12.2-GHz methanol
maser spots largely overlap, with several features showing a
one-to-one correspondence within milliarcseconds and the spectra of
the two transitions typically covering similar velocity ranges
(Menten et al. 1992, Norris et al. 1993, Minier et al. 2000). Since
6.7 GHz masers are almost always stronger, i.e., \textit{much}
stronger than their 12.2-GHz counterparts, they are also expected to
be a useful for probing distances to massive star-forming regions in
the Galaxy. Measuring accurate distances are critical for studying
the massive star-forming regions individually and understanding the
distribution of these regions in the context of our perception of
the Galaxy's spiral structure.

For phase-referenced VLBI observations (mandatory for high precision
astrometry), one usually employs strong masers as the
phase-reference, and synthesizes images of nearby extragalactic
continuum sources. The astrometric precision scales with the source
separation and, statistically, weaker sources are found closer to
masers. These sources can be detected, e.g., by VLA observations;
see Xu et al. 2006b). On the other hand, for a successful VLBI
astrometric measurement, one requires a position estimate of the
maser position that is accurate to at least 1$''$, as input in the
correlator. This means that a large number of masers in the
literature have positions determined with single dish observations
that are not accurate enough for VLBI observations. Here we report
absolute position measurements of 35 6.7-GHz methanol masers with
subarcsecond accuracy using the Australia Telescope Compact Array
(ATCA), MERLIN and the NRAO Very Large Array (VLA). Most of the
sources in our study are associated with 12.2-GHz counterparts.

\section{Observations and data reduction}
The ATCA observations were completed in 2006 April in the 6C
configuration, which produces baselines from 153 to 6000~m. The
observations were done in snap-shot mode. Each source was observed
in six 5-minute scans spread over a range of hour angles to ensure a
good uv-coverage. The correlator was configured to have a 4 MHz
bandwidth with 1024 spectral channels. Two orthogonal linear
polarizations were observed and averaged together during the data
processing. The full width at half-maximum (FWHM) of the primary
beam was $7.2'$. The default pointing model was used, giving an rms
pointing accuracy of around of 5--10$''$. The accuracy of the
pointing model affects the accuracy of the flux density
measurements, particularly for sources that are offset from the
pointing center. The absolute flux density scale was determined from
observations of PKS B1934--638. The accuracy of the flux density
calibration is expected to be approximately 3\% \footnote{For
details of the calibration using 1934--638, query this calibrator at
the ATCA calibrators webpage ({\it
http://www.narrabri.atnf.csiro.au/calibrators})}. The bandpass
calibration was carried out using observations of the continuum
source 1921--293. The data were reduced with the MIRIAD package
using standard procedures.

The MERLIN observations were carried out in 2007 February using six
telescopes, and in 2007 March using five telescopes. The correlator
was used in two modes. The phase reference sources and the primary
calibrator (3C84) were observed in wide-band mode with a bandwidth
of 16 MHz and 32 spectral channels. The bandpass calibrator (also
3C84) and the targets were observed in the narrow-band mode with 2
MHz bandwidth and 256 spectral channels. The total on-source
integration time per source was over 1 hour, divided into a number
of 5 min scans to achieve good uv-coverage. The flux density of 3C84
was assumed to be 14.5 Jy in February 2007 and 15 Jy in March 2007.
The initial calibration and conversion of the data to FITS format
was done using the local MERLIN software, and subsequent analysis
was done using the Astronomical Image Processing System (AIPS). The
instrumental phase offset between the wide band and narrow band data
was derived using 3C84 (see the MERLIN User Guide for details;
Diamond et al. 2003).

The VLA observations were conducted on 2009 January 27 in BnA
configuration using 18 EVLA antennas. The observations were done in
single IF mode (A1, RCP) with each source being observed in three
scans with an integration time of 2.75 minutes per scan. The full
3.125~MHz bandwidth  was divided into 256 channels. For all target
sources the bandwidth was always centered on the velocity of 14.75
km~s$^{-1}$ with respect to the local standard of rest. The primary
beam was 6.75$'$. The total flux density of the flux calibrator
3C\,286 was calculated to be 6.072 Jy. The source 2007+404 served as
a bandpass calibrator and 2084+431 was the phase reference
calibrator for all target sources.

The spectral resolution was 0.18, 0.35 and 0.55 km~s$^{-1}$ for the
ATCA, MERLIN and VLA observations, respectively. The rest frequency
of 6668.5192~MHz was assumed for all observations.

After imaging the targets, the AIPS package task ``JMFIT'' was used
to determine the fluxes and positions of each maser feature in all
observations.

\section{Results}
The accuracy of the absolute maser positions is limited by several
factors, such as source elevation, weather conditions, length and
type of observations, position accuracy of the phase calibrator, and
signal-to-noise ratio. For the ATCA data, the typical
signal-to-noise ratio is over 500, and the phase calibrators have
positions accurate to better than 0.15$''$. Hence, we assume that
most of the target sources have an absolute position accuracy of
0.5$''$ or higher, except for sources that are close to the
celestial equator. This is a typical position accuracy of ATCA data
(Caswell et al. 1995c; Walsh et al. 1998; Phillips et al. 1998;
Minier et al. 2001, Caswell 2009). For the MERLIN data, the typical
signal-to-noise ratio is at least 100, and all but one phase
calibrator are from the Jodrell Bank--VLA Astrometric Survey, which
has a position accuracy of higher than 5 milliarcseconds. Hence, we
estimate that the absolute positions of the target sources are
accurate to within 0.1$''$. The source $IRAS$ 20290+4052 was
observed by both the VLA and MERLIN, and the VLA positions deviate
from the MERLIN positions by 0.05$''$. On the other hand, our
previous observations with the VLA B configuration have a position
uncertainty of better than 0.1$''$ (Xu et al. 2006b). Therefore, a
position uncertainty of better than 0.1$''$ is expected for sources
observed with the VLA. A position uncertainty of better than $1''$
is sufficient for successful observations using the European VLBI
Network (EVN) and the Very Long Baseline Array (VLBA), for example
to determine parallax measurements.

Tables 1 - 3 list the properties of the 6.7-GHz methanol sources
observed with the ATCA, MERLIN, and VLA, respectively. For sources
that exhibit multiple masing spots, the properties of individual
spots are given in separate rows. In the three tables, the first
three columns show the source name and  J2000 equatorial
coordinates. Columns 4 and 5 give their Galactic coordinates.
Columns 6 and 7 show the radial velocity of the maser peak and that
of the molecular lines, respectively. Column 8 presents the peak
flux density. Columns 9 to 11 present the distance of the source
from the Galactic center and its heliocentric kinematical distance.
The kinematical distances were calculated using the velocities of
molecular lines, such as CO, CS, and NH$_{3}$ (where these data are
available; for other sources, the maser peak velocity was used) and
the Galactic rotation model of Wouterloot \& Brand (1989), assuming
$R_{0}$ = 8.5 kpc and $\Theta_{0}$ = 220 km s$^{-1}$. The
uncertainties in kinematic distances were estimated by applying
 $\pm$10 km s$^{-1}$ velocity offsets.

\subsection{Notes on selected sources}
The spectra of the 6.7-GHz methanol masers are shown in Figs. 3--5,
which is available on line, for sources observed with the ATCA,
MERLIN, and VLA, respectively . The ATCA data have a velocity
resolution of 0.18 km~s$^{-1}$ which is sufficient to resolve the
multiple velocity components of each source. The velocity resolution
of the VLA data is 0.55 \kms. The velocity resolution of the MERLIN
data is only 0.7 \kms{ } after Hanning smoothing (because of the
limitations of the correlator), which results in the blending of
individual components in the spectra. Here we present notes on some
sources.

\noindent\begin{bf}{G8.8316--0.0281}\end{bf}. \ There are at least
six features within the velocity range of 10 km~s$^{-1}$ that have
peak flux densities exceeding 10 Jy/beam. The strongest feature is
at the LSR velocity of $-$3.9 km~s$^{-1}$.

\noindent\begin{bf}{G8.8722--0.4928}\end{bf}. \ There are only two
features stronger than 10 Jy/beam, at $+$23.4 and $+$24.1 km~s$^{-1}$.
The two features are spectrally blended together.

\noindent\begin{bf}{G14.1014+0.0869}\end{bf}. \ There are a number
of features located within a 0.2$'' \times 0.2''$ region. The
strongest 6.7~GHz peak is at the velocity of the second strongest 12.2-GHz
feature (Blaszkiewicz \& Kus 2004).

\noindent\begin{bf}{G23.0099--0.4107}\end{bf}. \ The features span
the velocity range from 70.1 to 83.2 km~s$^{-1}$, as do the 12.2-GHz
maser features in this source, although they do not coincide
precisely. There are at least 5 features for which the peak flux
density exceeds 10 Jy/beam. The strongest feature is at the same
velocity as the strongest 12.2-GHz maser (Caswell et al. 1995b).
However, the 6.7-GHz feature corresponding to the second strongest
12.2-GHz feature at +76.6 km s$^{-1}$ is not clearly distinguishable
from the other lines.

\noindent\begin{bf}{G23.2068--0.3777}\end{bf}. \ This source has two
prominent peaks. The strongest feature is at the same velocity, 81.7
km s$^{-1}$, as the strongest feature at 12.2-GHz (Blaszkiewicz \&
Kus 2004).

\noindent \begin{bf}{G24.1480--0.0092}\end{bf}. \ The spectra at both
6.7-GHz and 12.2-GHz (Blaszkiewicz \& Kus 2004) are similar and
dominated by a single feature.

\noindent \begin{bf}{G27.3652--0.1659}\end{bf}. \ There are clearly
five features with peak flux densities exceeding 10 Jy/beam. The
emission has a spatial extent of 0.1$''$ and is confined to a narrow
velocity range of approximately 5 km~s$^{-1}$. The 12.2-GHz spectrum
is dominated by a single feature (Caswell et al. 1995b). The peak
velocity is the same for both 6.7 and 12.2-GHz masers. Our
observations detected an unresolved 8.6 GHz continuum source that is
offset by about 2.4$''$ from the maser.

\noindent\begin{bf}{G29.8630--0.0442}\end{bf}. \ There are three
strong features within an area of 0.1$'' \times 0.1''$. The
strongest feature at 6.7-GHz matches the velocity of the second
strongest 12.2-GHz feature and vice versa (Caswell et al. 1995b).

\noindent\begin{bf}{G30.1987--0.1687}\end{bf} and \begin{bf}
{G30.2251-0.1796}\end{bf}. \ The two sources are separated by
103$''$. The peak velocities are $+$108.6 km~s$^{-1}$ and $+$113.5
km~s$^{-1}$ for G30.1987--0.1687 and G30.2251--0.1796, respectively.
There are two corresponding 12.2-GHz features at $+$108.5 and +110.2
km s$^{-1}$, respectively (Caswell et al. 1995b), which are likely
to originate in the same two emission centres. However, a high
angular resolution study at 12.2-GHz is required to confirm this.
Although there are a number of features in both sources, only these
two peaks have peak flux densities that exceed 10 Jy/beam.

\noindent \begin{bf}{G30.8987+0.1616}\end{bf}. \ There are two
features with peak flux densities exceeding 10 Jy/beam. The emission
peaks at 12.2 and 6.7~GHz are close in velocity.

\noindent \begin{bf}S255\end{bf}. \ The spectrum of Szymczak et al.
(2000) shows multiple spectral features, which are not visible in
the VLA spectrum because of its poor velocity resolution. However,
imaging shows two maser sites separated by 0.2$''$.

\noindent \begin{bf}18556+0136\end{bf}. \ The spectrum of Szymczak
et al. (2000) shows multiple spectral features. However, in this
study we found only a single feature. The 12.2-GHz spectrum also has
multiple features, but is dominated by just two peaks (Caswell et
al. 1995b).

\noindent \begin{bf}G43.15+0.02\end{bf}. \ The spectrum of Caswell
et al. (1995a) indicated that multiple features were present. We
detected only one spectral feature. Its velocity corresponds to that
of the strongest 12.2~GHz feature (Caswell et al. 1995b).

\noindent \begin{bf}19120+0917\end{bf}. \ This sources exhibits
multiple features that coincide in velocity with the 12.2-GHz
features (Blaszkiewicz \& Kus 2004).

\noindent \begin{bf}19186+1440\end{bf}. \ This source displays
multiple features at 6.7-GHz in the range from -16 to -9
km~s$^{-1}$. Szymczak et al. (2000) also reported 6.7-GHz emission
in the velocity range from $-$31 to $-$25 km~s$^{-1}$, which was not
detected in our observations.

\noindent \begin{bf}19303+1651\end{bf} and
\begin{bf}20290+4052\end{bf}. \ The spectra of these sources are both dominated by
a single feature. Each velocity component has its 12.2-GHz
counterpart (Blaszkiewicz \& Kus 2004).

\noindent \begin{bf}ON1\end{bf}. \ This source consists of two
separate masing sites with a separation of around 1$''$. The
peak velocities  for these two sites are 15.7 and $-$0.1 km~s$^{-1}$.

\noindent \begin{bf}21381+5000\end{bf}. \ There is only one feature
detected in the MERLIN observations, while Szymczak et al. (2000)
detected multiple features.

\section{Methanol masers and spiral arms}
Since 6.7-GHz methanol masers appear to be exclusively associated
with massive star-forming regions, they are reliable tracers of the
spiral arms of the Galaxy. This is especially so since the lifetime
of methanol masers is understood to be about 10$^{4}$ yr (van der
Walt 2005). To investigate whether any information about the spiral
structure of the Galaxy can be inferred from the data of methanol
masers detected to date, we compiled a table of all known 6.7-GHz
methanol masers (Table 5, on line). The LSR velocities in Table 5
originate in molecular lines such as CS, CO, and NH$_{3}$, where
such data are available. For other sources, the velocity of the
maser peak was used. The kinematical distances were calculated using
the Galactic rotation model of Wouterloot \& Brand (1989), assuming
$R_{0}$ = 8.5 kpc and $\Theta_{0}$ = 220 km s$^{-1}$. We made no
attempt to calculate the distances for those sources, which are
located in the two Galactic longitude ranges $0^{\circ} \pm
10^{\circ}$ and $180^{\circ} \pm 10^{\circ}$, where the uncertainty
in the kinematic method is large.

A significant fraction of the sources in the first and the fourth
Galactic quadrants are affected by an ambiguity between two
distances, the near and the far distance. This kinematical distance
ambiguity has been resolved for only a small number of methanol
masers (Pandian et al. 2008). Sobolev et al. (2005) proposed that
statistically, it is preferable to assume a more nearby kinematic
distance than a far distance. The left panel of Fig. 1 shows a
face-on diagram of the Galaxy, where the near kinematic distance is
assumed for all sources affected by a distance ambiguity. Spiral arm
loci from the NE2001 model of Cordes \& Lazio (2002) are
superimposed. It can be seen that there is little if any correlation
between the location of methanol masers and the spiral arm model.
The right panel of Fig. 1 shows the same diagram, but with the far
distance being assumed for all masers affected by a distance
ambiguity. Qualitatively, there appears to be a stronger correlation
with the spiral arm loci in the right-hand panel than in the
left-hand one. Keeping in mind that in reality there are only a
fraction of sources located at the near distance, and that the
kinematic distances have relatively large uncertainties, it seems
possible to reconcile the spiral arm model with the distribution of
methanol masers in the Galaxy. However, this exercise does suggest
that the assumption of the majority of sources being at the near
kinematic distance may be flawed. This suggestion can be
corroborated by a general observation that the majority of young
massive star-forming regions associated with HII regions appear to
be at the far distance in the studies able to resolve the ambiguity
(e.g., Kolpak et al. 2003).

Figure 1 also shows that there is a poor correspondence between the
spiral arm model and the massive star-forming regions in the outer
Galaxy where there is no distance ambiguity. This is mostly caused
by a significant deviation from the circular rotation in the Perseus
arm region (Xu et al. 2006a). Based on the VLBI parallax
measurements for a number of massive star forming-regions, Reid et
al. (2009) found that these regions orbit the Galactic center $\sim$
15 km~s$^{-1}$ slower than the Galaxy itself, if one assumes
circular rotation. In addition, the motion of the Sun towards the
local standard of rest (LSR) was found to be consistent with that
derived by Dehnen \& Binney (1998) from Hipparcos data. We hence
recalculated kinematic distances using the methodology explained in
sect. 4 of Pandian et al. (2008) -- the radial velocities were
recalculated to the new frame of solar motion, and kinematic
distances were calculated using the Galactic rotation curve of
Wouterloot \& Brand (1989) with $R_{0}$ = 8.4 kpc and $\Theta_{0}$ =
254 km s$^{-1}$ (Reid et al. 2009), assuming that the massive star
forming-regions were rotating 15 km~s$^{-1}$ slower than predicted
by the rotation curve. The left-hand and the right-hand panels of
Fig. 2 show the equivalent of Fig. 1 for the new kinematic
distances. It can be seen that there is little difference between
Figs. 1 and 2 for the inner Galaxy, but there is now much closer
agreement between the model and the data in the Perseus arm region
of the outer Galaxy.

\section{Association with star formation tracers}
It is well established that the 6.7-GHz methanol masers are
associated with  high-mass stars (e.g., W3(OH); see Menten et al.
1992), which are able to pump the masers by heating a sufficient
amount of surrounding dust to temperatures higher than 100 K, or
producing hypercompact HII regions with extremely high emission
measures (Sobolev et al. 2007). No 6.7-GHz methanol masers have been
found to be associated with low-mass young stellar objects (Minier
et al. 2003, Bourke et al. 2005).

Maser surveys suggest that the 6.7-GHz methanol masers are
associated with different phases of development in the HII regions
(Ellingsen 2007). Almost all 6.7-GHz masers are found to be
associated with 1.2 mm emission (Hill et al. 2005), while many have
no associated 8.6-GHz continuum emission (Walsh et al. 1998).
Relevant cases can be found even within one star-forming region,
e.g., NGC6334 I, which possesses a maser cluster associated with a
prominent ultracompact HII region and another one associated with
the sub-mm core and a candidate  hypercompact HII region with very
weak radio continuum emission (Hunter et al. 2006).

Maser positions measured to subarcsecond accuracy allow us to study
the connection between the methanol masers and the other signposts
of massive star formation. However, there are no published high
resolution continuum surveys in radio or submillimeter wavelengths
that cover all or a significant fraction of the sources in our
sample. Hence, we focus on infrared counterparts from all sky or
Galactic plane surveys.

Table 4 presents the association with infrared sources. Among the 35
sources in our sample, 25 are in the inner Galaxy, while 10 are
located in the outer Galaxy. In the inner Galaxy, 19 of 25 sources
are covered by the GLIMPSE survey, which is limited to Galactic
longitudes, $|l| \leq 65^\circ$. Seventeen sources have a GLIMPSE
point source within 5 arcsec, dropping to 11 within 2 arcsec. Most
sources with no nearby point source in the GLIMPSE catalog or
archive data are associated with extended emission, and one source
(G23.2068-0.3777) is associated with an infrared dark cloud. Only
four sources have flux measurements in all four bands (often due to
extended emission in the other bands), and hence we do not attempt
to compare the properties of the sources with those published
previously (e.g., Ellingsen 2006).

Twenty-one masers have a 2MASS point-source counterpart within 5
arcsec, dropping to 9 within 2 arcsec. Most of the sources show an
infrared excess based on their JHK colors, which is indicative of an
association with protostars. Four sources show no infrared excess,
suggesting that they are either foreground stars or more evolved
objects. By Cross-correlating with the GLIMPSE catalog, only five
GLIMPSE sources are found to have 2MASS counterparts. This strongly
suggests that most of the 6.7-GHz methanol masers do not have 2MASS
counterparts, and that most of the nearby 2MASS sources are more
evolved young stellar objects in the star-forming region. This
supports the results of Ellingsen (2005, 2006).

Since the black-body emission of the warm dust, hypothesized to be
the pump source for the masers, peaks at around 25 $\mu$m
(Ostrovskii \& Sobolev, 2002), one expects all methanol masers to
have mid/far infrared counterparts. Two surveys with data at this
wavelength range are the all sky survey of $IRAS$, and the MIPSGAL
survey using the Spitzer space telescope. Keeping in mind that the
$IRAS$ point source catalog is limited by both source confusion and
poor resolution, 20 methanol masers have an $IRAS$ point source
within 30 arcsec.  We note that 9 of 10 sources located in the outer
Galaxy, where source confusion is not as severe as in the inner
Galaxy, have an $IRAS$ point source with an infrared luminosity
greater than 10$^{3}$ $L_{\odot}$. However, due to the poor
resolution of the $IRAS$ satellite, it is possible that a single
point source in the $IRAS$ catalog may correspond to multiple
star-forming sites in the molecular cloud. Hence, higher spatial
resolution data is required to infer properties such as the
luminosity and mass of the source associated with the maser.

MIPSGAL is a Galactic plane survey at 24 and 70~$\mu$m using the
MIPS camera of the Spitzer Space Telescope (Rieke et al. 2004, Carey
et al. 2005). The survey is limited to Galactic longitudes between 5
and 63 degrees in the first Galactic quadrant and 298 and 355
degrees in the fourth quadrant for Galactic latitudes $|b| \leq
1^\circ$. Eighteen sources in our sample are covered by the survey,
and all sources are associated with 24 $\mu$m emission, as shown in
Fig. 6 (available on line). An association with point sources (which
are occasionally saturated) is evident for 16 sources, while for 2
sources (G43.15+0.02 and $W51e2$) the images are completely
saturated. It is thus reasonable to expect that all the 6.7-GHz
methanol masers have MIPSGAL counterparts. However, image artifacts
such as saturation make it difficult to determine 24 $\mu$m fluxes.
Further observations will be required before we will be able to
determine their spectral energy distributions and dust properties.
Fifteen sources in our sample have an MSX point source within 5$''$.
When restricted to sources that are covered by the MIPSGAL survey,
only six sources have a nearby MSX source. The far poorer statistics
of the associations with MSX sources is probably caused by the
coarser spatial resolution of the MSX satellite and its poorer
sensitivity.

\section{Conclusions}

Absolute positions with an accuracy of ~1 arcsecond  or higher have
been determined for 35 6.7-GHz methanol masers. Our measurements are
essential to a future VLBI astrometric follow-up observations.
Kinematic distances to the masers imply that they do not trace the
spiral arms well irrespective of whether they are at the near or far
kinematic distances, although there is a small improvement if  the
rotation curve of Reid et al. (2009) is used. Although our sample is
not statistically complete, the number of associations with infrared
sources is consistent with the expectation that the 6.7~GHz masers
are associated with the early phases of massive star formation.

\begin{acknowledgements}
We would like to thank the referee, Simon Ellingsen, for many useful
suggestions and comments which help us to improve this paper. The
Australia Telescope is funded by the Commonwealth of Australia for
operation as a National Facility managed by CSIRO. We thank Drs. A.
M. S. Richards and R. Beswick for help in reducing MERLIN data. This
work was supported by the Chinese NSF through grants NSF 10673024,
NSF 10733030, NSF 10703010 and NSF 10621303, and NBRPC (973 Program)
under grant 2007CB815403. AMS was supported by RFBR grants
07-02-00628-a and 08-02-00933-a. This work used the NASA/IPAC
Infrared Science Archive.
\end{acknowledgements}


\begin{table*}
\begin{flushleft}
\normalsize  \caption{6.7-GHz methanol masers observed with ATCA.
The first column lists the source name. The next two columns give
their J2000 equatorial coordinates. Columns. (4) and (5) give their
galactic coordinates. Columns. (6) and (7) show the radial velocity
of peak emission from the maser and molecular lines. Col. (8)
presents the peak flux density. Columns. (9) - (11) present the
distance from the Galactic center, and the far and near kinematic
distances.}
\end{flushleft}
         \label{Tabiras}
      \[
           \begin{array} {lccrrrrccrr}

             \hline \hline
      \noalign{\smallskip}
{\parbox[t]{19mm}{\centering Source\\Name}}&
{\parbox[t]{19mm}{\centering R.A.(2000)\\
\mbox{$\mathrm{(^h\;\;\;^m\;\;\;^s)}$}}}&
{\parbox[t]{19mm}{\centering DEC(2000) \\
\mbox{$(\degr\;\;\;\arcmin\;\;\;\arcsec)$}}}&
{\parbox[t]{16mm}{\centering l \\ \mbox{$(\degr)$}}}&
{\parbox[t]{16mm}{\centering b \\ \mbox{$(\degr)$}}}&
{\parbox[t]{10mm}{\centering $\rm V_{LSR}$\\ \mbox{\scriptsize (km
s$^{-1}$)}}}& {\parbox[t]{10mm}{\centering $\rm V_{mol}$\\
\mbox{\scriptsize (km s$^{-1}$)}}}&
{\parbox[t]{11mm}{\centering $\rm S_{peak}$\\
\mbox{\scriptsize (Jy/beam)}}}&
{\parbox[t]{9mm}{\centering $\rm R $\\
\mbox{\scriptsize (kpc)}}}&
{\parbox[t]{14mm}{\centering $\rm d_{far}$\\
\mbox{\scriptsize (kpc)}}}&
{\parbox[t]{14mm}{\centering $\rm d_{near}$\\
\mbox{\scriptsize (kpc)}}}
\\
       \noalign{\smallskip}
\hline
      \noalign{\smallskip}
\small

  $G$8.832-0.028      &18\ 05\ 25.66   &-21\ 19\ 25.5  &  8.832 & -0.028 &  -3.9 &   0.5(1) & 132.0 & 8.4 & 16.7(\pm3.6) & 0.1(\pm2.0) \\
  $G$8.872-0.493      &18\ 07\ 15.32   &-21\ 30\ 54.4  &  8.872 & -0.493 &  23.4 &          &  30.9 & 4.9 & 13.2(\pm1.1) & 3.6(\pm1.1) \\
  $G$14.101+0.087     &18\ 15\ 45.80   &-16\ 39\ 09.7  & 14.101 &  0.087 &  15.2 &   9.3(2) &  90.2 & 7.2 & 15.1(\pm1.5) & 1.3(\pm1.1) \\
                      &18\ 15\ 45.80   &-16\ 39\ 09.5  & 14.101 &  0.087 &   5.9 &          &  15.4 &     &              &             \\
                      &18\ 15\ 45.81   &-16\ 39\ 09.5  & 14.102 &  0.087 &  10.1 &          &  25.1 &     &              &             \\
                      &18\ 15\ 45.81   &-16\ 39\ 09.6  & 14.101 &  0.087 &  11.0 &          &  22.8 &     &              &             \\
                      &18\ 15\ 45.80   &-16\ 39\ 09.7  & 14.101 &  0.087 &  13.1 &          &  16.3 &     &              &             \\
                      &18\ 15\ 45.80   &-16\ 39\ 09.7  & 14.101 &  0.087 &  13.6 &          &  22.6 &     &              &             \\
  $G$23.010-0.411     &18\ 34\ 40.29   &-09\ 00\ 38.1  & 23.010 & -0.411 &  74.8 &  74.8(2) & 415.4 & 4.4 & 10.8(\pm0.5) & 4.9(\pm0.5) \\
                      &18\ 34\ 40.29   &-09\ 00\ 38.2  & 23.010 & -0.411 &  72.7 &          &  43.4 &     &              &             \\
                      &18\ 34\ 40.28   &-09\ 00\ 38.4  & 23.010 & -0.411 &  80.6 &          &  49.1 &     &              &             \\
                      &18\ 34\ 40.28   &-09\ 00\ 38.4  & 23.010 & -0.411 &  81.6 &          &  52.0 &     &              &             \\
                      &18\ 34\ 40.27   &-09\ 00\ 38.5  & 23.010 & -0.411 &  82.3 &          &  43.4 &     &              &             \\
  $G$23.207-0.378     &18\ 34\ 55.20   &-08\ 49\ 14.2  & 23.207 & -0.378 &  81.7 &          &  38.2 & 4.3 & 10.5(\pm0.5) & 5.2(\pm0.5) \\
                      &18\ 34\ 55.21   &-08\ 49\ 14.4  & 23.207 & -0.378 &  76.5 &          &  19.4 &     &              &             \\
                      &18\ 34\ 55.21   &-08\ 49\ 14.6  & 23.207 & -0.378 &  77.0 &          &  35.4 &     &              &             \\
  $G$24.148-0.009     &18\ 35\ 20.94   &-07\ 48\ 55.6  & 24.148 & -0.009 &  17.7 &  23.1(3) &  26.8 & 6.7 & 13.5(\pm0.8) & 2.0(\pm0.8) \\
  $G$24.329+0.144     &18\ 35\ 08.14   &-07\ 35\ 04.0  & 24.329 &  0.144 & 110.2 & 112.0(2) &   5.0 & 3.7 &  8.9(\pm0.9) & 6.6(\pm0.9) \\
  $G$27.220+0.260     &18\ 40\ 03.72   &-04\ 57\ 45.6  & 27.220 &  0.260 &   9.3 &          &   6.2 & 7.8 & 14.3(\pm0.9) & 0.8(\pm0.8) \\
  $G$27.365-0.166     &18\ 41\ 51.06   &-05\ 01\ 42.8  & 27.365 & -0.166 & 100.0 &  92.2(4) &  28.0 & 4.2 &  9.2(\pm0.7) & 5.9(\pm0.7) \\
                      &18\ 41\ 51.06   &-05\ 01\ 42.8  & 27.365 & -0.166 &  98.0 &          &  10.2 &     &              &             \\
                      &18\ 41\ 51.06   &-05\ 01\ 42.8  & 27.365 & -0.166 &  98.9 &          &  13.7 &     &              &             \\
                      &18\ 41\ 51.06   &-05\ 01\ 42.8  & 27.365 & -0.166 & 100.7 &          &  14.2 &     &              &             \\
                      &18\ 41\ 51.06   &-05\ 01\ 42.7  & 27.365 & -0.166 & 101.5 &          &  10.8 &     &              &             \\
  $G$29.863-0.044     &18\ 45\ 59.57   &-02\ 45\ 04.4  & 29.863 & -0.044 & 101.4 & 100.4(4) &  76.5 & 4.3 &  8.2(\pm0.8) & 6.6(\pm0.8) \\
                      &18\ 45\ 59.57   &-02\ 45\ 04.5  & 29.863 & -0.044 & 100.3 &          &  52.0 &     &              &             \\
                      &18\ 45\ 59.57   &-02\ 45\ 04.5  & 29.863 & -0.044 & 101.7 &          &  54.2 &     &              &             \\
  $G$30.199-0.169     &18\ 47\ 03.07   &-02\ 30\ 33.6  & 30.199 & -0.169 & 108.6 & 103.3(4) &  16.0 & 4.3 &  7.8(\pm1.0) & 6.9(\pm1.0) \\
  $G$30.225-0.180     &18\ 47\ 08.30   &-02\ 29\ 27.1  & 30.225 & -0.180 & 113.5 & 104.5(4) &  10.8 & 4.3 &  7.7(\pm1.1) & 7.0(\pm1.1) \\
  $G$30.899+0.162     &18\ 47\ 09.13   &-01\ 44\ 08.8  & 30.899 &  0.162 & 101.8 &          &  25.7 & 4.4 &  7.7(\pm1.1) & 6.9(\pm1.1) \\
                      &18\ 47\ 09.13   &-01\ 44\ 08.7  & 30.899 &  0.162 & 103.0 &          &  14.2 &     &              &             \\

\noalign{\smallskip} \hline

         \end{array}
      \]

References for the velocities: \\
1 \  Zhang et al. 2005.   \ 2 \  Larionov et al. 1999.   \ 3 \  Szymczak et al. 2007.  \ 4 \  van der Walt et al. 2007. \\

\end{table*}

\begin{table*}
\begin{flushleft}
\normalsize  \caption{Same as Table 1 for the 6.7-GHz methanol
masers observed with MERLIN.}
\end{flushleft}
         \label{Tabiras}
      \[
           \begin{array} {lccrrrrccrr}

             \hline \hline
      \noalign{\smallskip}
{\parbox[t]{19mm}{\centering Source\\Name}}&
{\parbox[t]{19mm}{\centering R.A.(2000)\\
\mbox{$\mathrm{(^h\;\;\;^m\;\;\;^s)}$}}}&
{\parbox[t]{19mm}{\centering DEC(2000) \\
\mbox{$(\degr\;\;\;\arcmin\;\;\;\arcsec)$}}}&
{\parbox[t]{16mm}{\centering l \\ \mbox{$(\degr)$}}}&
{\parbox[t]{16mm}{\centering b \\ \mbox{$(\degr)$}}}&
{\parbox[t]{10mm}{\centering $\rm V_{LSR}$\\ \mbox{\scriptsize (km
s$^{-1}$)}}}& {\parbox[t]{10mm}{\centering $\rm V_{mol}$\\
\mbox{\scriptsize (km s$^{-1}$)}}}&
{\parbox[t]{11mm}{\centering $\rm S_{peak}$\\
\mbox{\scriptsize (Jy/beam)}}}&
{\parbox[t]{9mm}{\centering $\rm R $\\
\mbox{\scriptsize (kpc)}}}&
{\parbox[t]{14mm}{\centering $\rm d_{far}$\\
\mbox{\scriptsize (kpc)}}}&
{\parbox[t]{14mm}{\centering $\rm d_{near}$\\
\mbox{\scriptsize (kpc)}}}
\\
       \noalign{\smallskip}
\hline
      \noalign{\smallskip}
\small

  $L1287$        &00\ 36\ 47.358   &63\ 29\ 02.18  & 121.298 &  0.659 & -23.2 & -17.6(3) &  7.5   &  9.4  &  1.6(\pm0.9) &             \\
  $NGC281-N$     &00\ 52\ 24.196   &56\ 33\ 43.17  & 123.066 & -6.309 & -29.3 & -31.8(3) &  32.4  & 10.4  &  2.9(\pm1.0) &             \\
  $S231$         &05\ 39\ 13.066   &35\ 45\ 51.29  & 173.482 &  2.446 & -12.8 & -15.8(3) &  32.1  & 24.1  & 15.6(\pm12.9)&             \\
  $AFGL5180$     &06\ 08\ 53.342   &21\ 38\ 29.09  & 188.946 &  0.886 & 10.7  &   3.1(1) &  194   &  9.4  &  0.9(\pm4.7) &             \\
  $AFGL6366$     &06\ 08\ 40.671   &21\ 31\ 06.89  & 189.030 &  0.784 & 8.8   &   2.5(1) &  1.6   &  9.2  &  0.7(\pm4.4) &             \\
  $S255$         &06\ 12\ 54.006   &17\ 59\ 23.21  & 192.600 & -0.048 & 5.5   &   8.2(3) &  10.0  & 10.3  &  1.9(\pm3.6) &             \\
                 &06\ 12\ 54.020   &17\ 59\ 23.27  & 192.600 & -0.048 & 4.4   &          &  8.5   &       &              &             \\
  $S269$         &06\ 14\ 37.051   &13\ 49\ 36.16  & 196.454 & -1.677 & 15.3  &  17.9(1) &  3.6   & 12.1  &  3.7(\pm3.7) &             \\
  18556+0136     &18\ 58\ 13.053   &01\ 40\ 35.68  &  35.197 & -0.743 & 28.5  &  35.0(3) &  72.8  &  6.6  & 11.4(\pm0.6) & 2.5(\pm0.6) \\
  $G43.15+0.02$  &19\ 10\ 11.049   &09\ 05\ 20.49  &  43.149 &  0.013 & 13.4  &   2.9(1) &  6.0   &  8.3  & 12.2(\pm0.8) & 0.2(\pm0.8) \\
  19120+0917     &19\ 14\ 26.393   &09\ 22\ 36.53  &  43.890 & -0.784 & 52.0  &  54.2(1) &  3.0   &  6.2  &  8.1(\pm1.6) & 4.2(\pm1.6) \\
                 &19\ 14\ 26.392   &09\ 22\ 36.62  &  43.890 & -0.784 & 47.7  &          &  4.5   &       &              &             \\
  19186+1440     &19\ 20\ 59.212   &14\ 46\ 49.65  &  49.416 &  0.326 & -12.1 &          &  7.0   &  9.2  & 12.1(\pm0.9) &             \\
  W51e2          &19\ 23\ 43.949   &14\ 30\ 34.44  &  49.490 & -0.388 & 59.2  &  56.5(3) &  217   &       &              &             \\
  19303+1651     &19\ 32\ 36.071   &16\ 57\ 38.46  &  52.663 & -1.092 & 65.7  &  59.7(1) &  1.3   &       &              &             \\
  $ON1A$         &20\ 10\ 09.047   &31\ 31\ 35.06  &  69.540 & -0.976 & 14.7  &  11.7(3) &  31.0  &  8.0  &  4.0(\pm1.7) & 2.0(\pm1.8) \\
  $ON1B$         &20\ 10\ 09.073   &31\ 31\ 35.94  &  69.540 & -0.976 & -0.1  &  11.7(3) &  8.8   &  8.0  &  4.0(\pm1.7) & 2.0(\pm1.8) \\
  20290+4052     &20\ 30\ 50.673   &41\ 02\ 27.55  &  79.736 &  0.991 & -5.2  &  -1.4(1) &  18.2  &  8.6  &  3.3(\pm1.5) &             \\
  21381+5000     &21\ 39\ 58.263   &50\ 14\ 20.96  &  94.602 & -1.796 & -40.7 & -45.6(3) &  4.2   & 10.8  &  6.1(\pm1.1) &             \\
                 &21\ 39\ 58.262   &50\ 14\ 21.05  &  94.603 & -1.796 & -43.6 &          &  4.0   &       &              &             \\
  $L1206$        &22\ 28\ 51.408   &64\ 13\ 41.30  & 108.184 &  5.519 & -11.0 &  -9.9(2) &  29.4  &  8.9  &  1.2(\pm1.1) &             \\

\noalign{\smallskip} \hline

         \end{array}
      \]

References for the velocities: \\
1 \  Bronfman et al. 1996. \ 2 \  Molinari et al. 1996.   \ 3 \  Plume et al. 1992. \\

\end{table*}

\begin{table*}
\begin{flushleft}
\normalsize  \caption{Same as Table 1 for the 6.7-GHz methanol
masers observed with VLA.}
\end{flushleft}
         \label{Tabiras}
      \[
           \begin{array} {lccrrrrccrr}

             \hline \hline
      \noalign{\smallskip}
{\parbox[t]{19mm}{\centering Source\\Name}}&
{\parbox[t]{19mm}{\centering R.A.(2000)\\
\mbox{$\mathrm{(^h\;\;\;^m\;\;\;^s)}$}}}&
{\parbox[t]{19mm}{\centering DEC(2000) \\
\mbox{$(\degr\;\;\;\arcmin\;\;\;\arcsec)$}}}&
{\parbox[t]{16mm}{\centering l \\ \mbox{$(\degr)$}}}&
{\parbox[t]{16mm}{\centering b \\ \mbox{$(\degr)$}}}&
{\parbox[t]{10mm}{\centering $\rm V_{LSR}$\\ \mbox{\scriptsize (km
s$^{-1}$)}}}& {\parbox[t]{10mm}{\centering $\rm V_{mol}$\\
\mbox{\scriptsize (km s$^{-1}$)}}}&
{\parbox[t]{11mm}{\centering $\rm S_{peak}$\\
\mbox{\scriptsize (Jy/beam)}}}&
{\parbox[t]{9mm}{\centering $\rm R $\\
\mbox{\scriptsize (kpc)}}}&
{\parbox[t]{14mm}{\centering $\rm d_{far}$\\
\mbox{\scriptsize (kpc)}}}&
{\parbox[t]{14mm}{\centering $\rm d_{near}$\\
\mbox{\scriptsize (kpc)}}}
\\
       \noalign{\smallskip}
\hline
      \noalign{\smallskip}
\small

$DR20$     &20\ 37\ 00.960 & 41\ 34\ 55.74 & 80.861 & 0.383 & -4.44 & 3.98(1) &   1.24 & 8.4 & 1.7(\pm2.1) & 1.0(\pm2.1) \\
$W75$      &20\ 38\ 36.428 & 42\ 37\ 34.82 & 81.871 & 0.781 &  6.76 &  9.8(2) & 152.28 &     &             &             \\
$DR21B\_1$ &20\ 39\ 01.989 & 42\ 24\ 59.30 & 81.752 & 0.591 & -9.07 & -2.5(2) &   4.00 & 8.6 & 3.0(\pm1.5) &             \\
$DR21B\_2$ &20\ 39\ 01.057 & 42\ 22\ 49.18 & 81.722 & 0.571 & -3.03 & -2.5(2) &   2.65 & 8.6 & 3.0(\pm1.5) &             \\
$DR21B\_3$ &20\ 39\ 00.374 & 42\ 24\ 37.13 & 81.744 & 0.591 &  3.56 & -2.5(2) &   3.16 & 8.6 & 3.0(\pm1.5) &             \\

\noalign{\smallskip} \hline

         \end{array}
      \]

References for the velocities: \\
1 \ Schneider et al. 2006.  \ 2 \  Plume et al. 1992.
\end{table*}

\begin{landscape}
\begin{table}
\footnotesize
\begin{flushleft}
\caption{Association of the 6.7-GHz methanol masers with the
GLIMPSE, 2MASS, MSX, and $IRAS$ point sources.}
\end{flushleft}
\label{tab4} \centering
\begin{tabular}{llrclrcccccr}
\hline\hline
Source & \multicolumn{2}{c}{2MASS} & & \multicolumn{2}{c}{GLIMPSE} & & \multicolumn{2}{c}{MSX} & & \multicolumn{2}{c}{$IRAS$} \\
\cline{2-3}\cline{5-6}\cline{8-9}\cline{11-12}
Name & Name & Separation &  & Name & Separation &  & Name & Separation &  & Name & Separation \\
     &    & (arcsec) &  &  & (arcsec)&  &  & (arcsec) &  &  & (arcsec)\\
\hline
G8.832$-$0.028    &                    &     &  & GLMC G008.8315$-$00.0278 & 0.5 &  &                     &      &  & 18024$-$2119 & 15.9 \\
G8.872$-$0.493    & 18071532$-$2130540 & 0.4 &  & GLMA G008.8721$-$00.4926 & 0.3 &  & G008.8725$-$00.4929 &  2.4 &  & 18042$-$2131 &  6.9 \\
G14.101$+$0.087   &                    &     &  & GLMC G014.1017$+$00.0872 & 1.7 &  &                     &      &  & 18128$-$1640 & 11.8 \\
G23.010$-$0.411   &                    &     &  & GLMC G023.0097$-$00.4105 & 0.3 &  &                     &      &  &              &      \\
G23.207$-$0.378   &                    &     &  & GLMC G023.2075$-$00.3772 & 3.1 &  &                     &      &  &              &      \\
G24.148$-$0.009   &                    &     &  & GLMC G024.1479$-$00.0091 & 0.3 &  &                     &      &  & 18326$-$0751 & 11.3 \\
G24.329$+$0.144   &                    &     &  & GLMC G024.3285$+$00.1440 & 0.4 &  &                     &      &  & 18324$-$0737 &  5.2 \\
G27.220$+$0.260   &                    &     &  & GLMC G027.2197$+$00.2607 & 0.9 &  &                     &      &  &              &      \\
G27.365$-$0.166   &                    &     &  & GLMA G027.3650$-$00.1656 & 0.8 &  &                     &      &  & 18391$-$0504 &  2.3 \\
G29.863$-$0.044   & 18455955$-$0245061 & 1.8 &  & GLMA G029.8623$-$00.0442 & 2.1 &  & G029.8620$-$00.0444 &  3.5 &  &              &      \\
G30.199$-$0.169   & 18470306$-$0230361 & 2.6 &  & GLMA G030.1981$-$00.1691 & 2.5 &  & G030.1981$-$00.1691 &  2.4 &  &              &      \\
G30.225$-$0.180   & 18470824$-$0229302 & 3.3 &  & extended emission        &     &  &                     &      &  &              &      \\
G30.899$+$0.162   &                    &     &  & GLMC G030.8988$+$00.1615 & 1.1 &  &                     &      &  &              &      \\
L1287             & 00364719$+$6329058 & 3.8 &  &                          &     &  &                     &      &  & 00338$+$6312 &  1.0 \\
NGC281-N          & 00522425$+$5633471 & 4.0 &  &                          &     &  &                     &      &  & 00494$+$5617 &  4.3 \\
S231              & 05391330$+$3545538 & 3.8 &  &                          &     &  & G173.4815$+$02.4459 &  1.4 &  &              &      \\
AFGL5180          & 06085340$+$2138281 & 1.3 &  &                          &     &  &                     &      &  & 06058$+$2138 & 12.0 \\
AFGL6366          & 06084091$+$2131056 & 3.6 &  &                          &     &  &                     &      &  & 06056$+$2131 &  7.6 \\
S255              & 06125385$+$1759242 & 2.5 &  &                          &     &  & G192.6005$-$00.0479 &  0.8 &  & 06099$+$1800 & 10.0 \\
S269              & 06143706$+$1349364 & 0.4 &  &                          &     &  & G196.4542$-$01.6777 &  1.8 &  & 06117$+$1350 &  7.1 \\
18556$+$0136      &                    &     &  & GLMA G035.1973$-$00.7430 & 1.1 &  & G035.1979$-$00.7427 &  3.7 &  & 18556$+$0136 &  1.6 \\
G43.15$+$0.02     & 19101091$+$0905176 & 3.5 &  & GLMC G043.1480$+$00.0131 & 3.5 &  & G043.1492$+$00.0130 &  1.1 &  &              &      \\
19120$+$0917      & 19142616$+$0922346 & 3.8 &  & extended emission        &     &  & G043.8896$-$00.7835 &  3.9 &  & 19120$+$0917 &  3.8 \\
19186$+$1440      &                    &     &  & GLMC G049.4152$+$00.3253 & 2.1 &  &                     &      &  &              &      \\
W51e2             &                    &     &  & GLMC G049.4892$-$00.3879 & 2.3 &  &                     &      &  &              &      \\
19303$+$1651      & 19323607$+$1657384 & 0.1 &  & GLMC G052.6625$-$01.0919 & 0.4 &  &                     &      &  & 19303$+$1651 & 12.1 \\
ON1A              & 20100886$+$3131392 & 4.8 &  &                          &     &  & G069.5395$-$00.9754 &  1.2 &  & 20081$+$3122 &  1.4 \\
ON1B              & 20100886$+$3131392 & 4.2 &  &                          &     &  & G069.5395$-$00.9754 &  2.0 &  & 20081$+$3122 &  1.8 \\
20290$+$4052      & 20305058$+$4102298 & 2.5 &  &                          &     &  & G079.7358$+$00.9905 &  0.9 &  & 20290$+$4052 &  2.9 \\
21381$+$5000      & 21395825$+$5014209 & 0.1 &  &                          &     &  & G094.6028$-$01.7966 &  2.7 &  & 21381$+$5000 &  7.0 \\
L1206             &                    &     &  &                          &     &  &                     &      &  & 22272$+$6358A&  5.8 \\
DR20              & 20370116$+$4134545 & 2.6 &  &                          &     &  & G080.8624$+$00.3827 &  4.6 &  & 20352$+$4124 &  2.2 \\
W75               &                    &     &  &                          &     &  &                     &      &  &              &      \\
DR21B\_1          & 20390200$+$4225008 & 1.6 &  &                          &     &  & G081.7522$+$00.5906 &  0.7 &  &              &      \\
DR21B\_2          & 20390101$+$4222502 & 1.2 &  &                          &     &  & G081.7220$+$00.5699 &  4.1 &  &              &      \\
DR21B\_3          & 20390047$+$4224369 & 1.2 &  &                          &     &  &                     &      &  &              &      \\

\hline
\end{tabular}
\end{table}
\end{landscape}

\clearpage

\clearpage
\begin{figure*}
\centering
\includegraphics[angle=0,width=19cm]{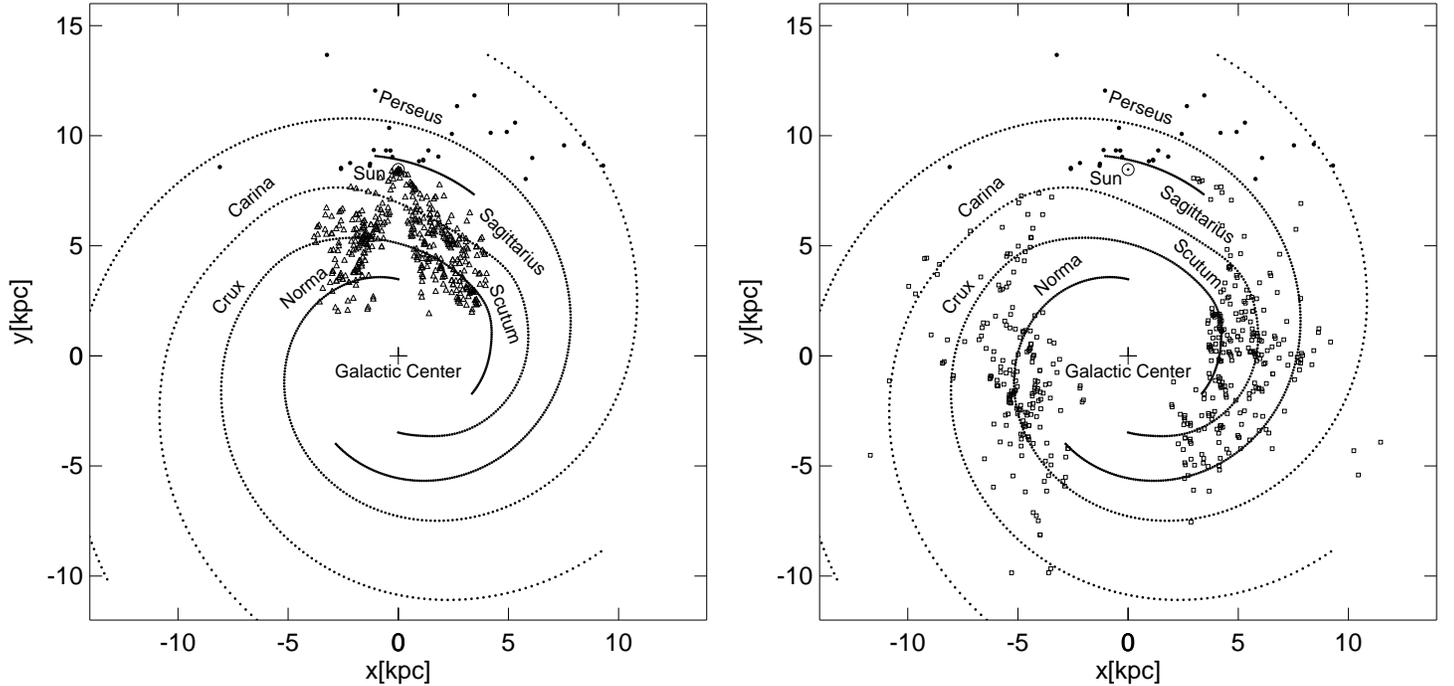}
\caption{Positions of 495 6.7-GHz masers at the near (open
triangles), far (open squares) kinematic distances, and of the outer
Galaxy (filled circles) in the Galactic plane. The distances
calculated using the Galactic rotation model of Wouterloot \& Brand
(1989), assuming $R_{0}$ = 8.5 kpc and $\Theta_{0}$ = 220 km
s$^{-1}$. There is a poor correspondence between the spiral arm
models and massive star-forming regions for both the near and far
distances and also in the outer Galaxy. The positions of the spiral
arms are taken from Cordes \& Lazio (2002). We do not show error
bars for clarity.}
\end{figure*}

\begin{figure*}
\centering
\includegraphics[angle=0,width=19cm]{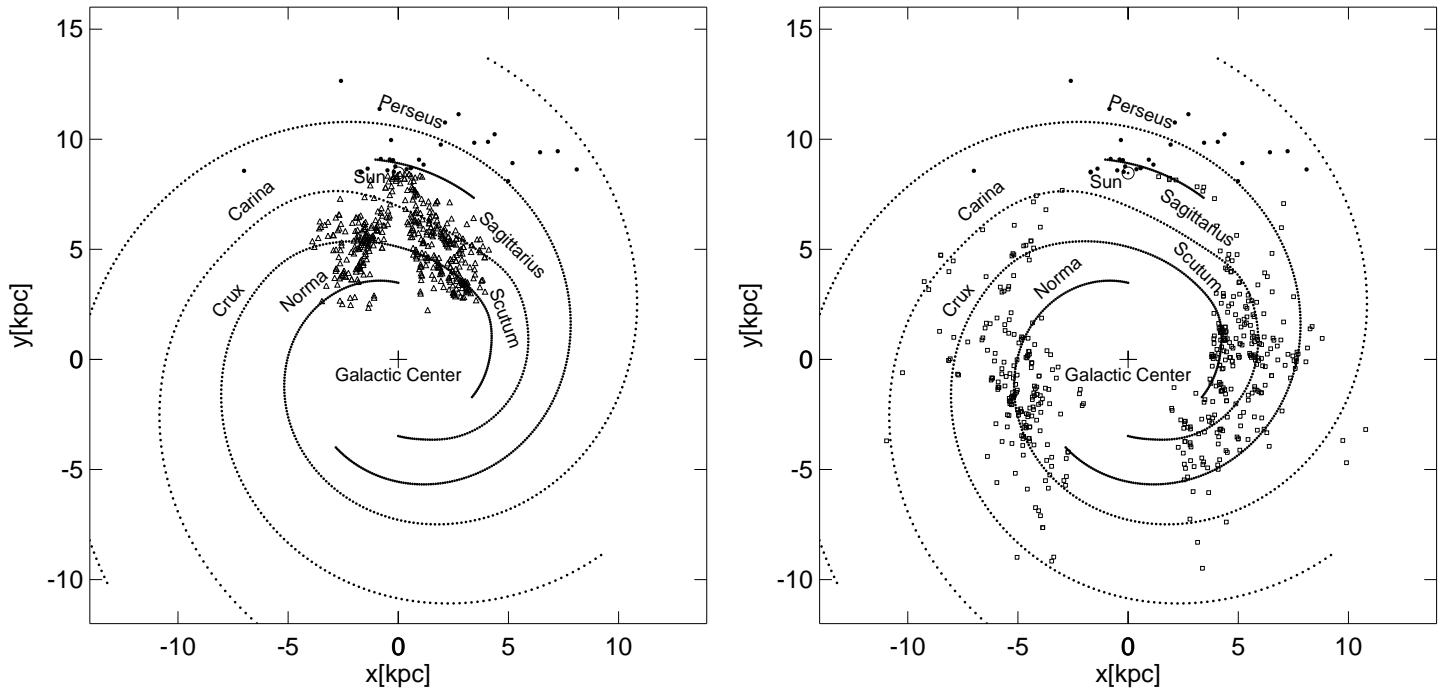}
\caption{Same as in Fig. 1, but the revised distances from Reid et
al. (2009) have been used. There is little difference between Figs.
1 and 2 in the inner Galaxy, but there is a far closer agreement
between the model and the data in the Perseus arm region of the
outer Galaxy.}
\end{figure*}


\clearpage
\Online
\begin{figure*}
\begin{tabular}{cccc}
\includegraphics[width=5cm]{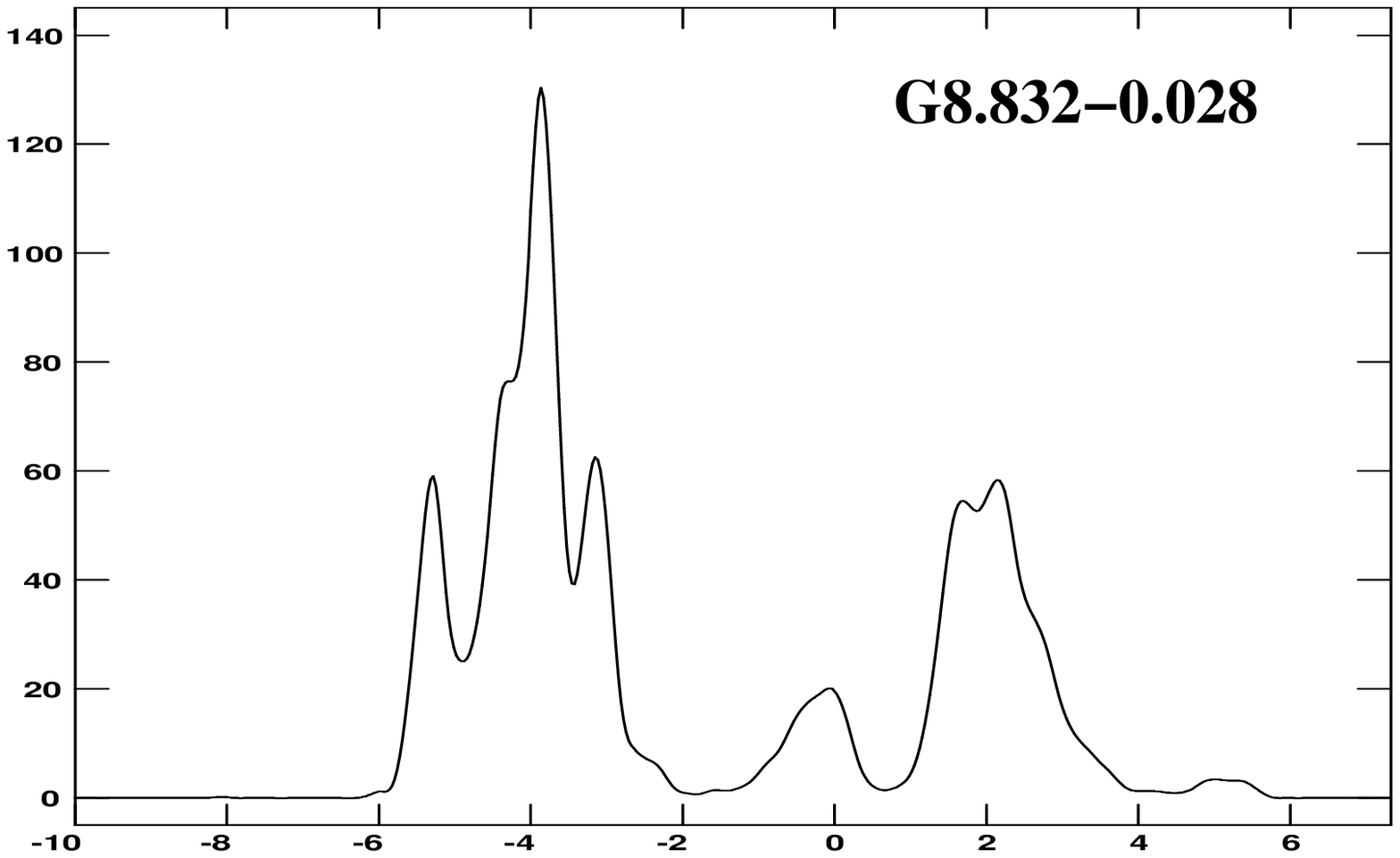} &
\includegraphics[width=5cm]{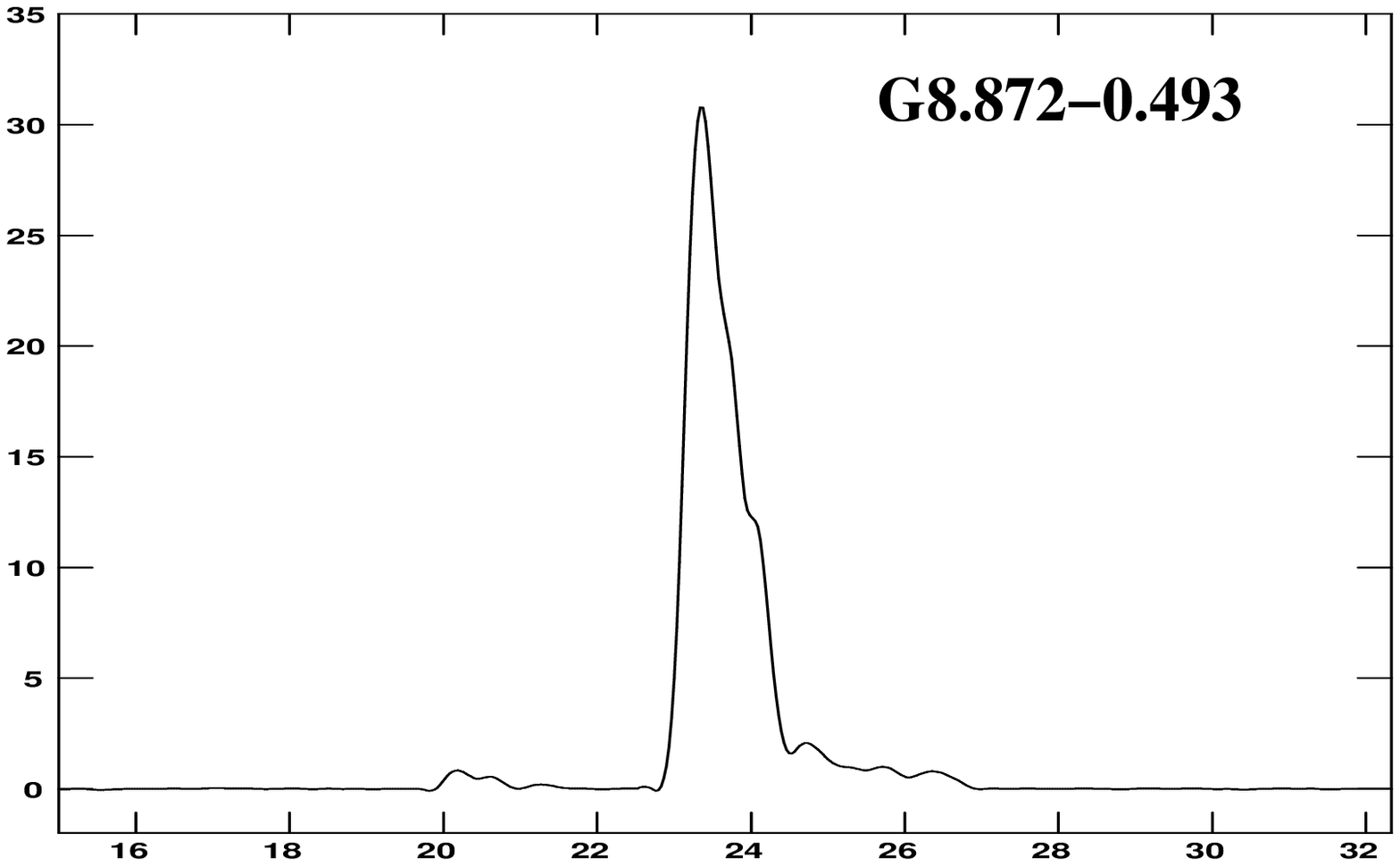} &
\includegraphics[width=5cm]{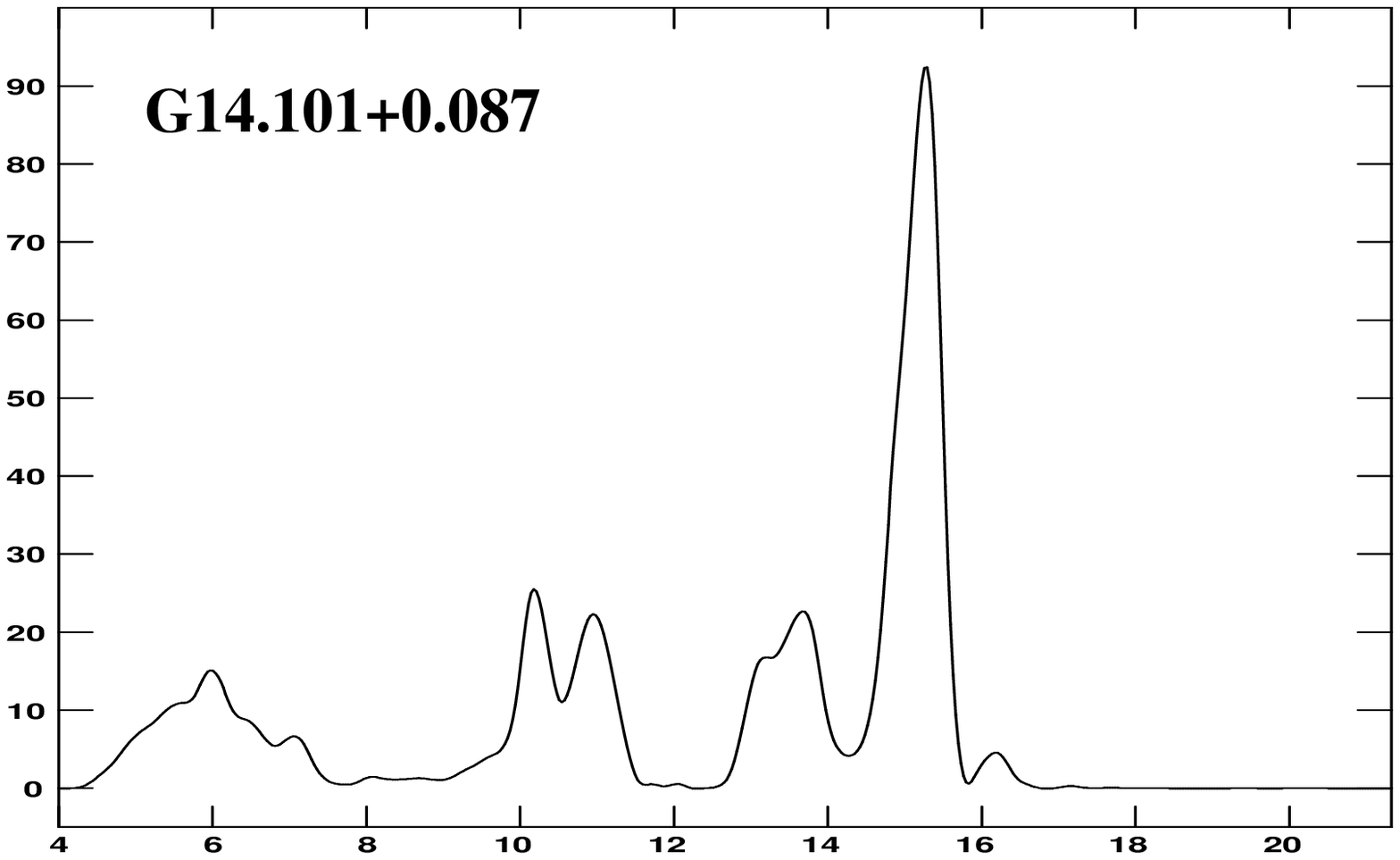} &
\\
\includegraphics[width=5cm]{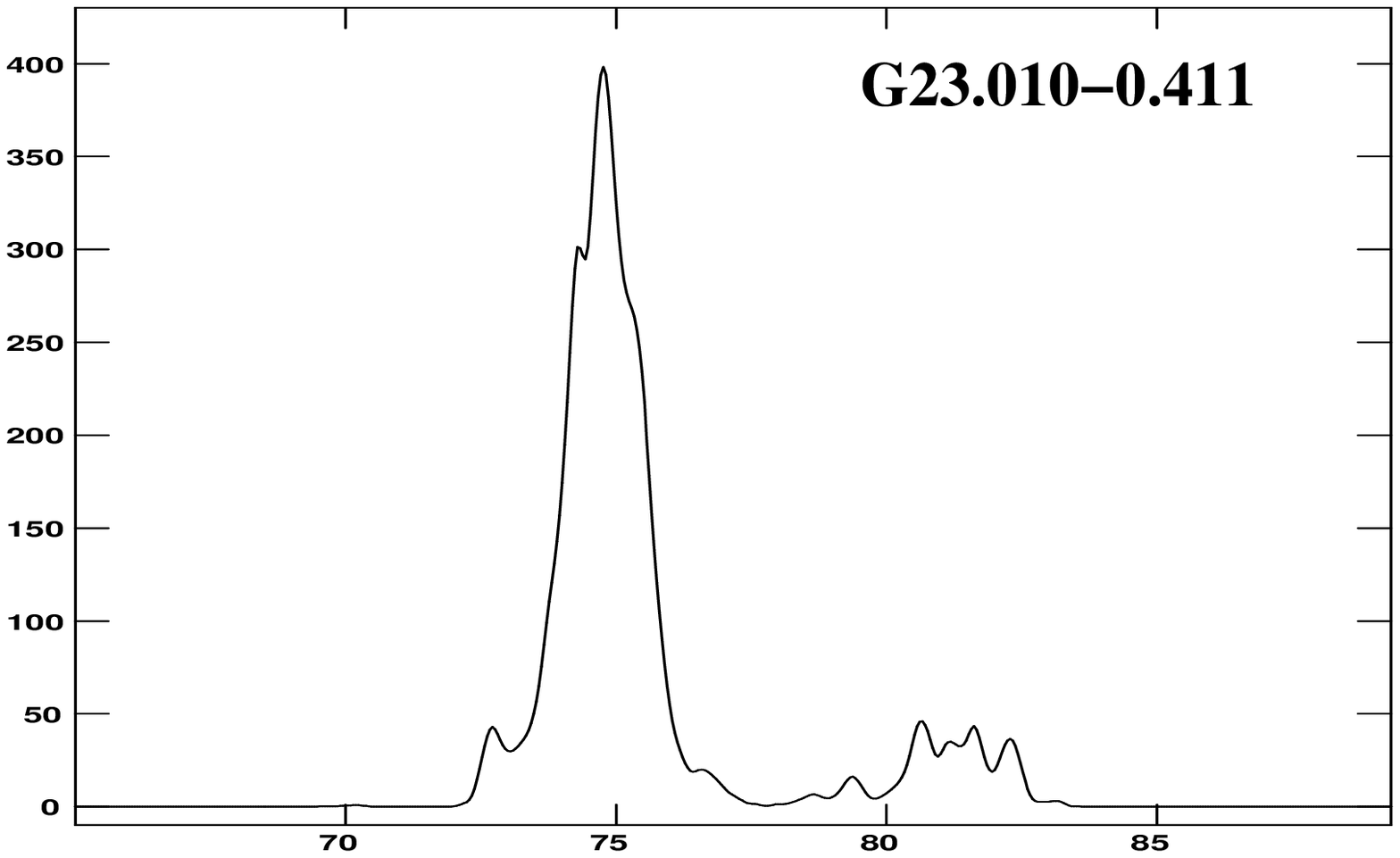} &
\includegraphics[width=5cm]{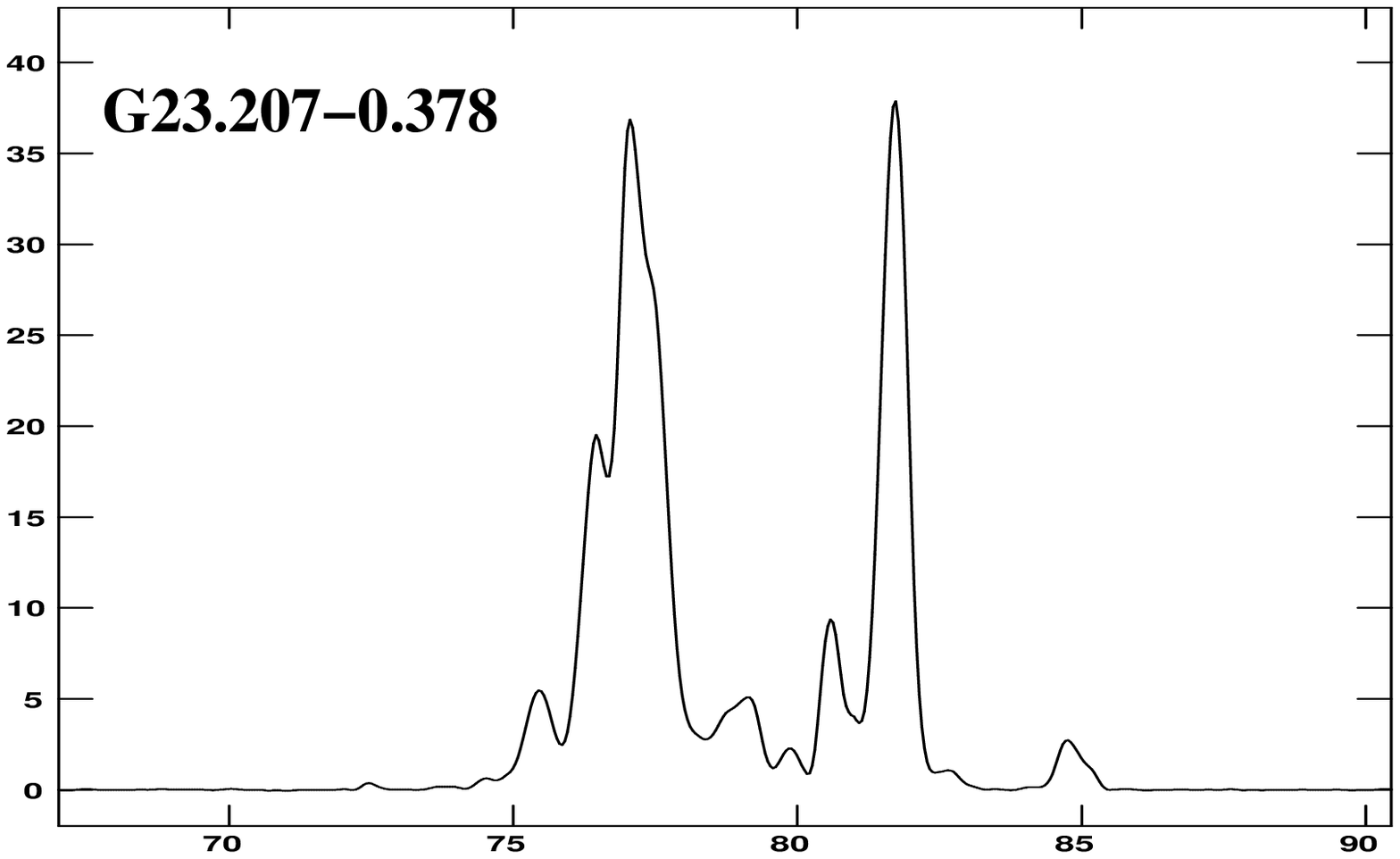} &
\includegraphics[width=5cm]{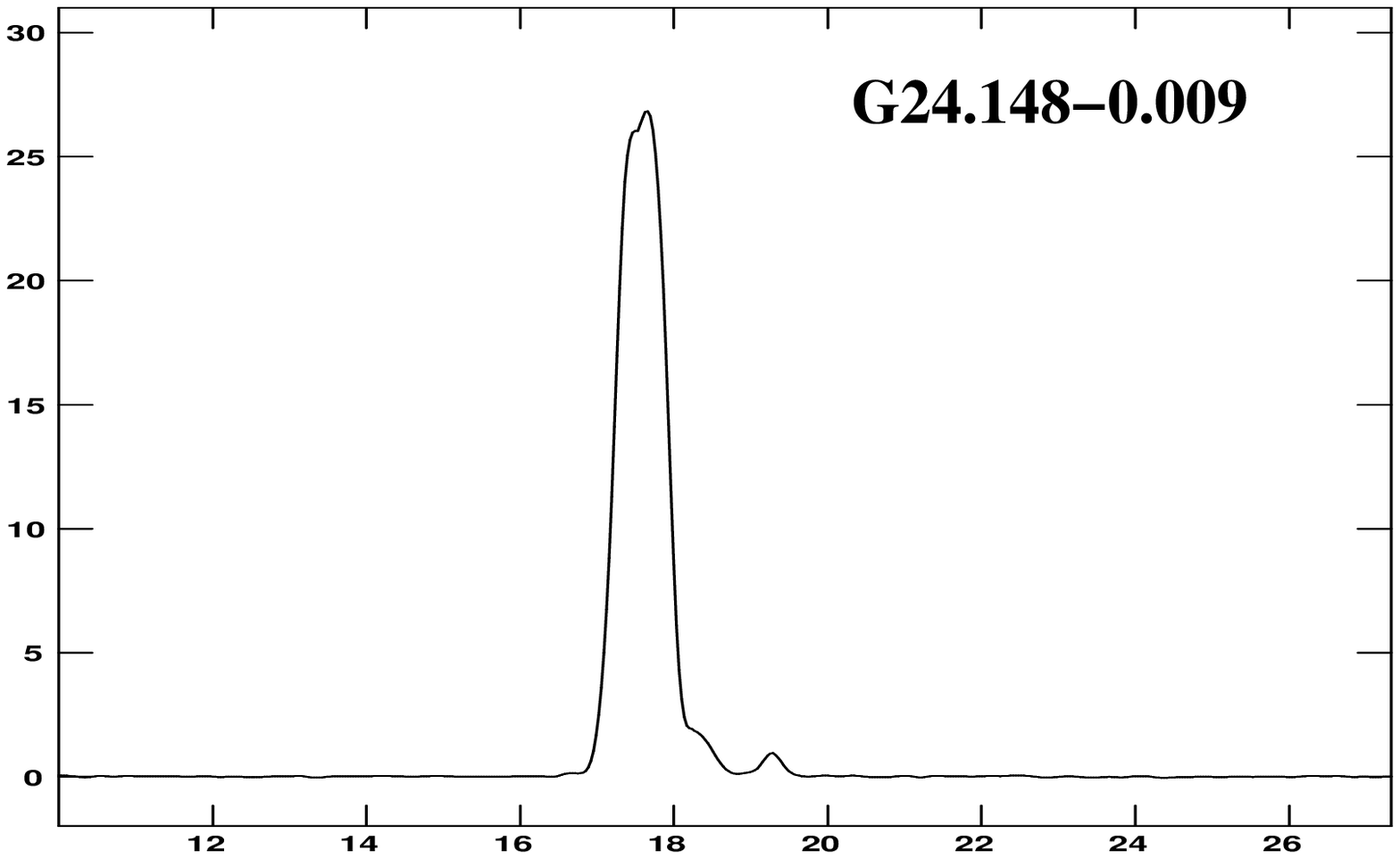} &
\\
\includegraphics[width=5cm]{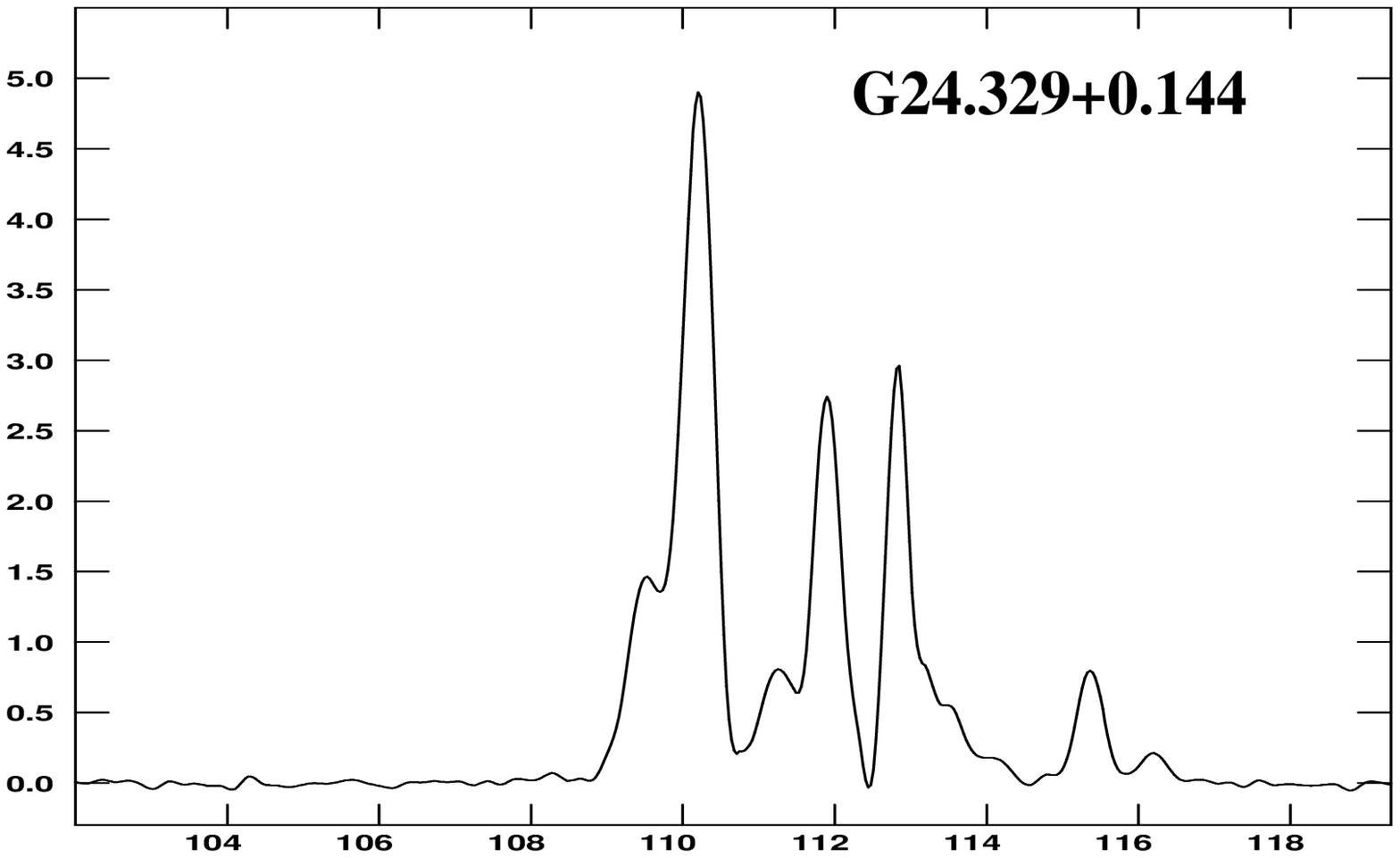} &
\includegraphics[width=5cm]{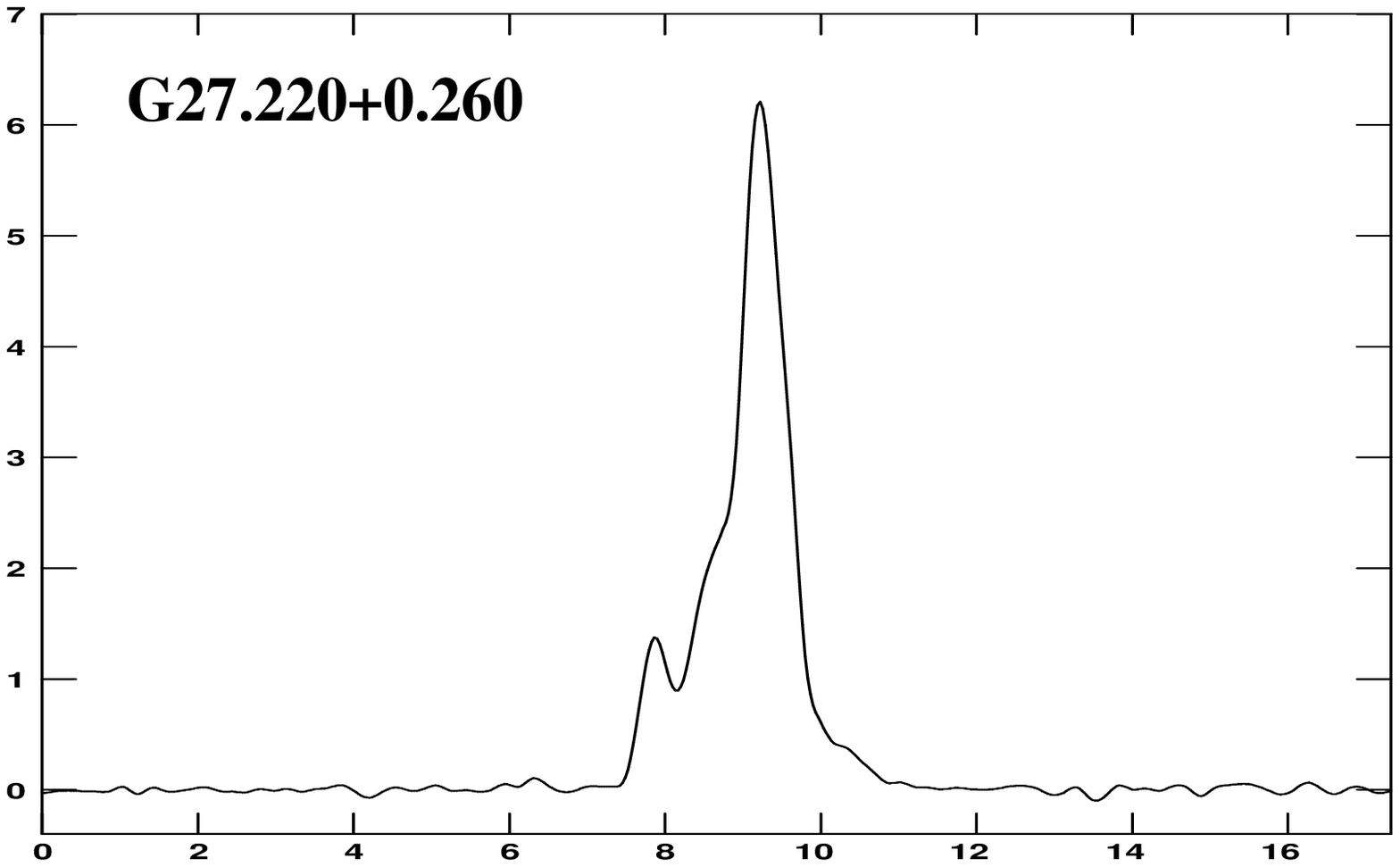} &
\includegraphics[width=5cm]{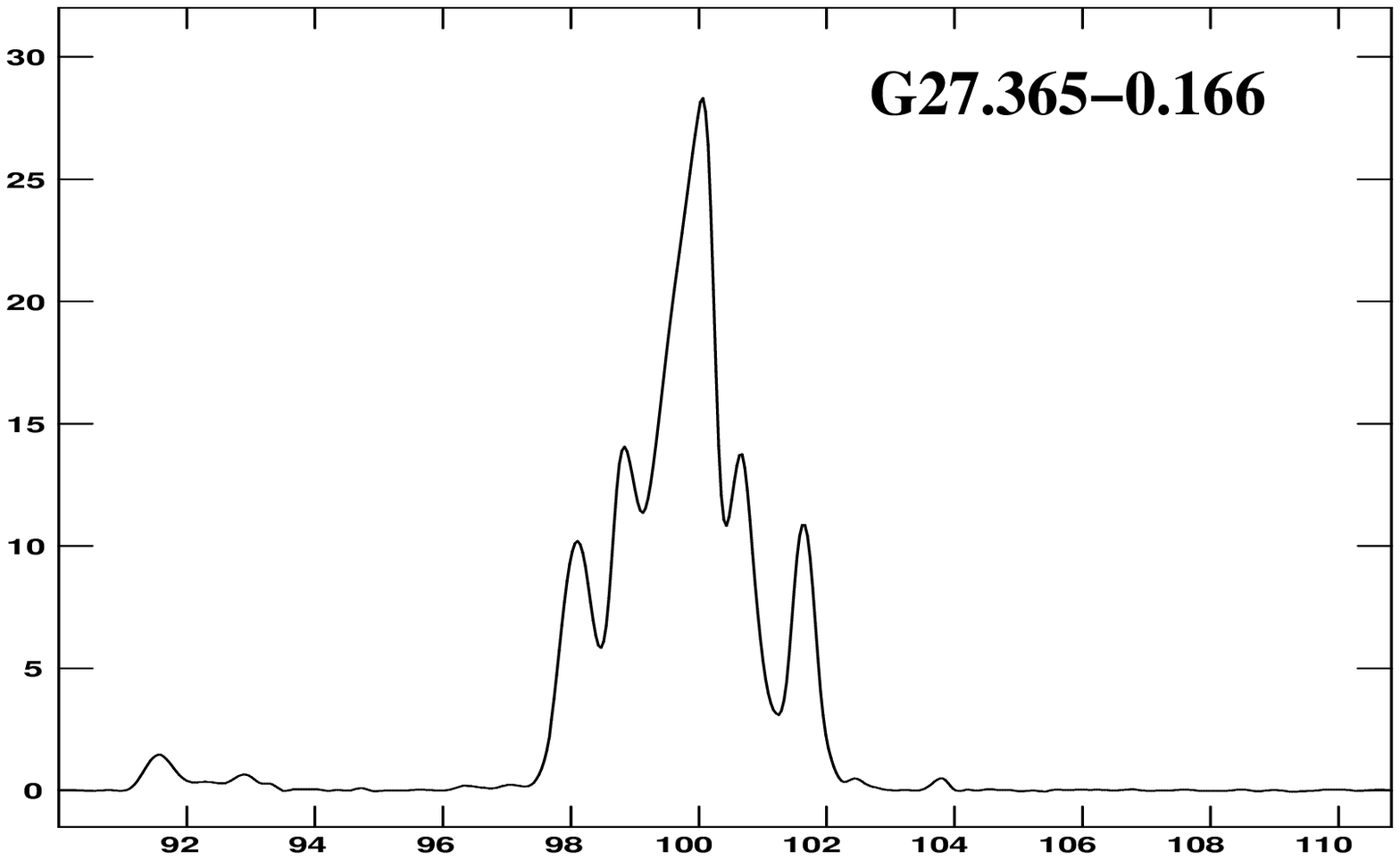} &
\\
\includegraphics[width=5cm]{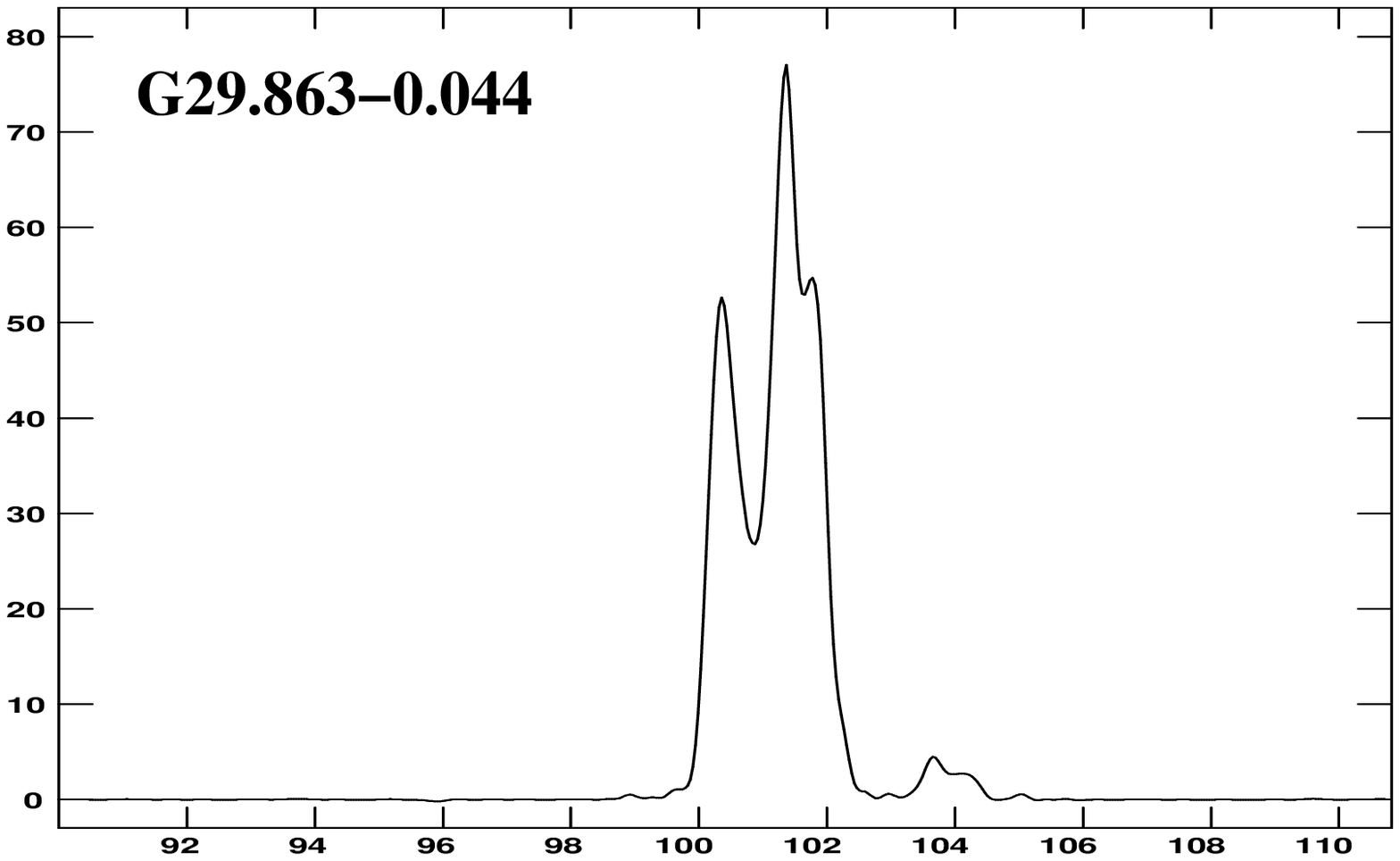} &
\includegraphics[width=5cm]{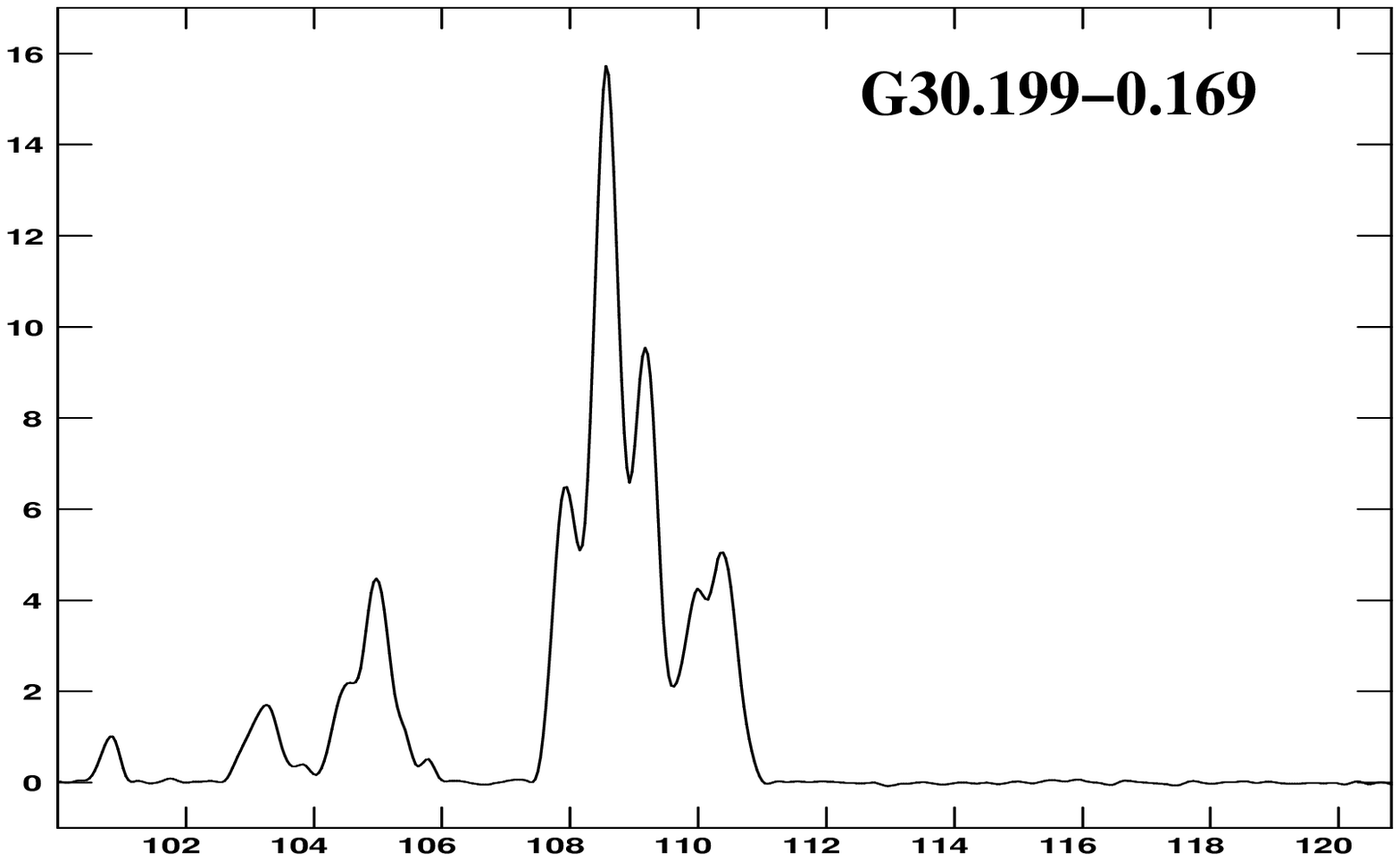} &
\includegraphics[width=5cm]{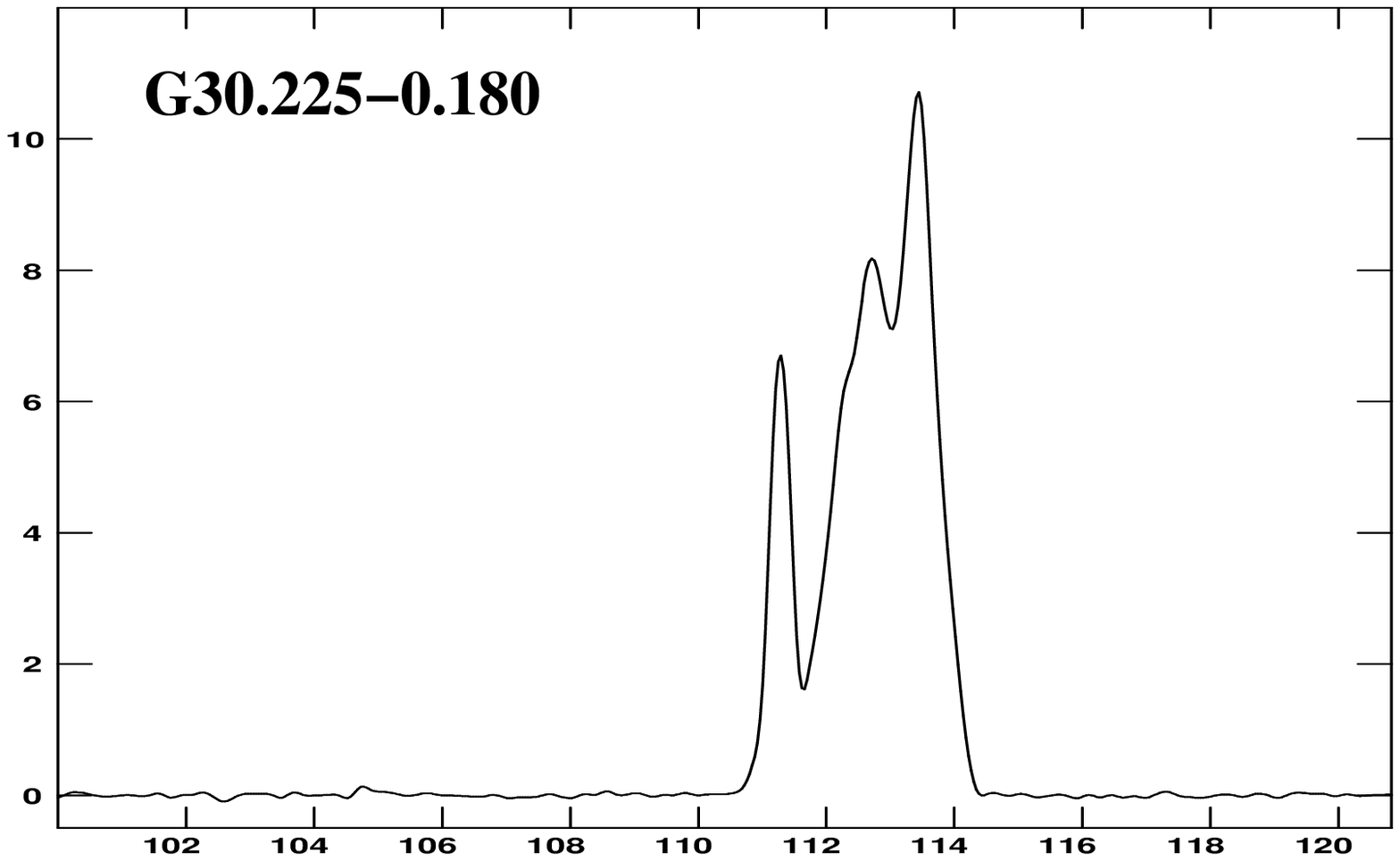} &
\\
\includegraphics[width=5cm]{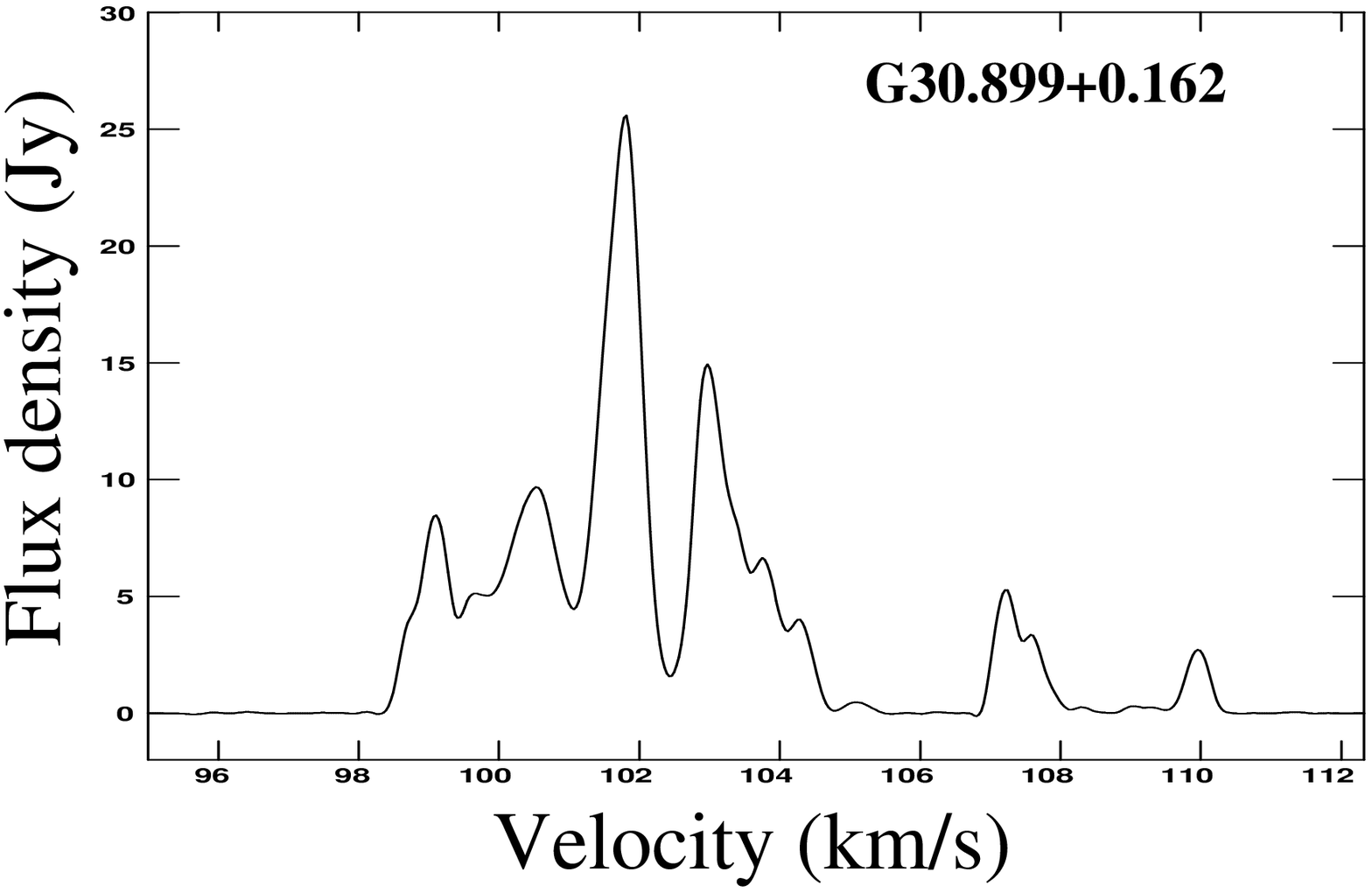} &
\end{tabular}
\caption{Spectra of the 6.7-GHz methanol masers from the ATCA
observations. The spectral resolution is approximately 0.18
km~s$^{-1}$.}
\end{figure*}

\begin{figure*}
\begin{tabular}{cccc}
\includegraphics[width=5cm]{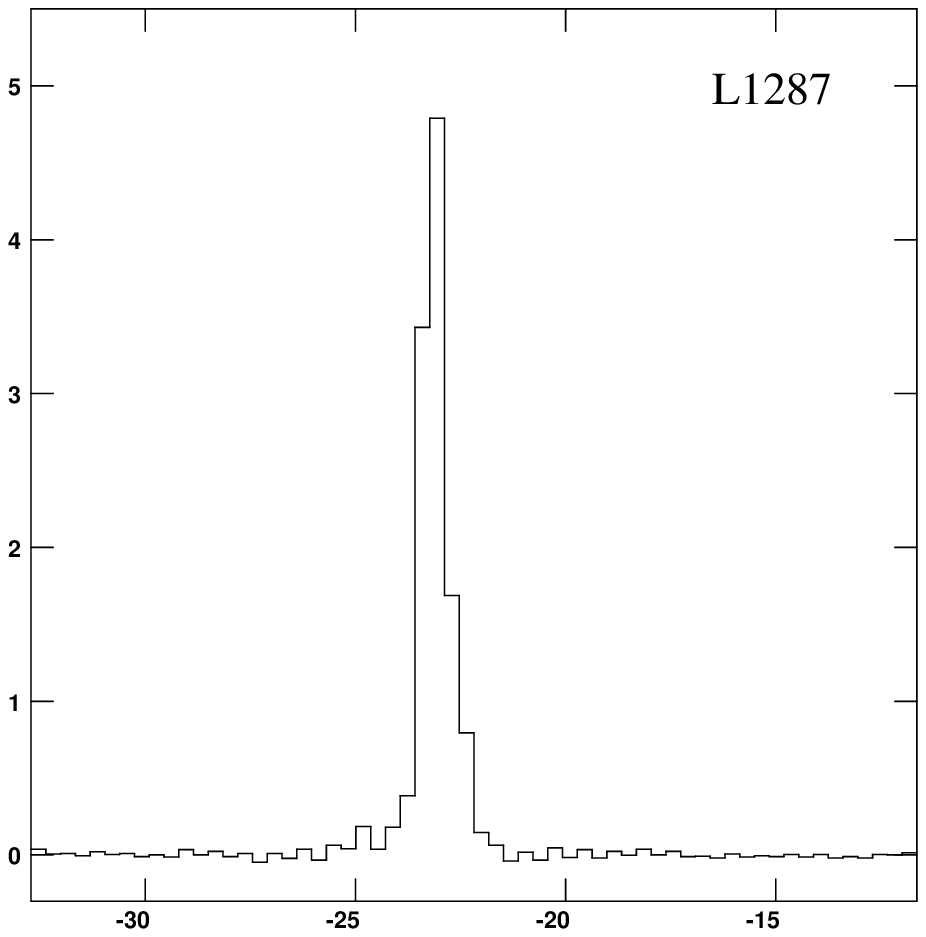} &
\includegraphics[width=5cm]{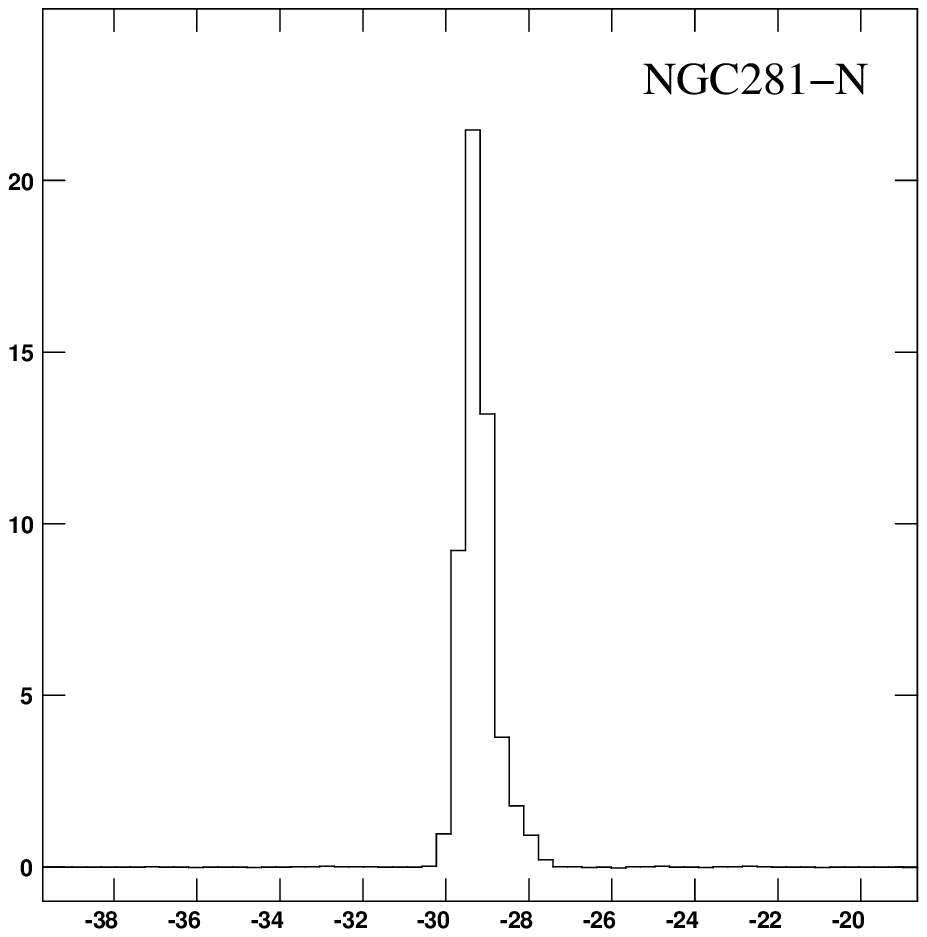} &
\includegraphics[width=5cm]{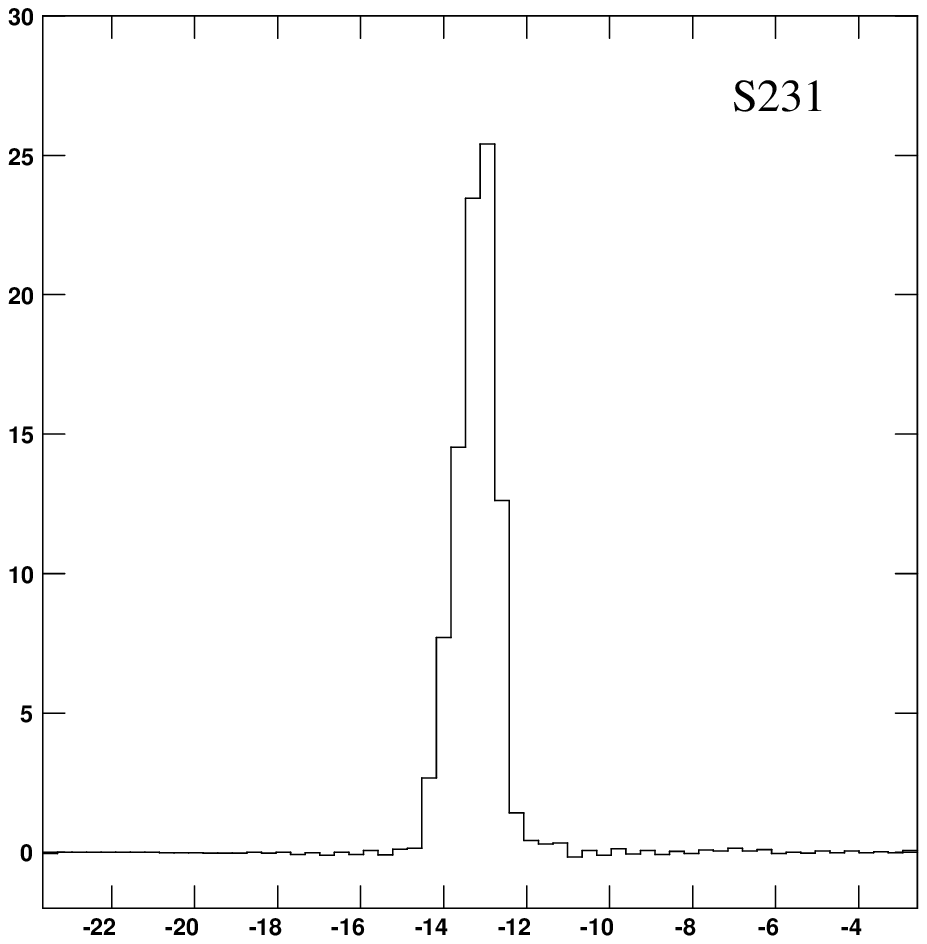} &
\\
\includegraphics[width=5cm]{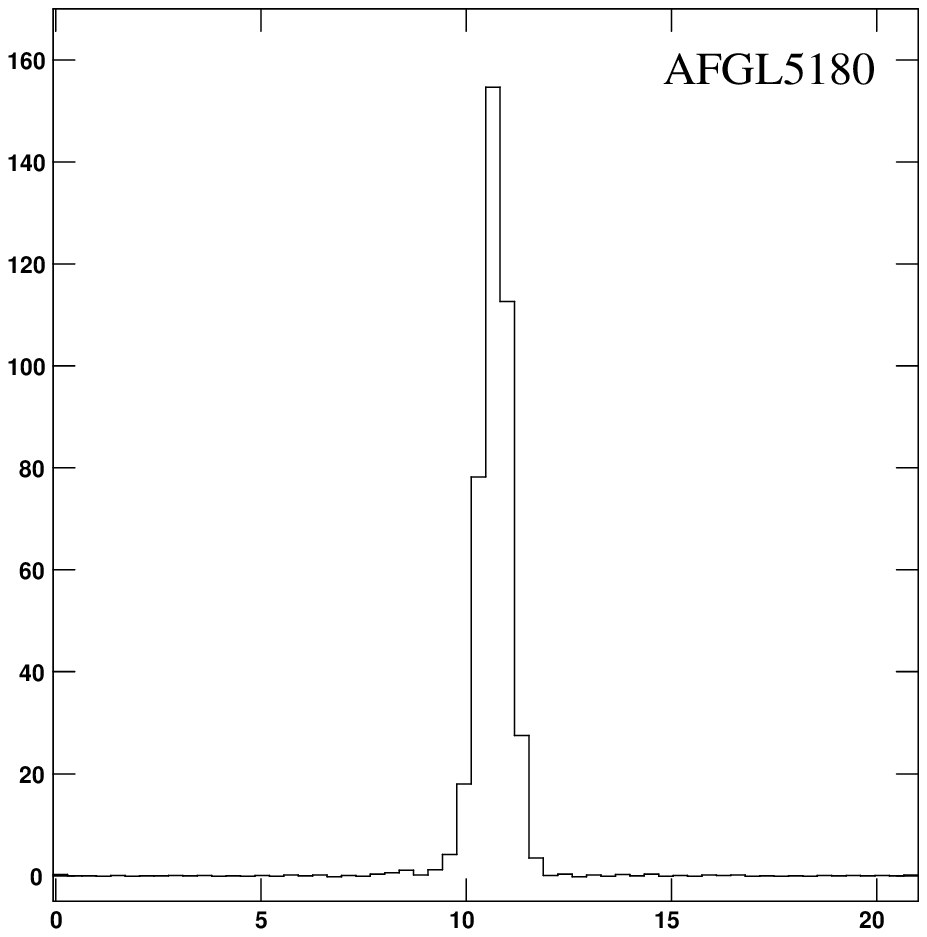} &
\includegraphics[width=5cm]{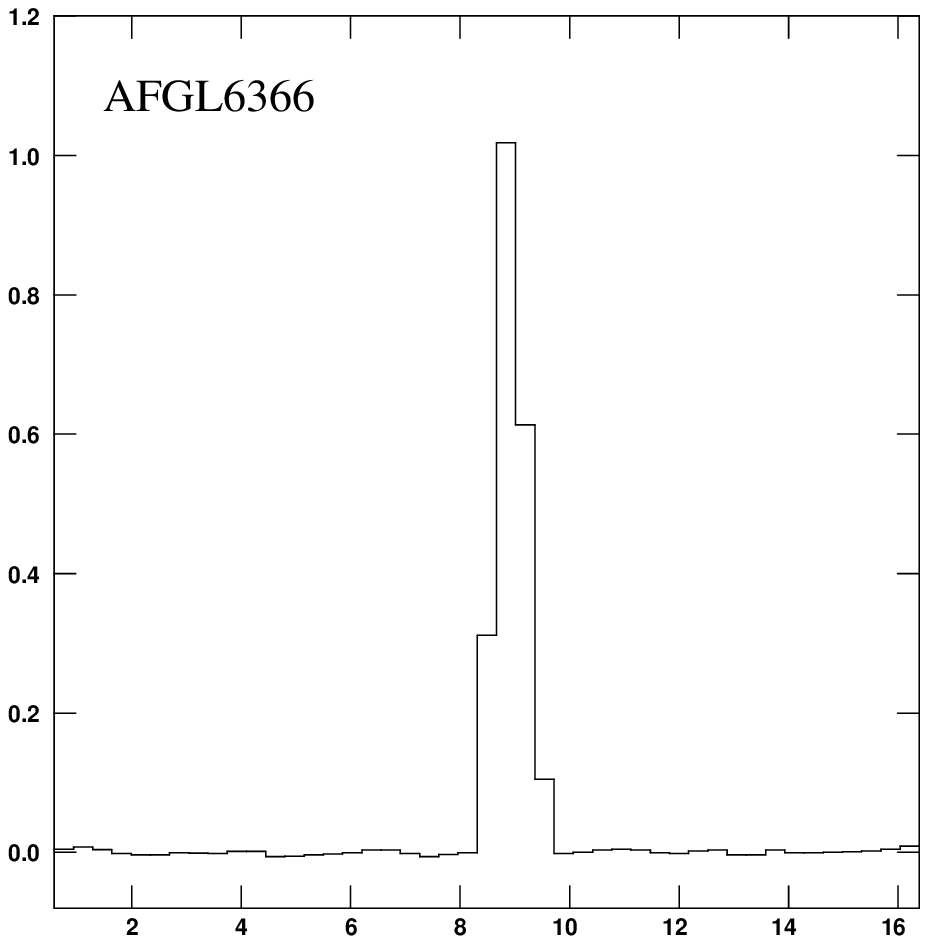} &
\includegraphics[width=5cm]{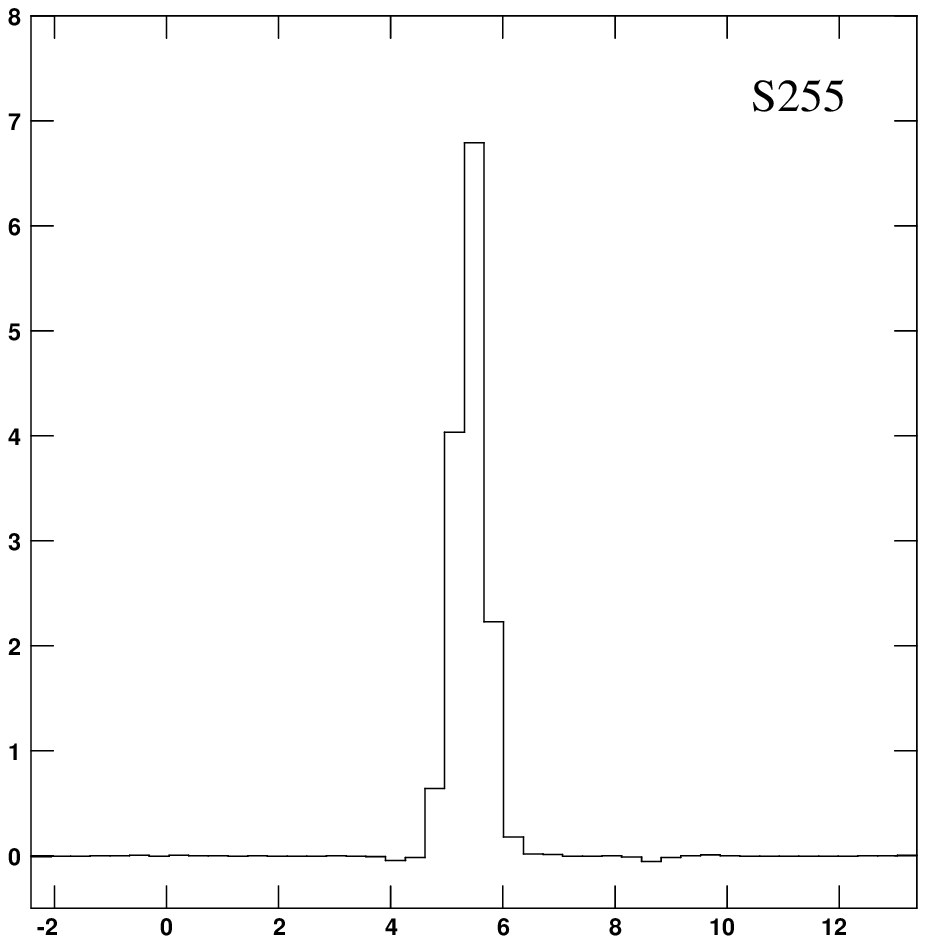} &
\\
\includegraphics[width=5cm]{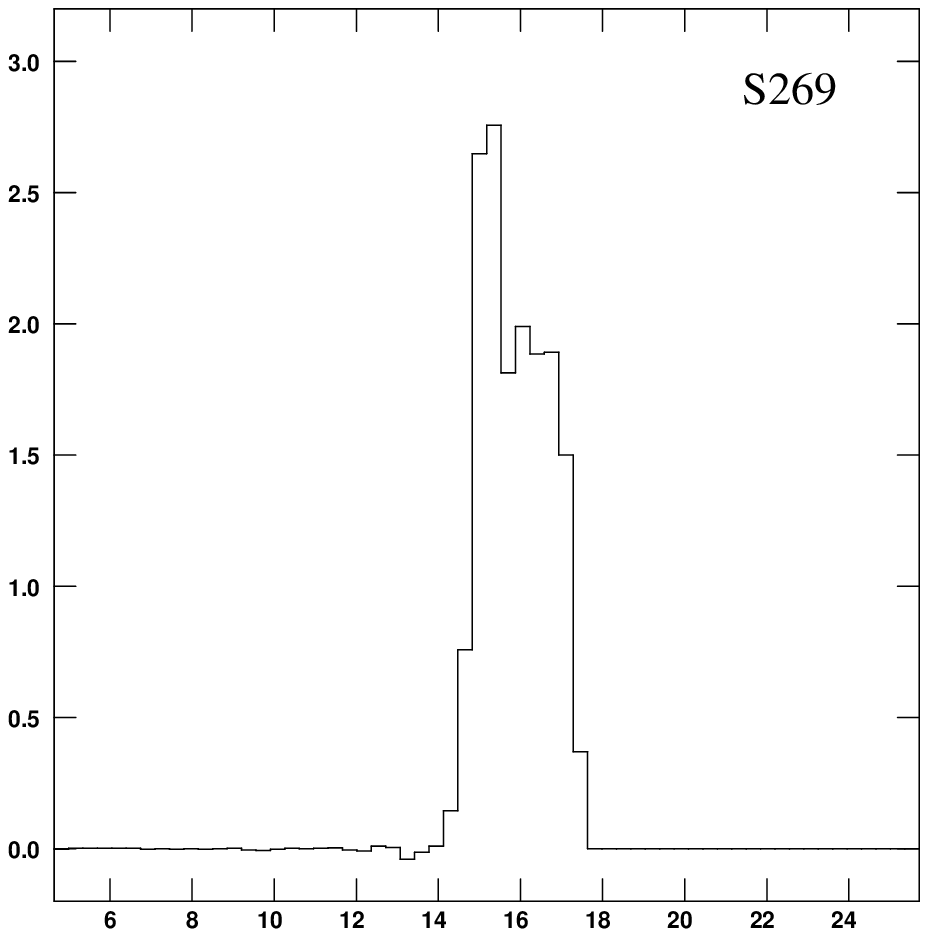} &
\includegraphics[width=5cm]{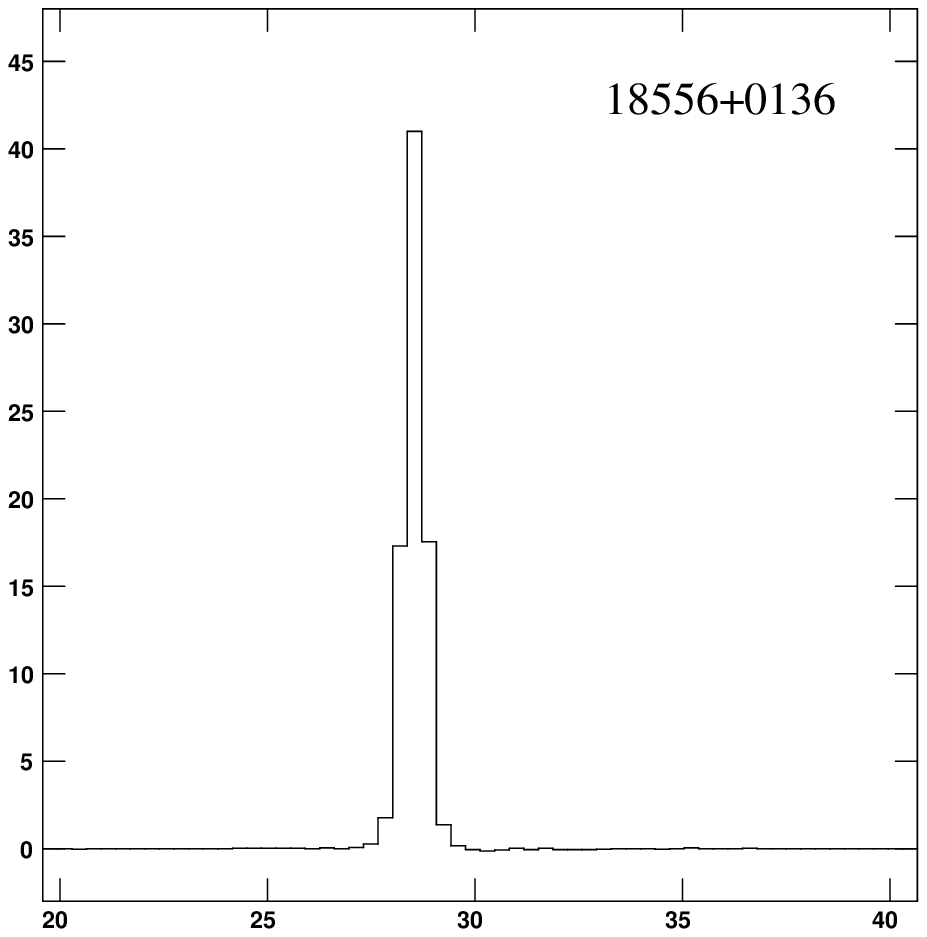} &
\includegraphics[width=5cm]{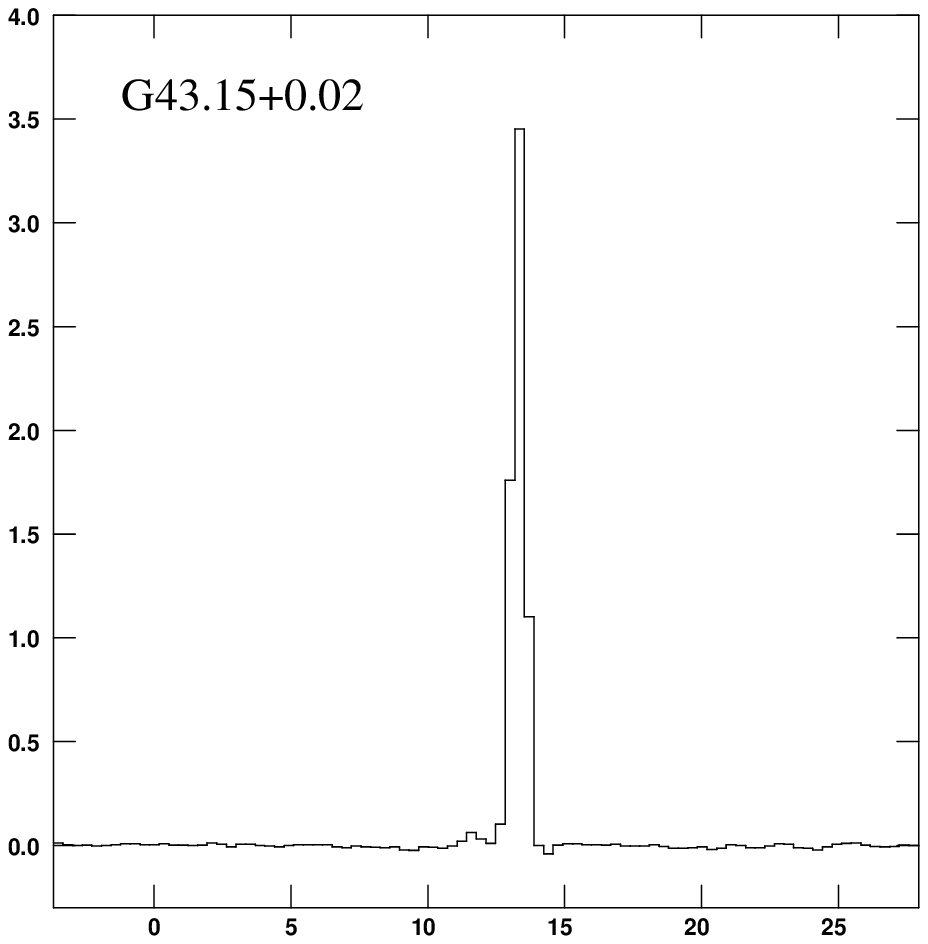} &
\\
\includegraphics[width=5cm]{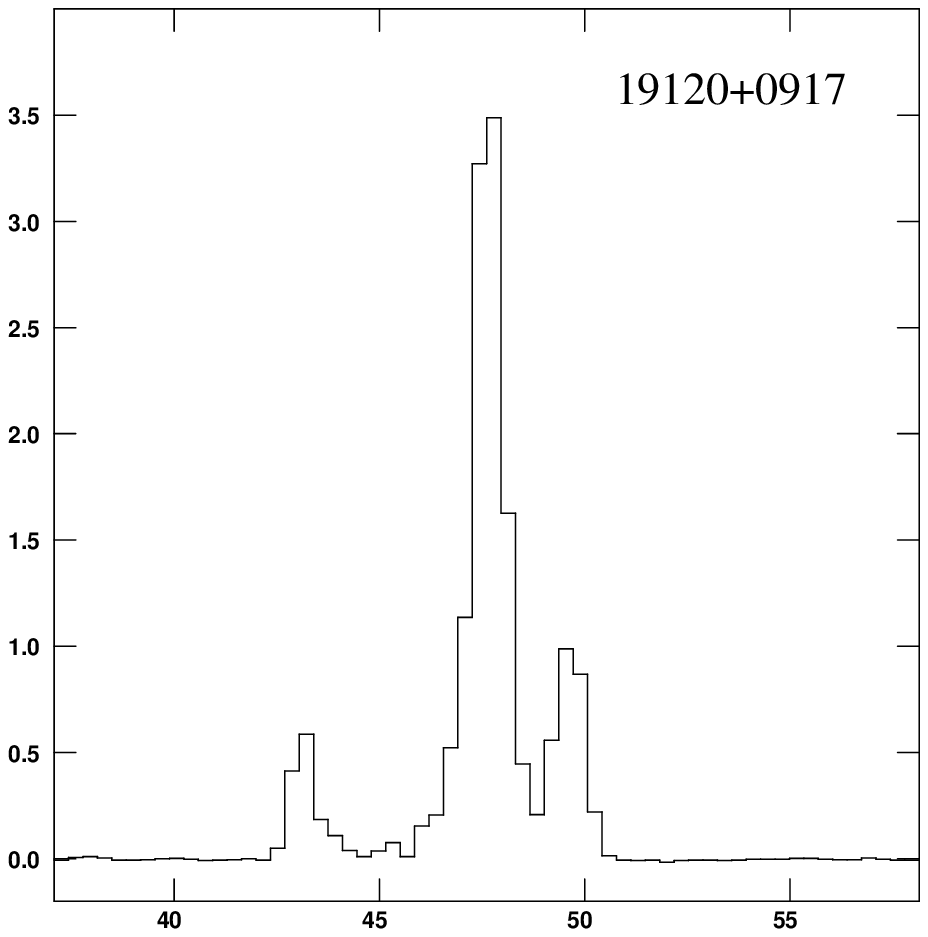} &
\includegraphics[width=5cm]{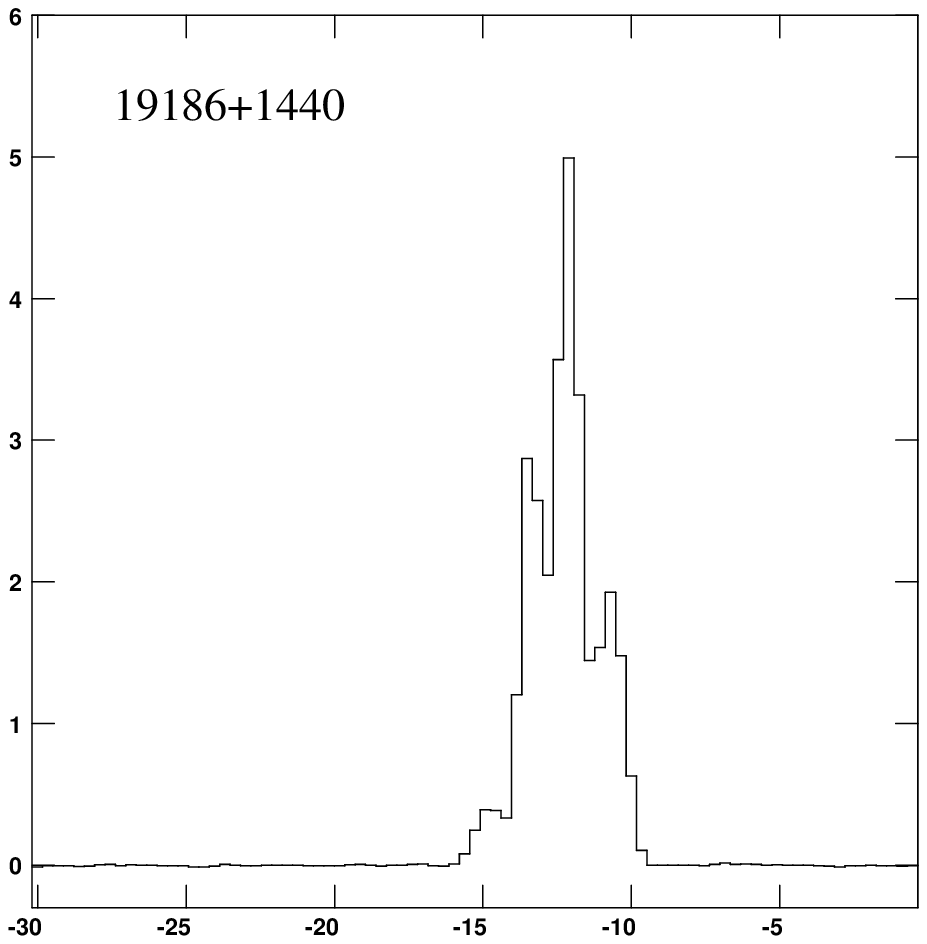} &
\includegraphics[width=5cm]{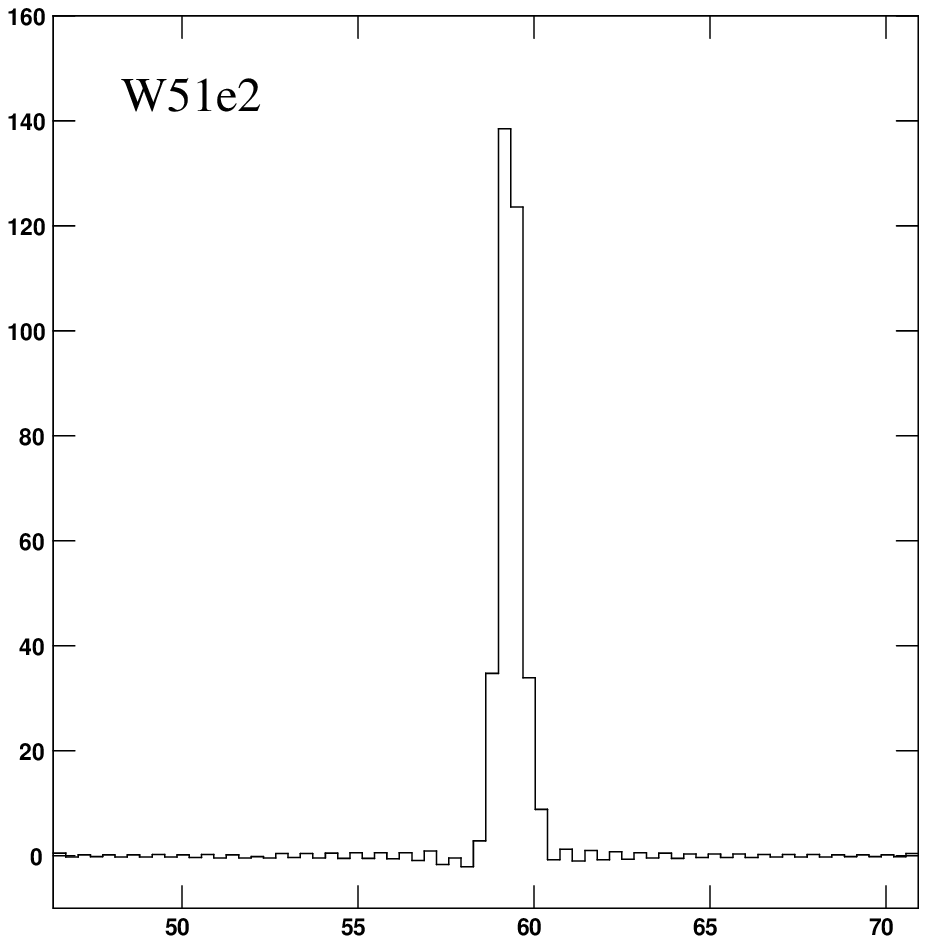} &
\end{tabular}
\end{figure*}

\begin{figure*}
\begin{tabular}{cccc}
\includegraphics[width=5cm]{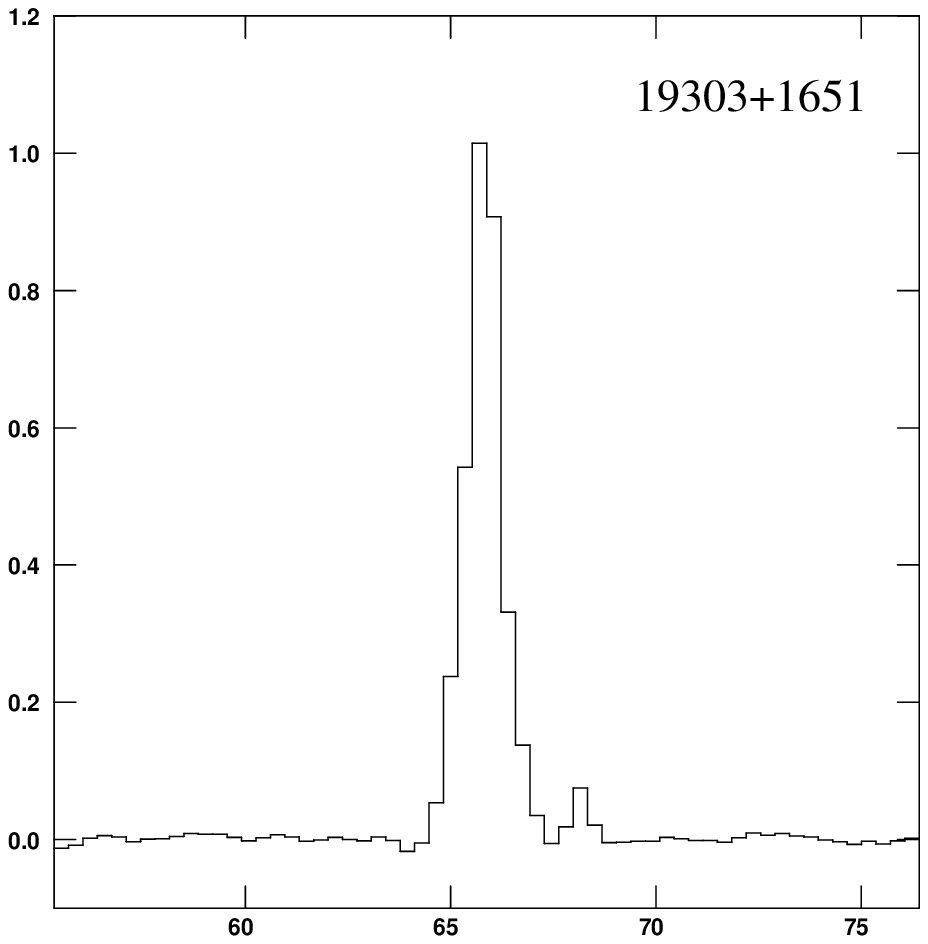} &
\includegraphics[width=5cm]{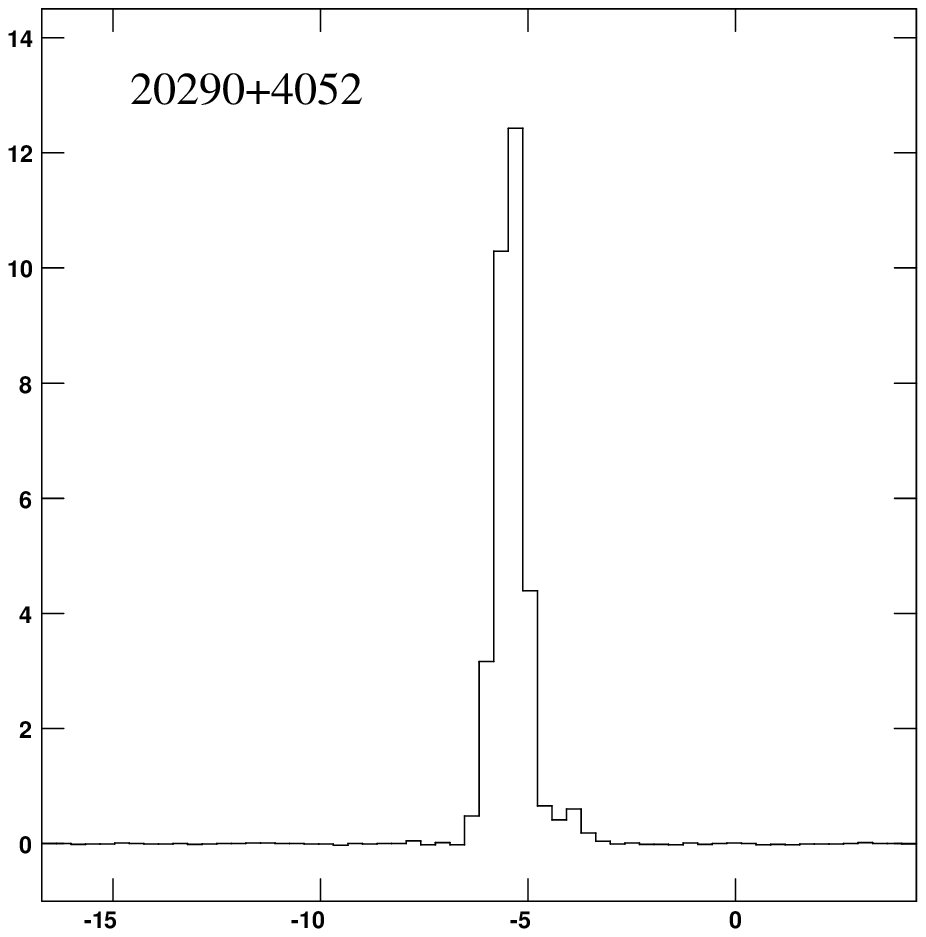} &
\includegraphics[width=5cm]{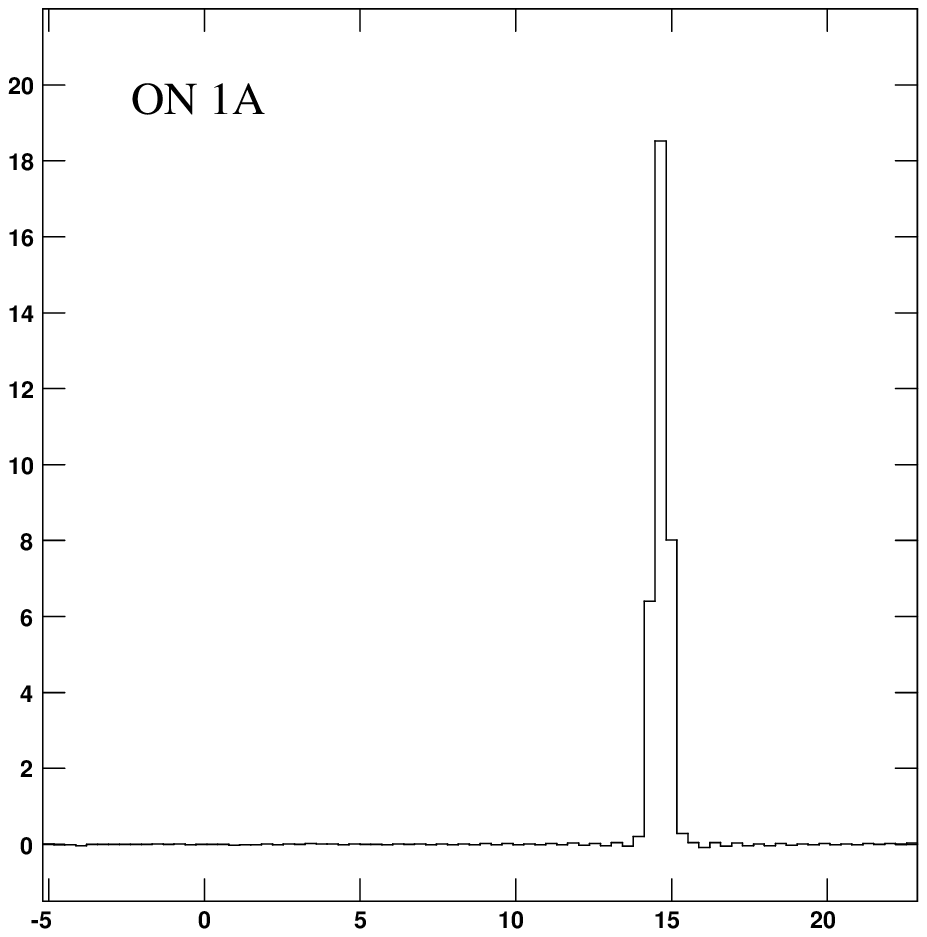}  &
\\
\includegraphics[width=5cm]{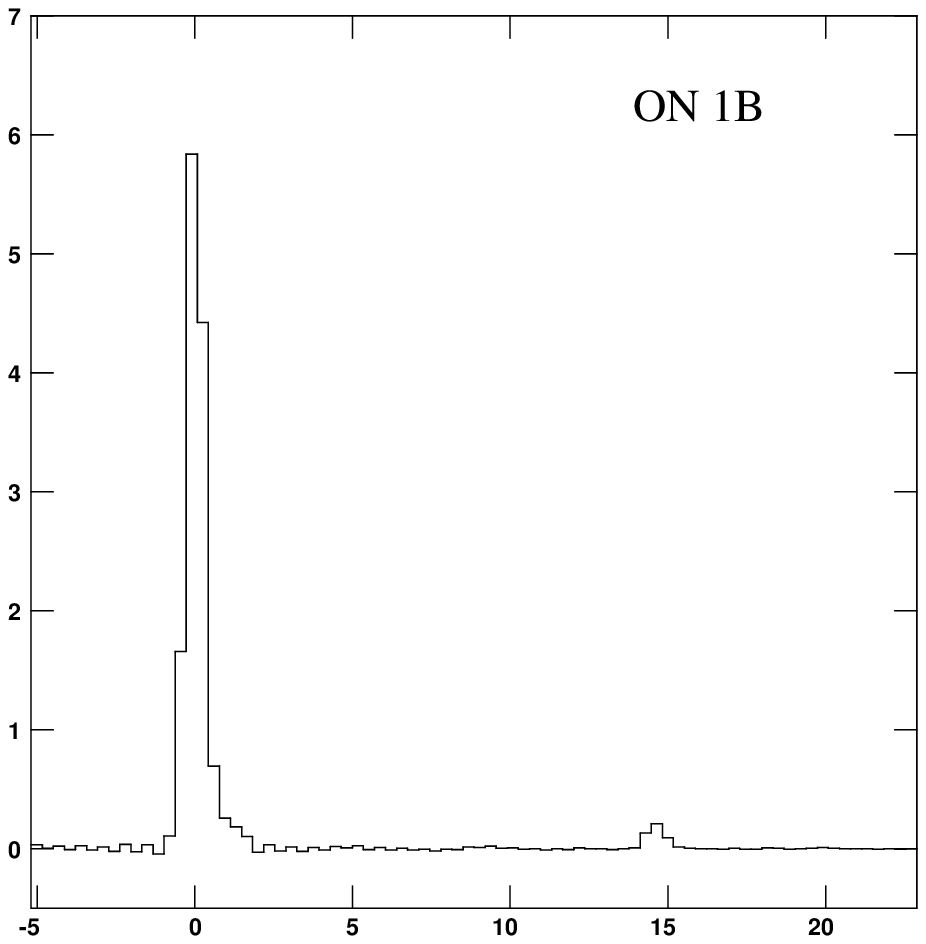}  &
\includegraphics[width=5cm]{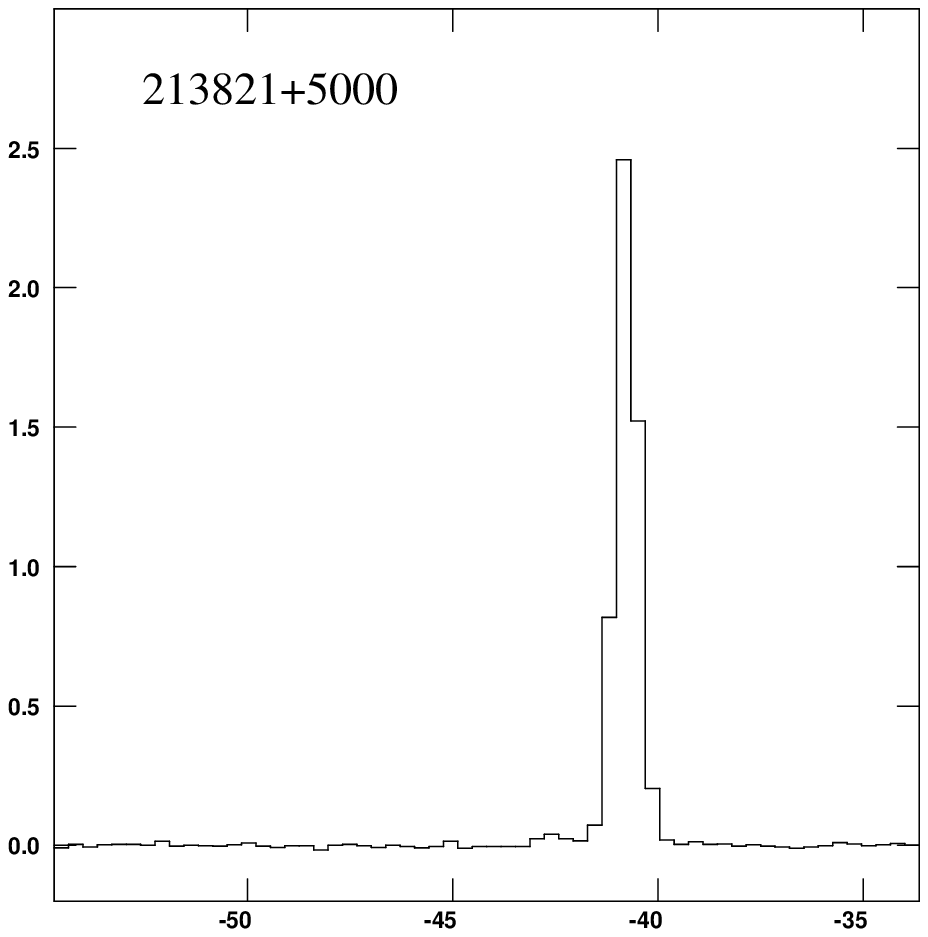} &
\includegraphics[width=5cm]{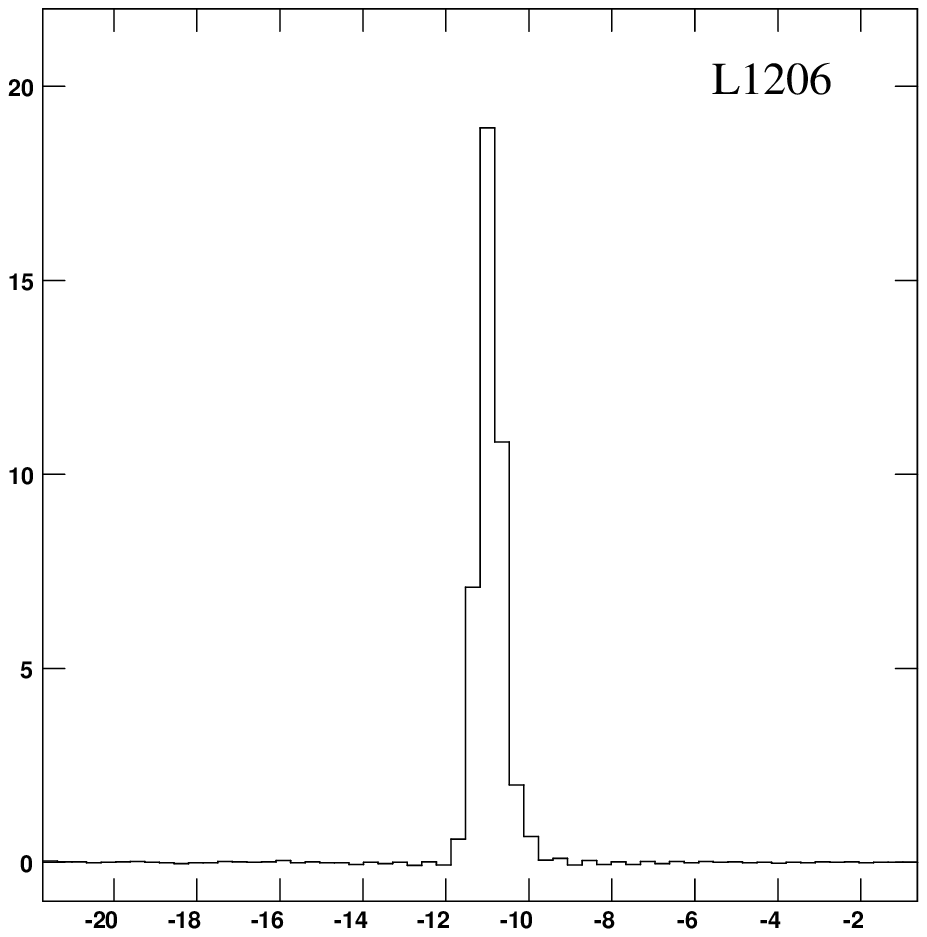} &
\end{tabular}
\caption{Same as fig. 3, but for the spectra obtained with the
MERLIN. The spectral resolution is approximately 0.7 km~s$^{-1}$.}
\end{figure*}

\begin{figure*}
\begin{tabular}{cccc}
\includegraphics[width=5cm]{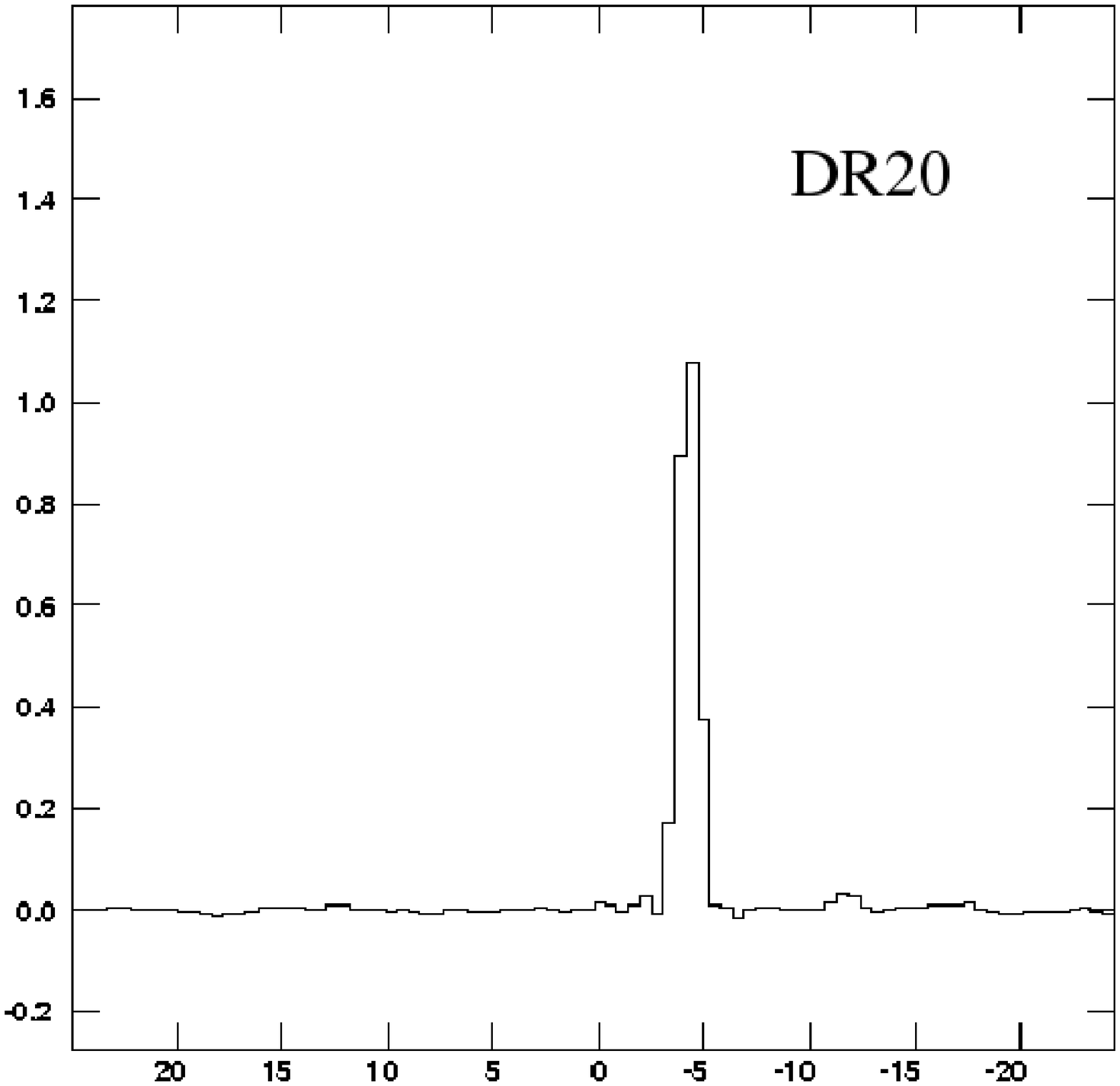} &
\includegraphics[width=5cm]{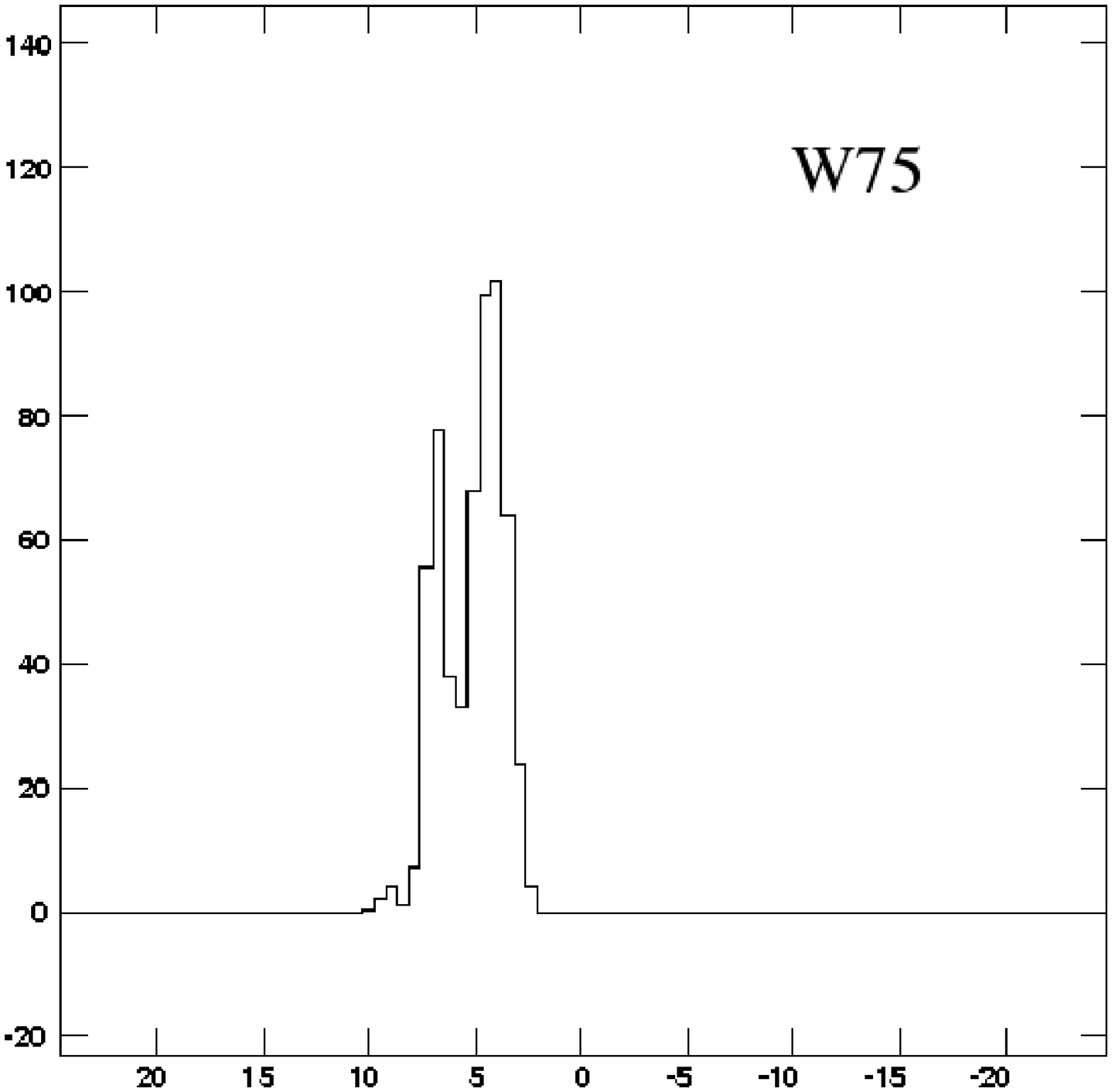} &
\includegraphics[width=5cm]{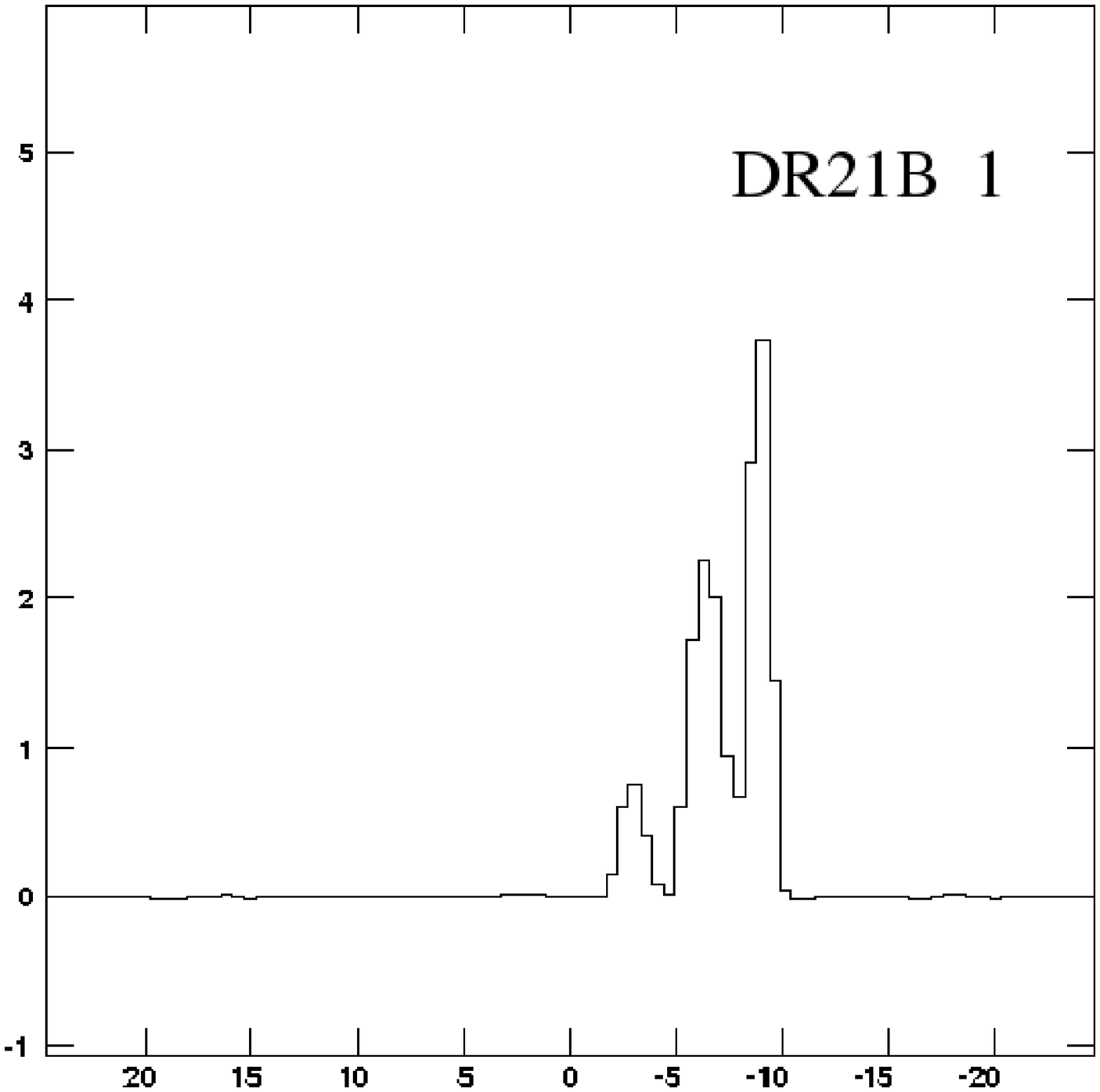} &
\\
\includegraphics[width=5cm]{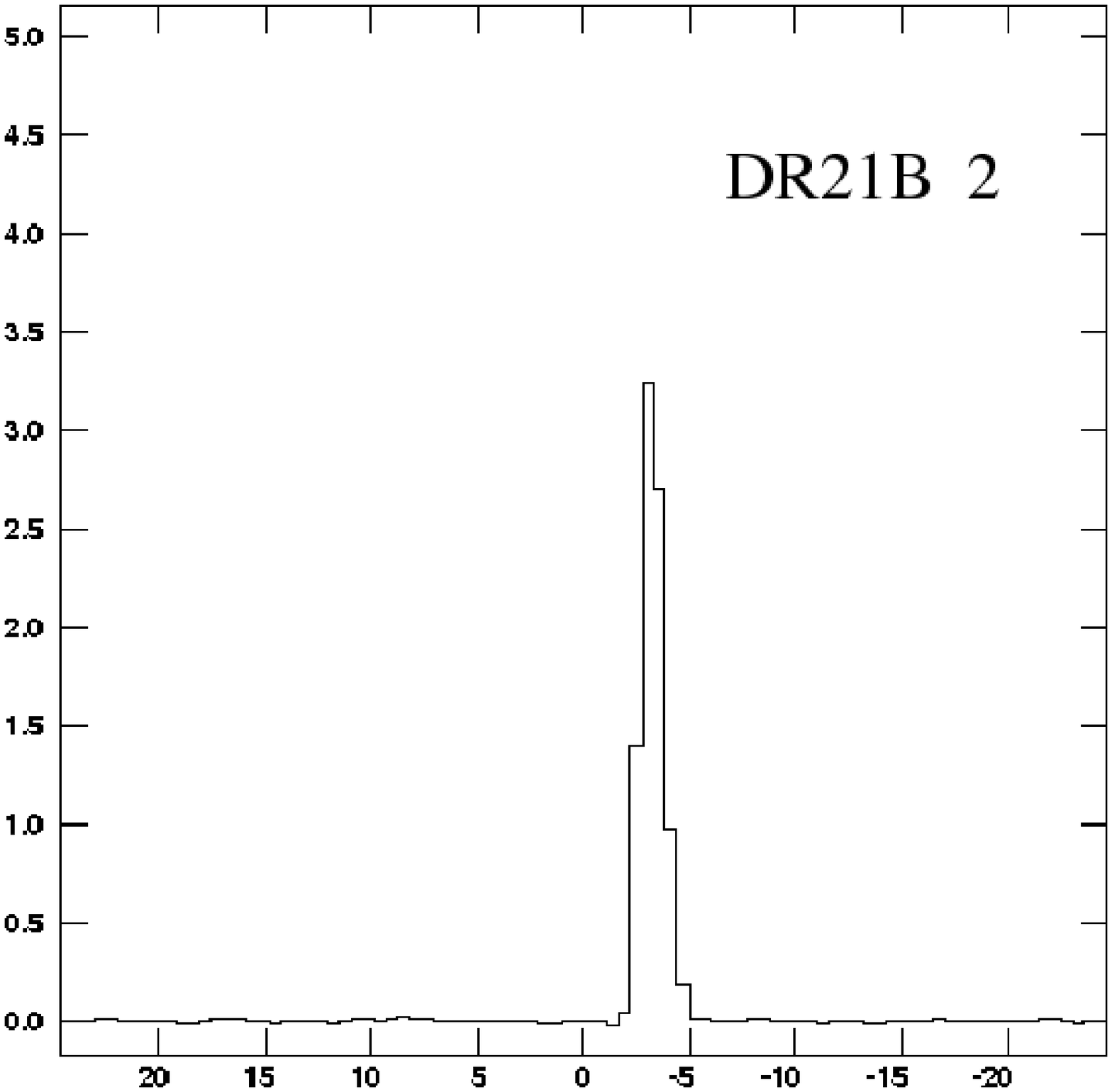} &
\includegraphics[width=5cm]{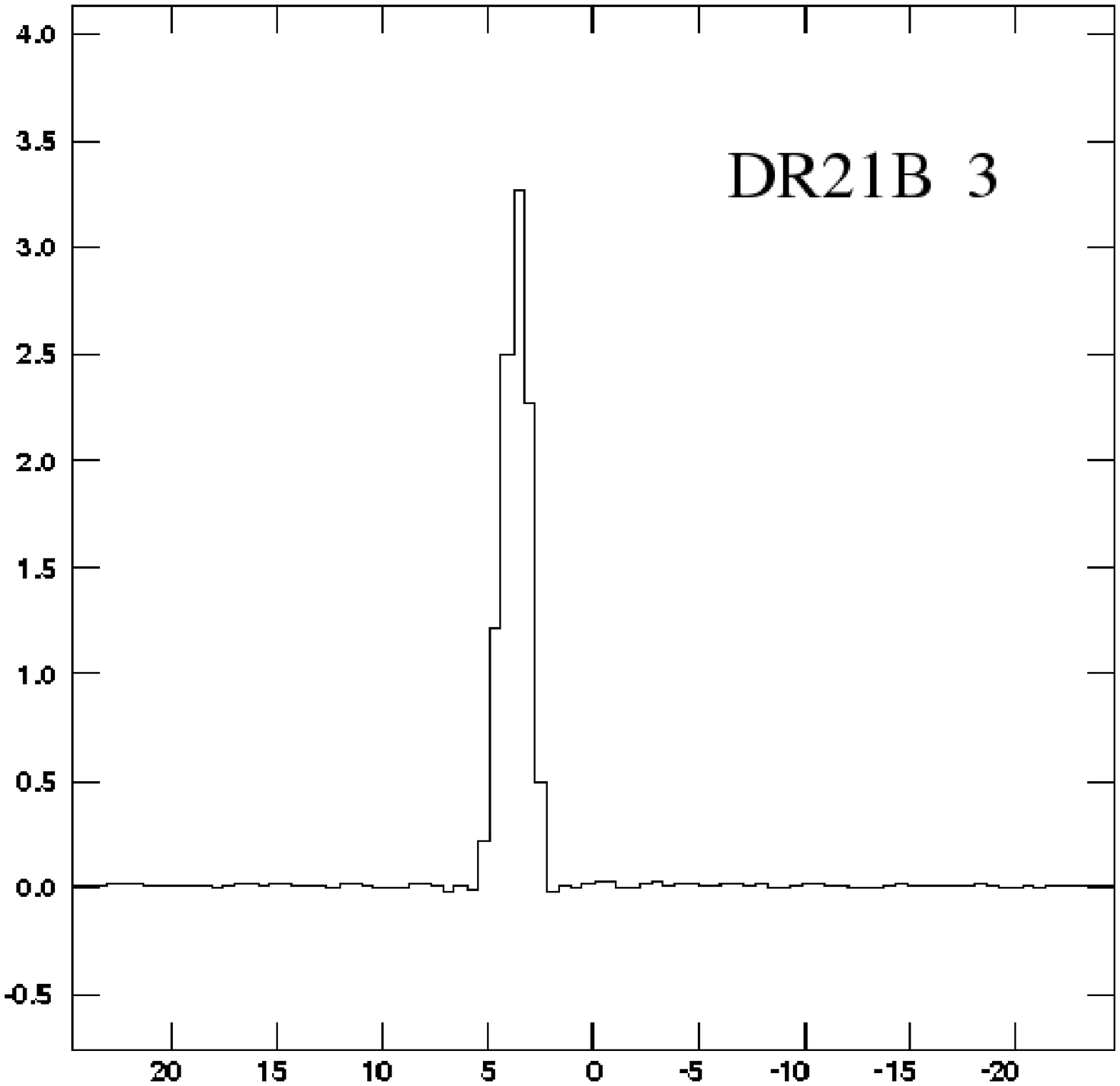} &
\end{tabular}
\caption{Same as fig. 3, but for the VLA spectra. The spectral
resolution is approximately 0.55 km~s$^{-1}$.}
\end{figure*}

\begin{figure*}
\begin{tabular}{cccc}
\includegraphics[bb=0 260 550 842,angle=90,width=6cm]{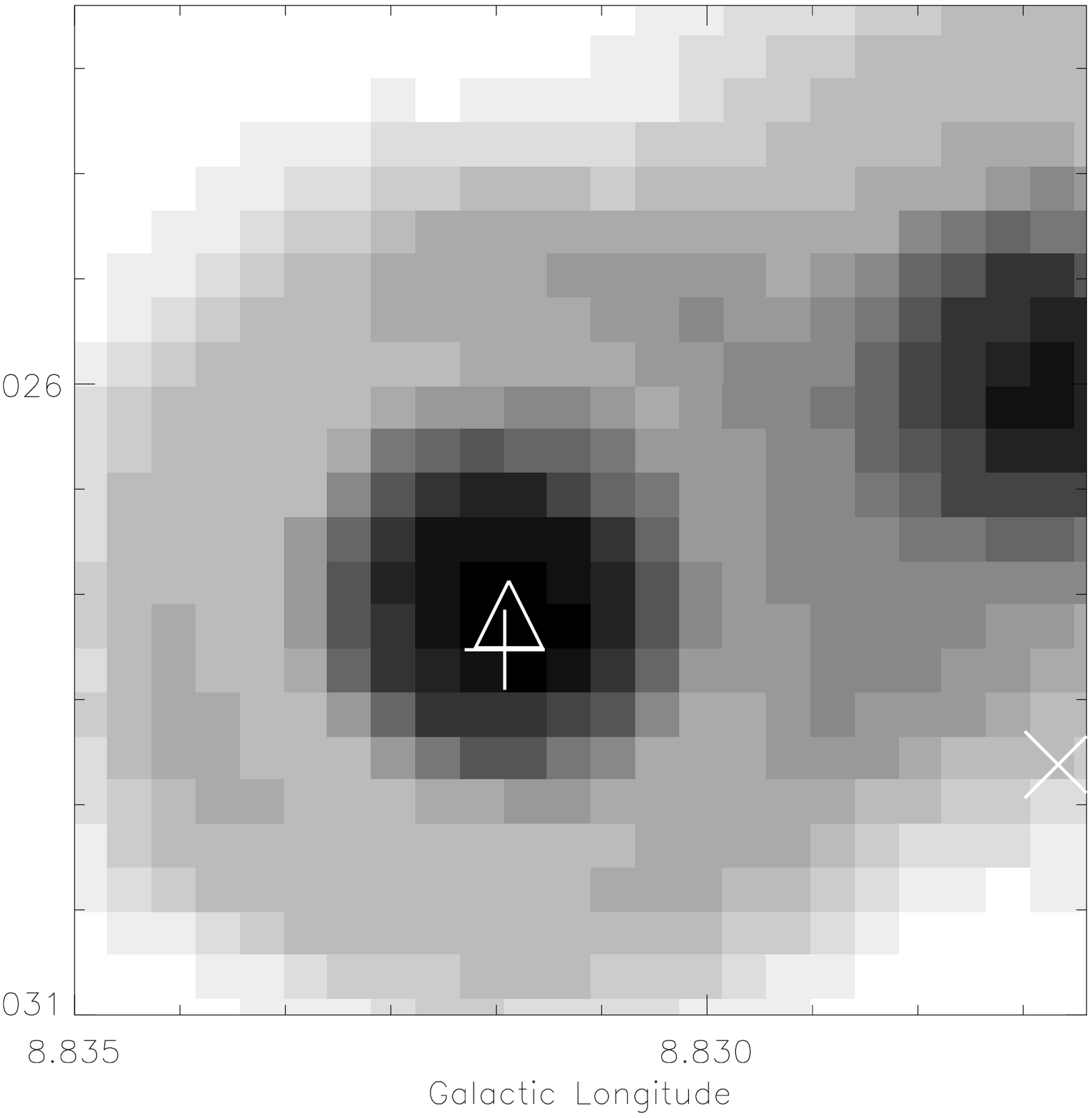}
\includegraphics[bb=0 260 550 842,angle=90,width=6cm]{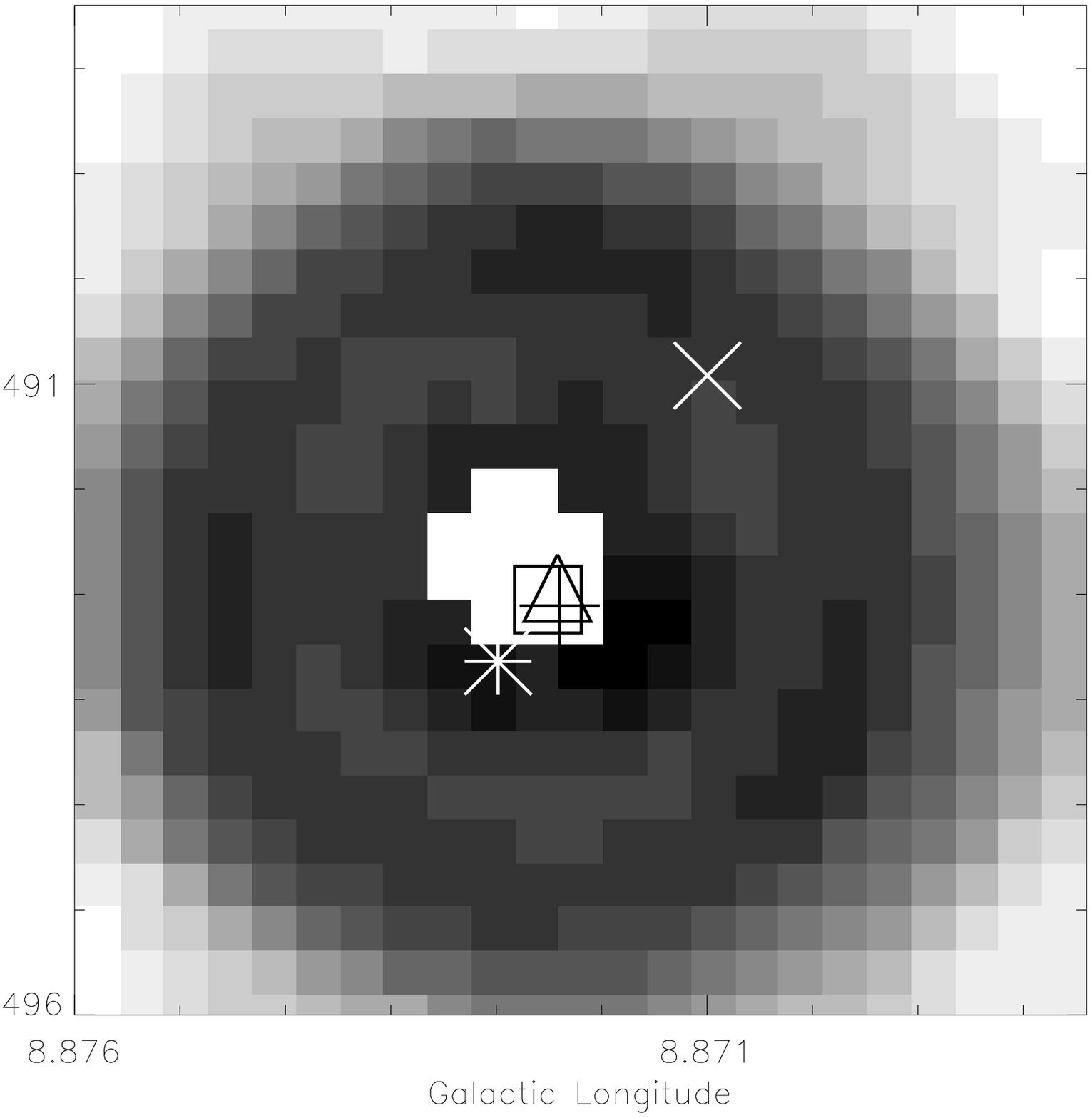}
\includegraphics[bb=0 260 550 842,angle=90,width=6cm]{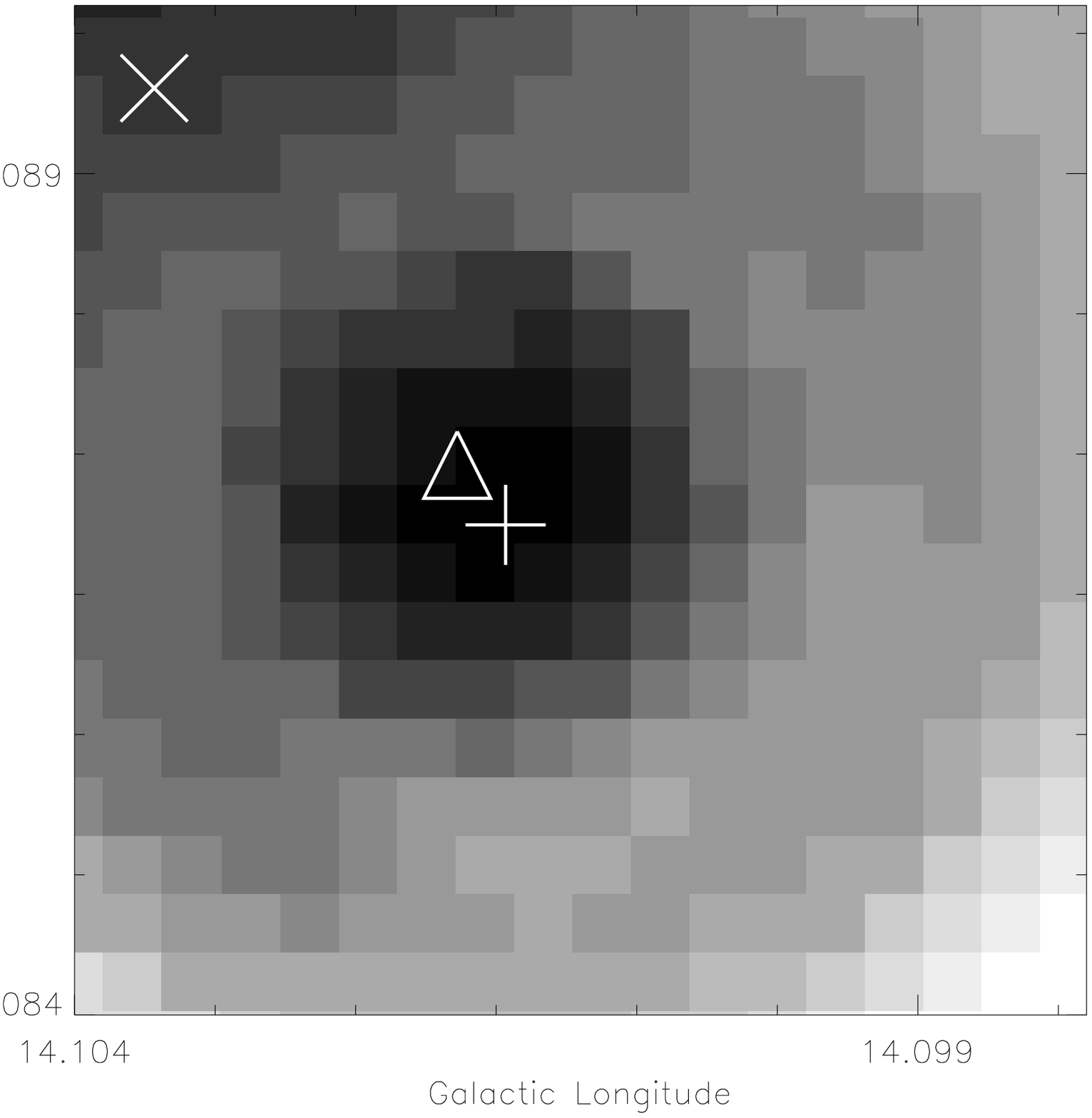}
\\
\includegraphics[bb=0 260 550 842,angle=90,width=6cm]{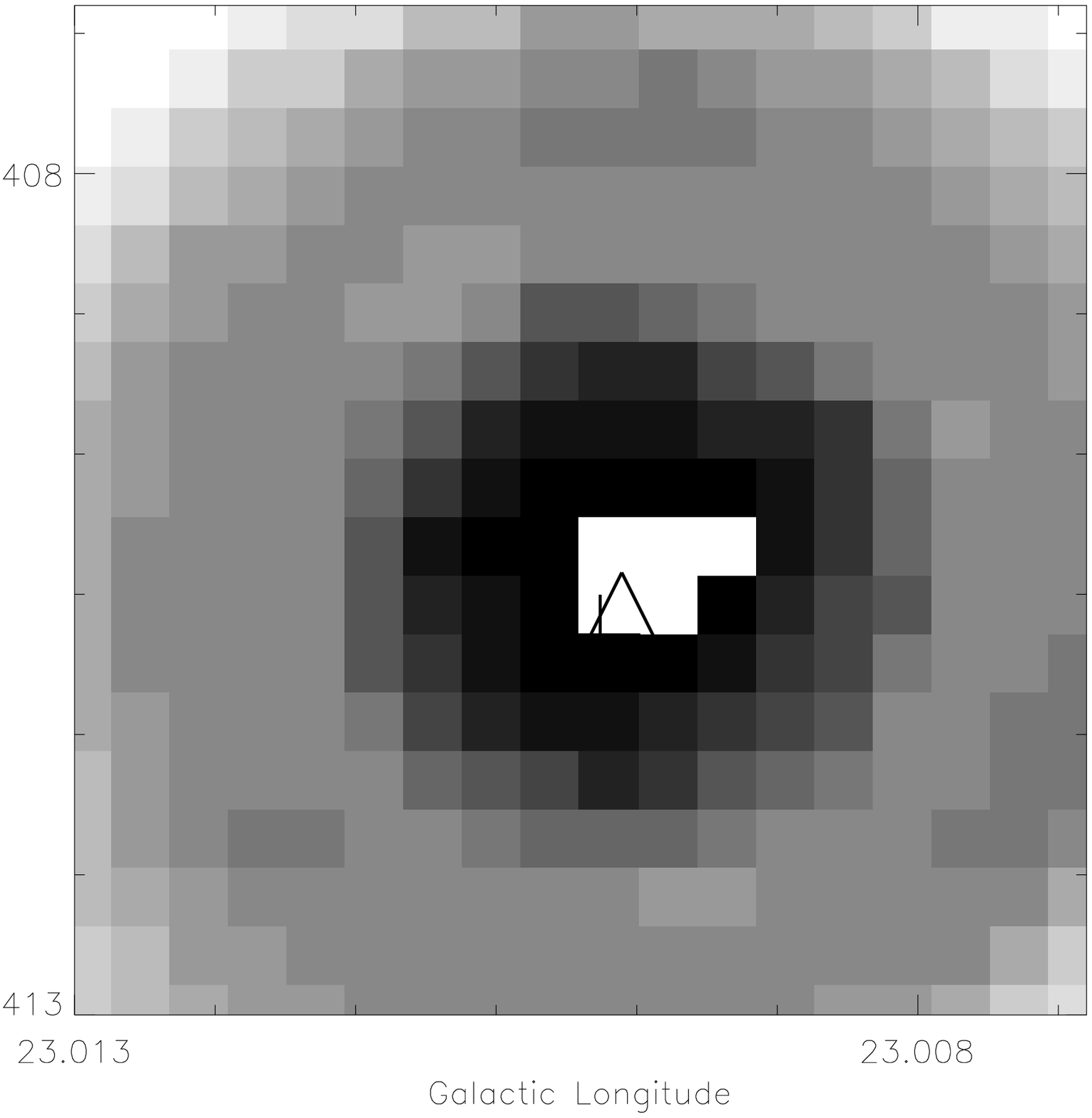}
\includegraphics[bb=0 260 550 842,angle=90,width=6cm]{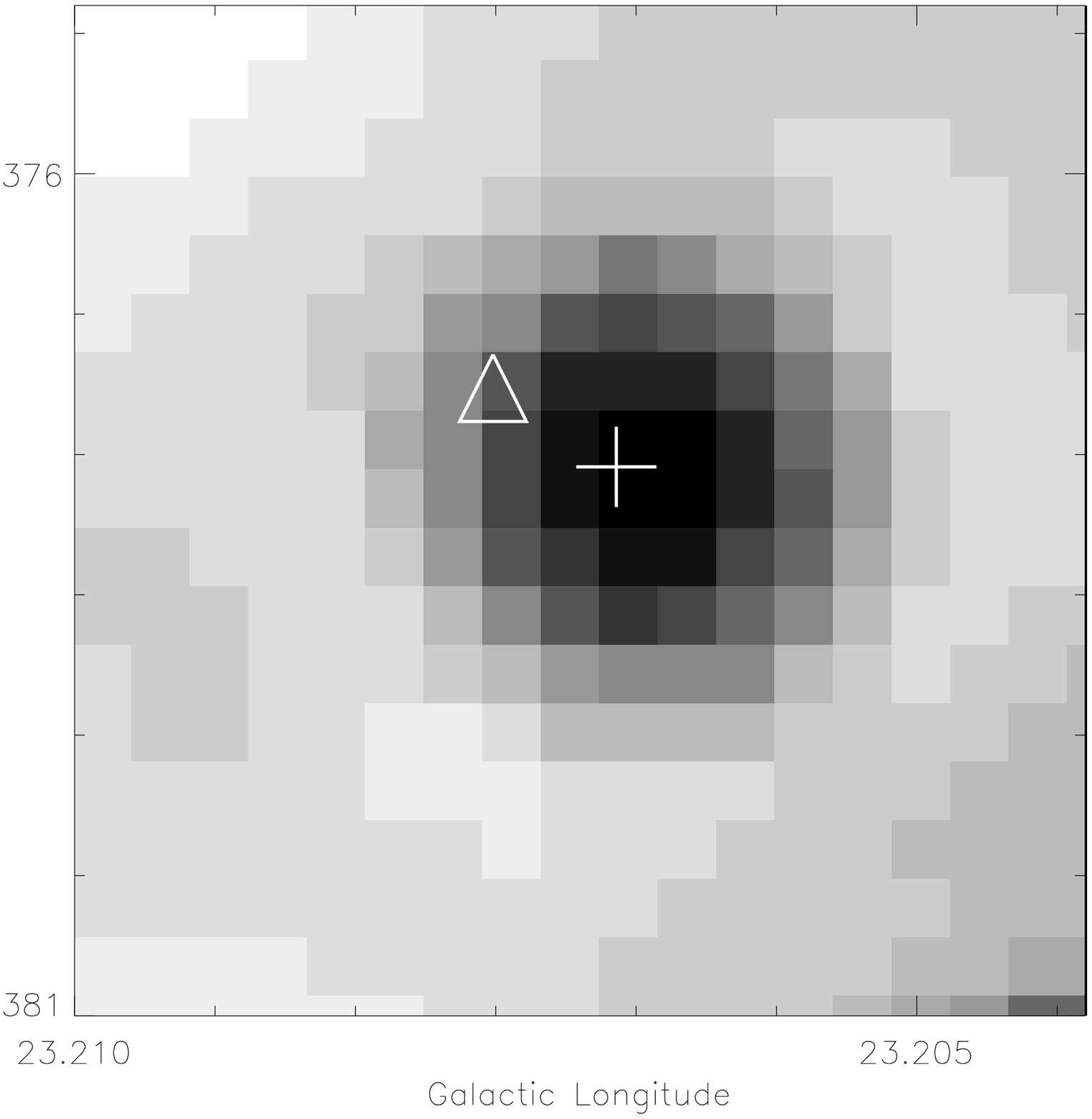}
\includegraphics[bb=0 260 550 842,angle=90,width=6cm]{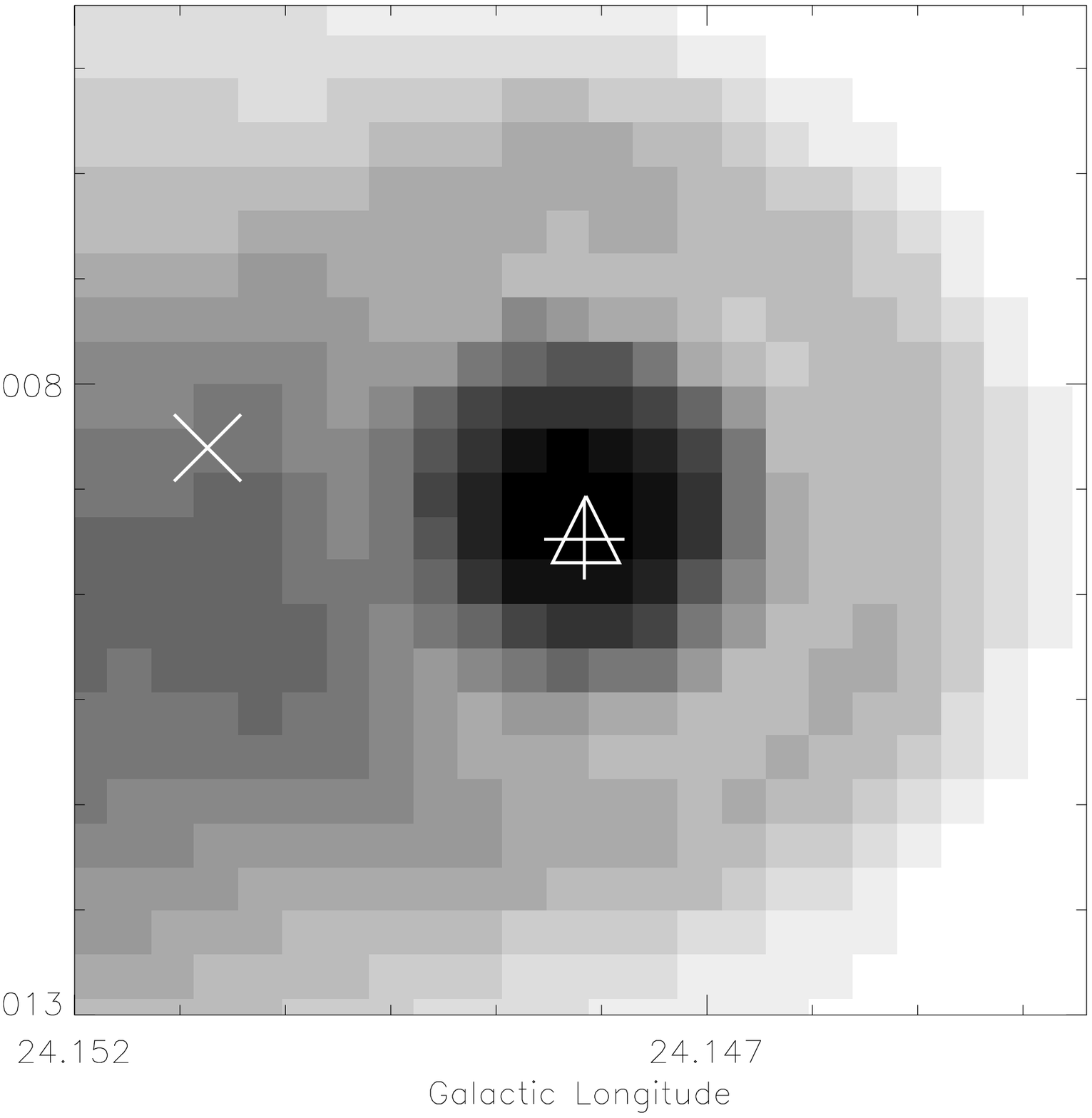}
\\
\includegraphics[bb=0 260 550 842,angle=90,width=6cm]{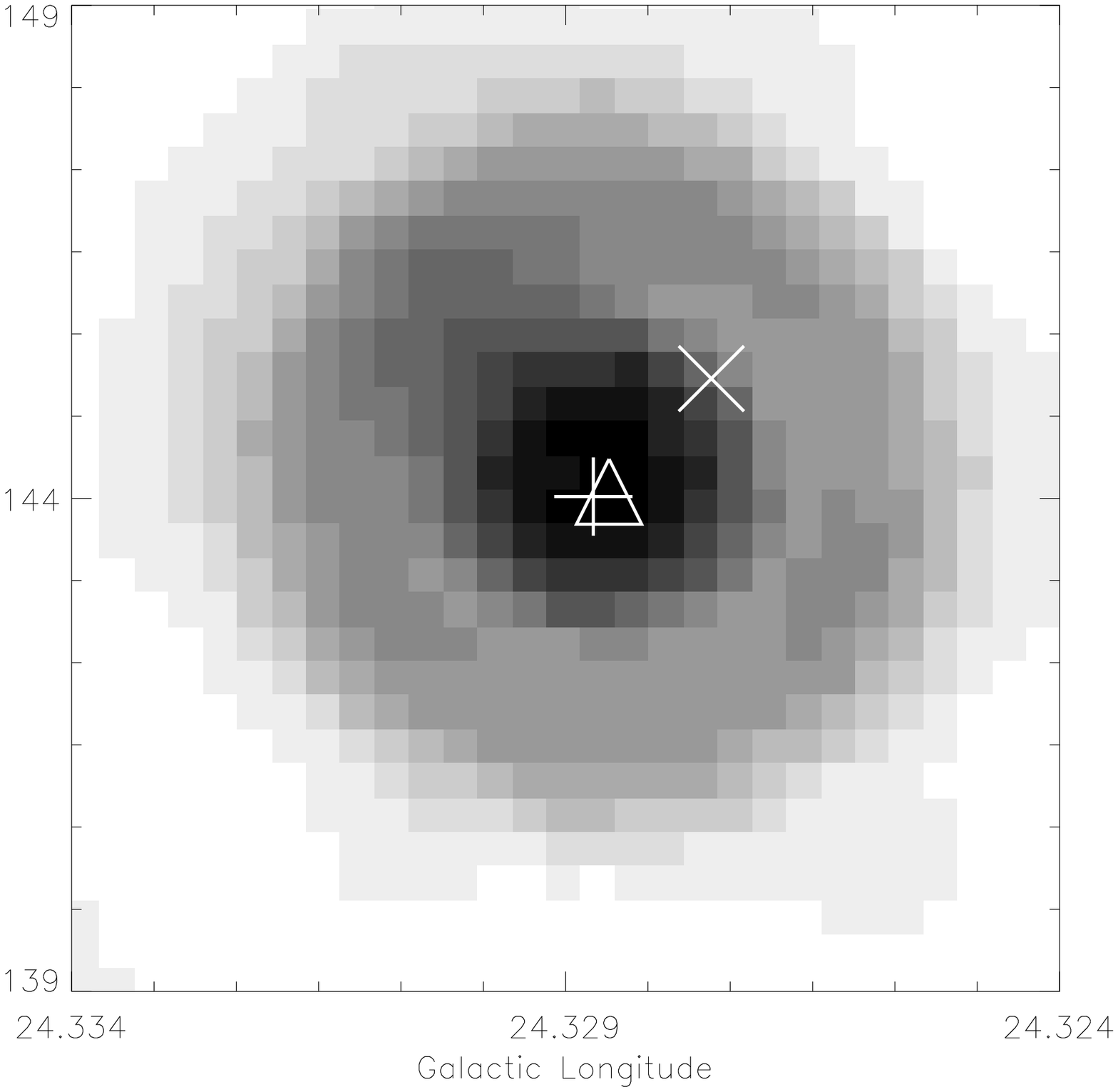}
\includegraphics[bb=0 260 550 842,angle=90,width=6cm]{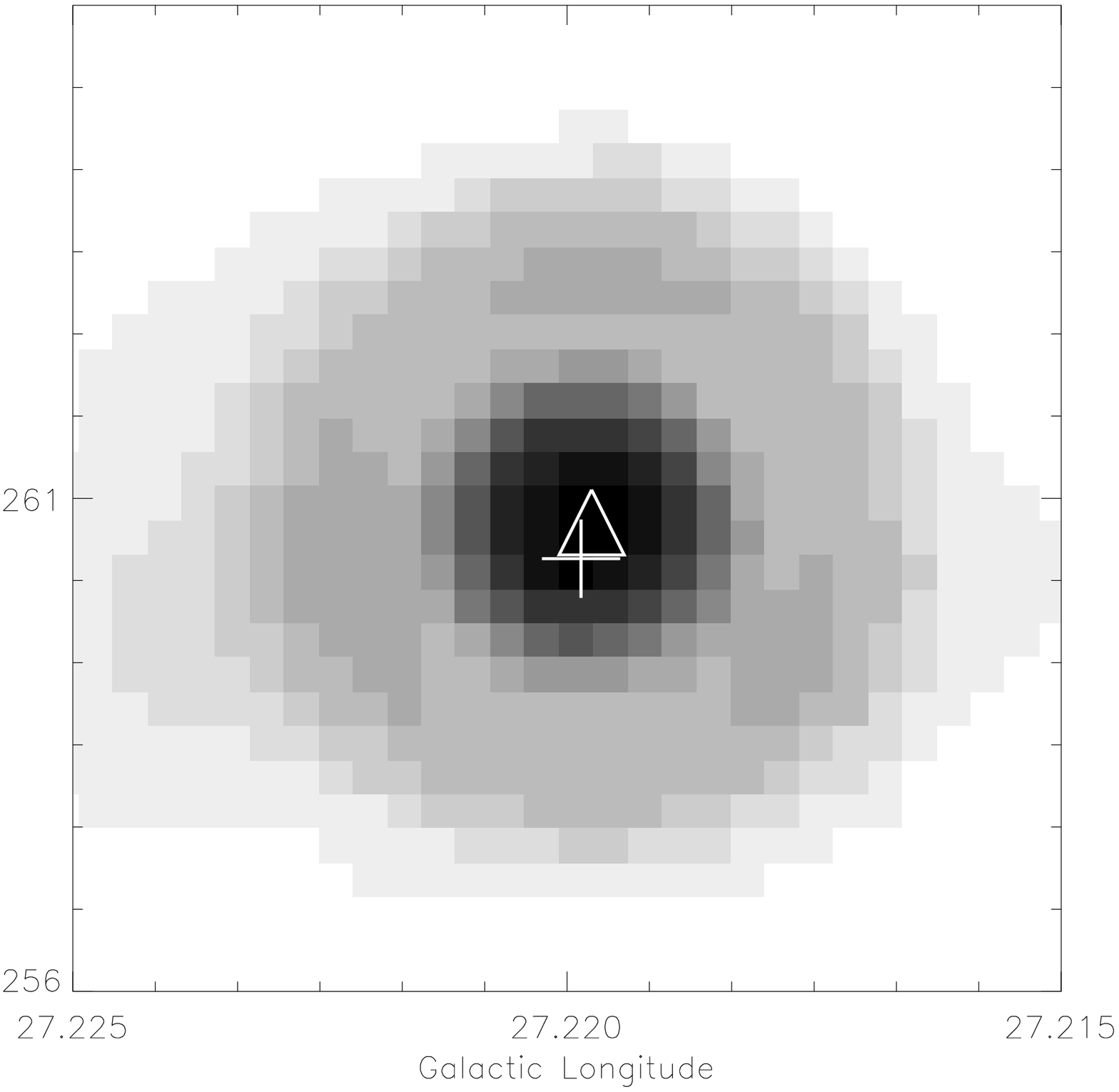}
\includegraphics[bb=0 260 550 842,angle=90,width=6cm]{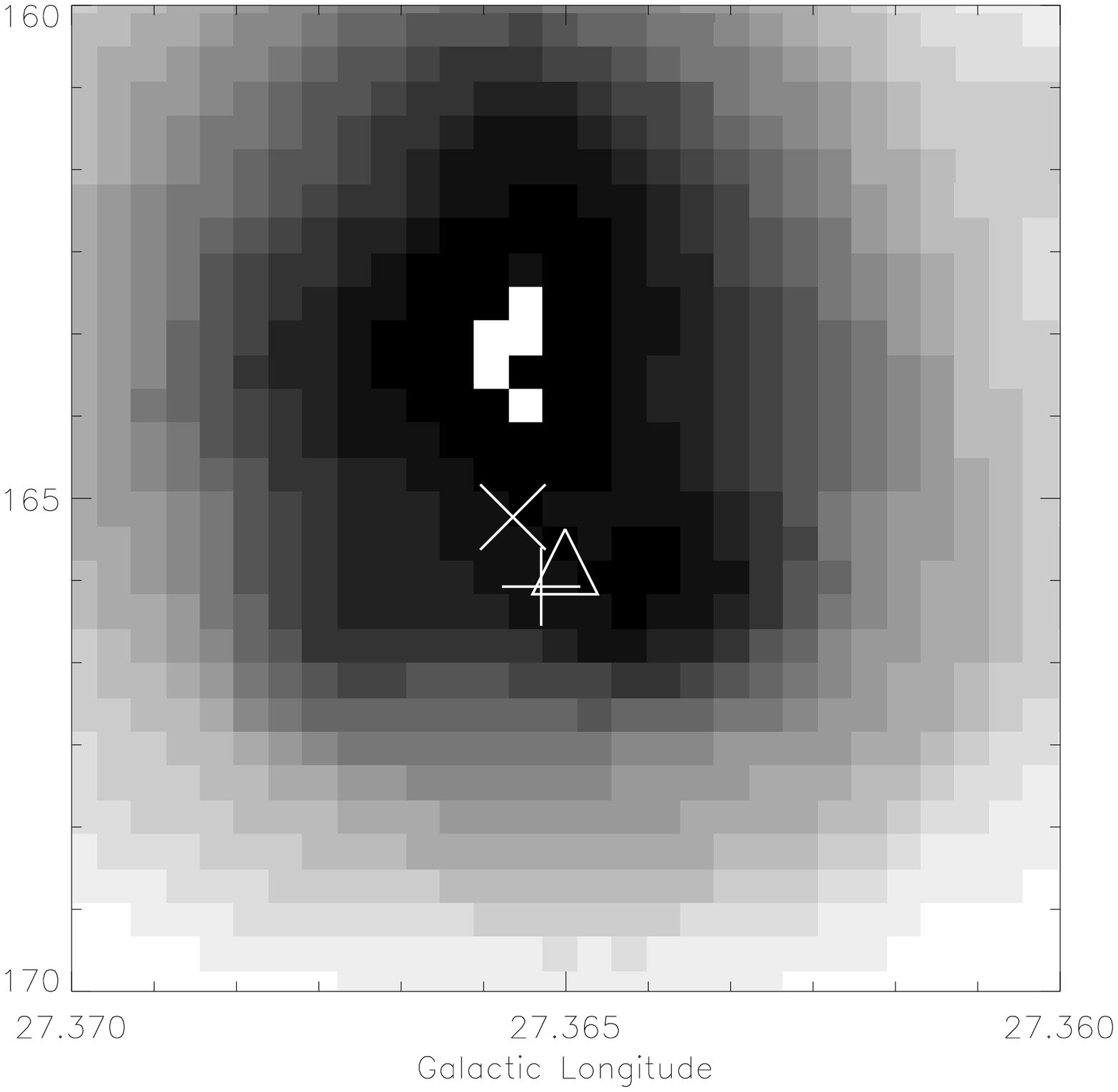}
\\
\includegraphics[bb=0 260 550 842,angle=90,width=6cm]{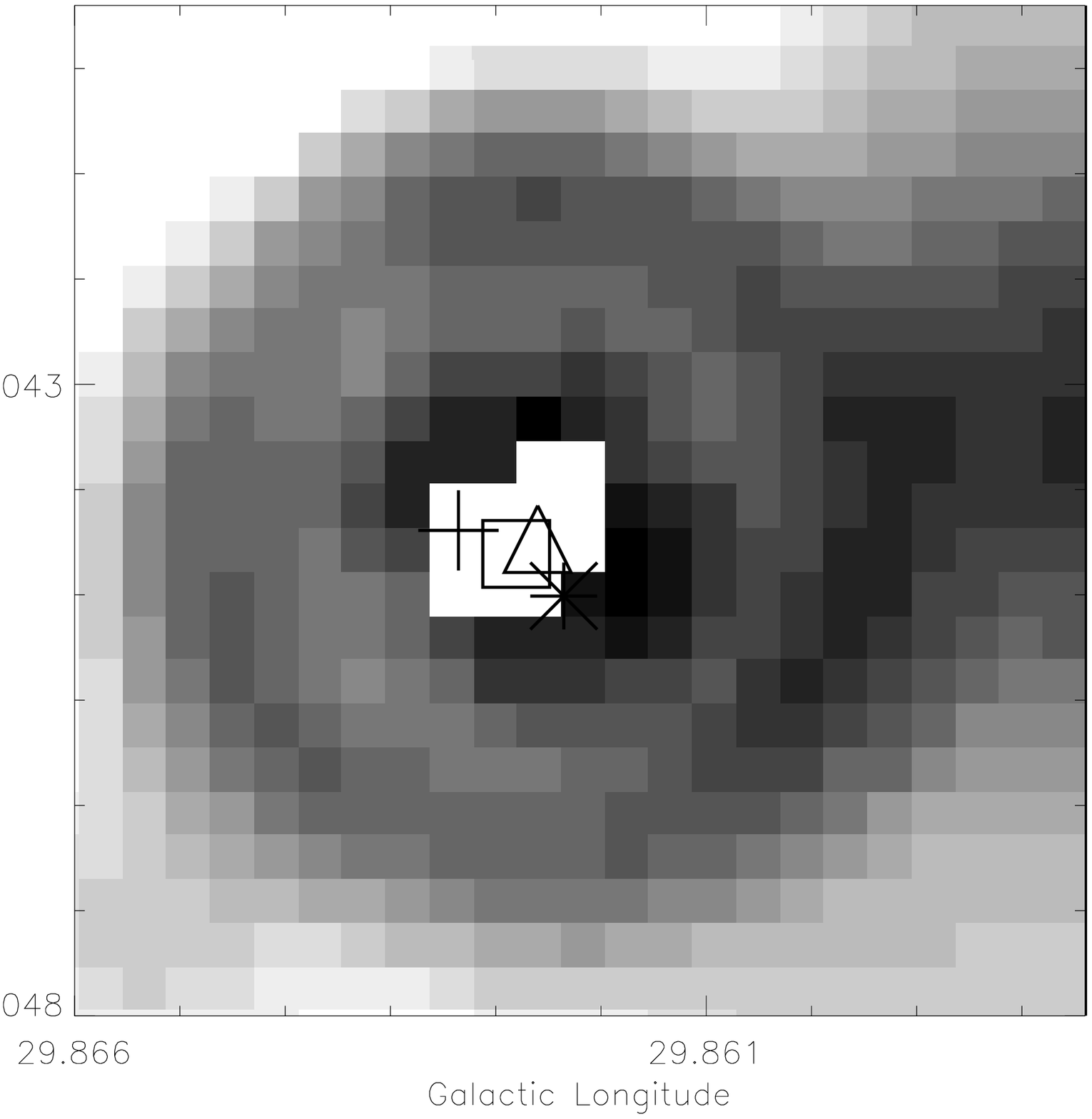}
\includegraphics[bb=0 260 550 842,angle=90,width=6cm]{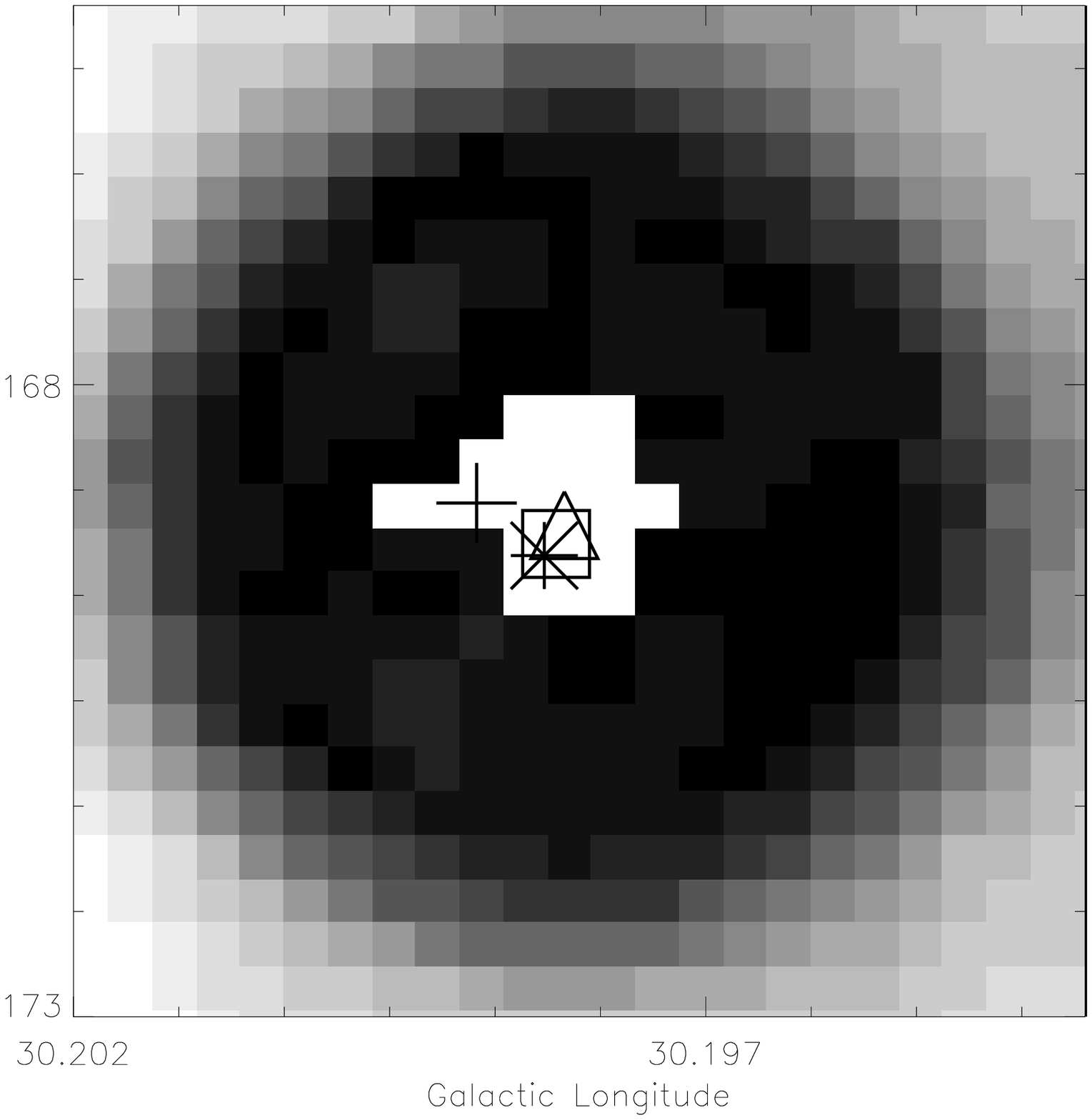}
\includegraphics[bb=0 260 550 842,angle=90,width=6cm]{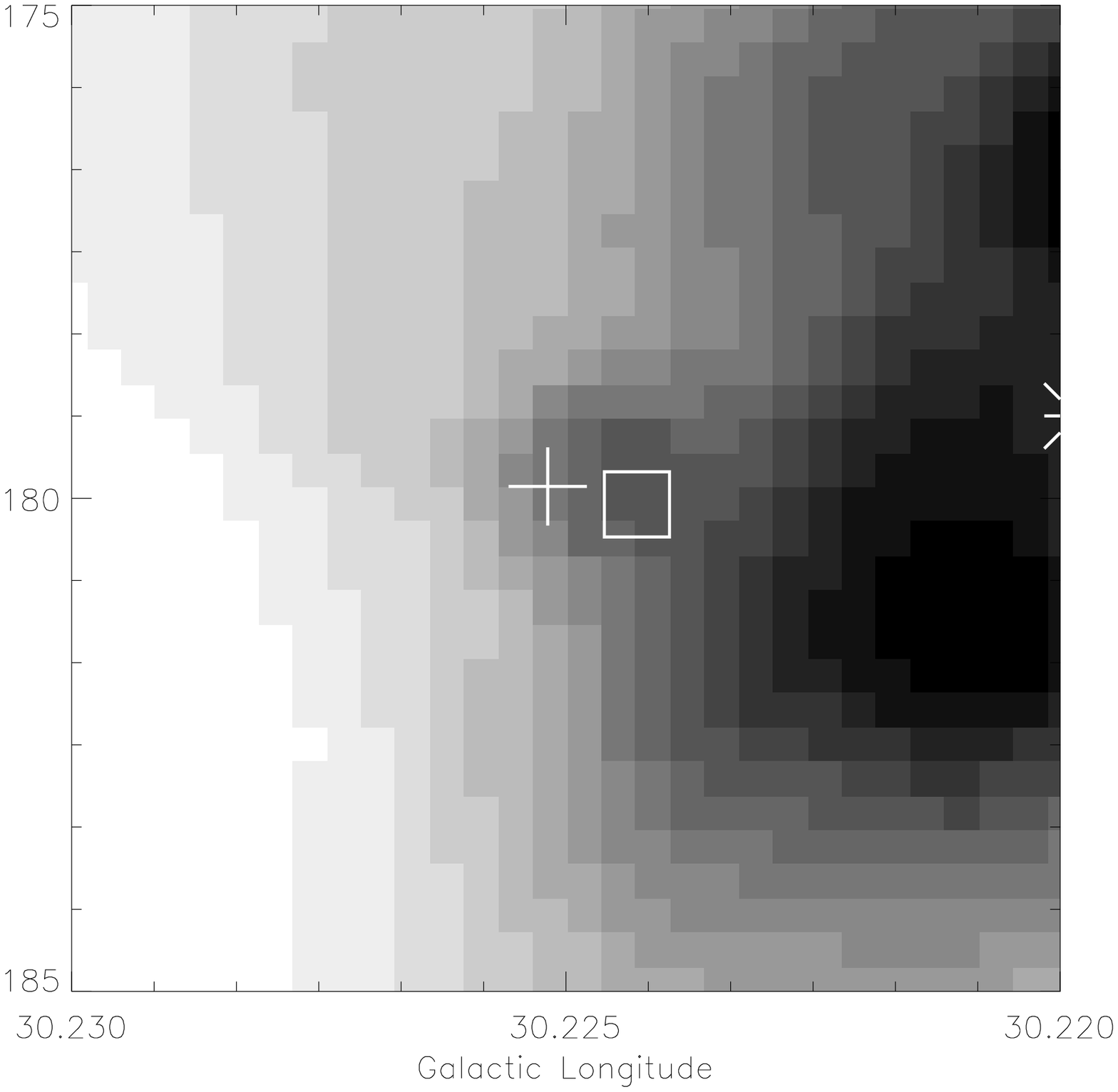}
\end{tabular}
\end{figure*}

\clearpage
\begin{figure*}
\begin{tabular}{cccc}
\includegraphics[bb=0 260 550 842,angle=90,width=6cm]{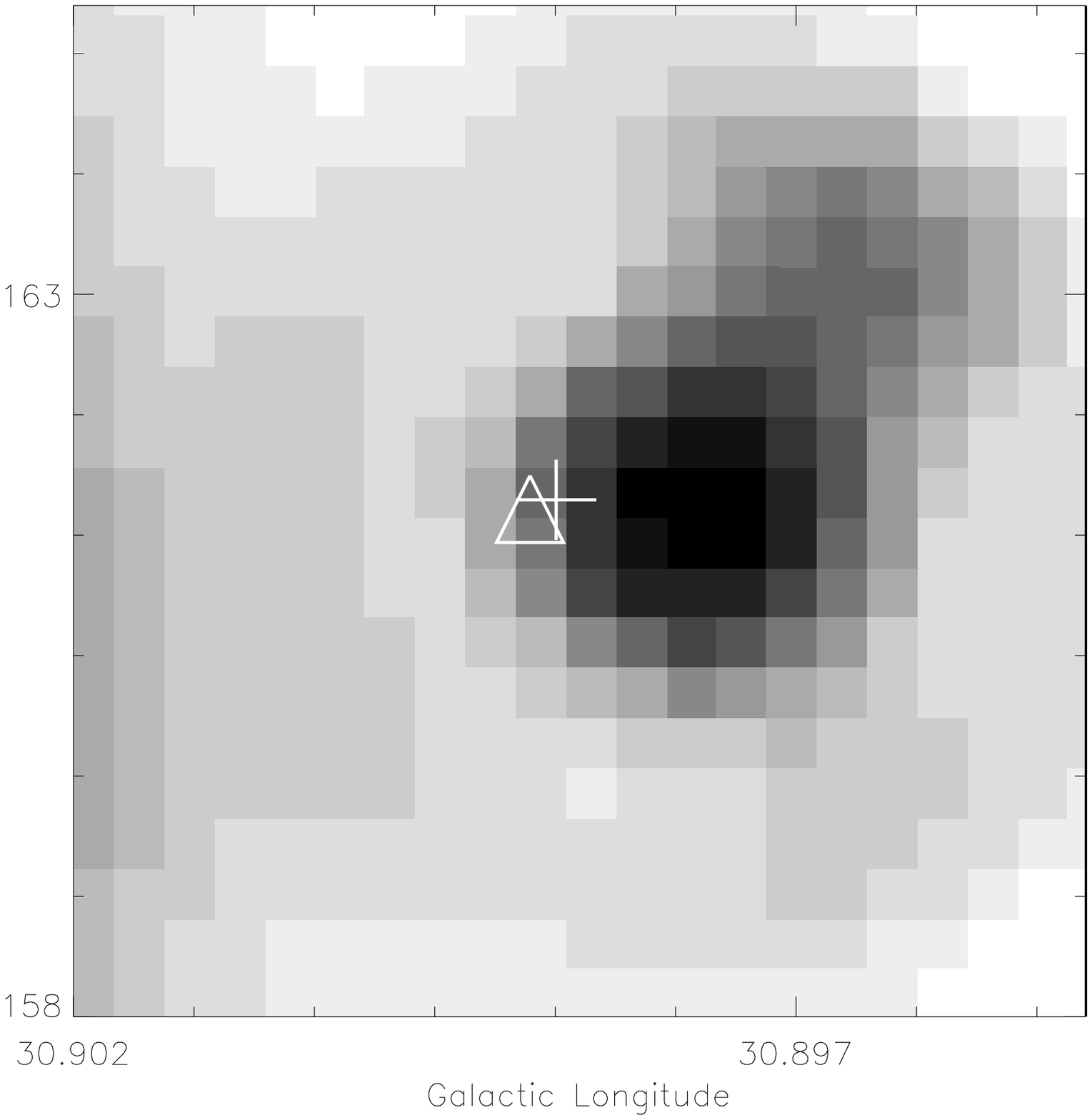}
\includegraphics[bb=0 260 550 842,angle=90,width=6cm]{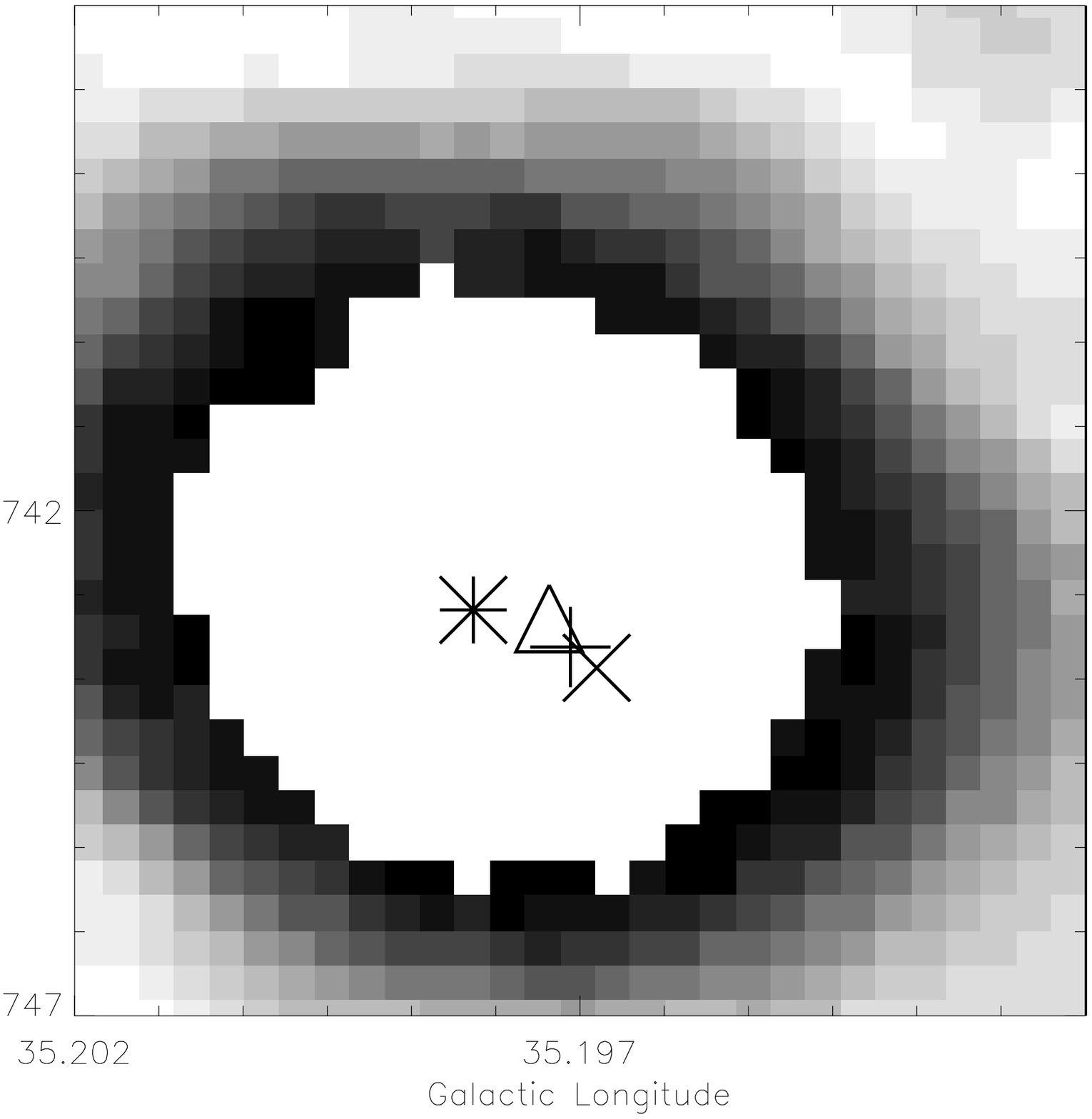}
\includegraphics[bb=0 260 550 842,angle=90,width=6cm]{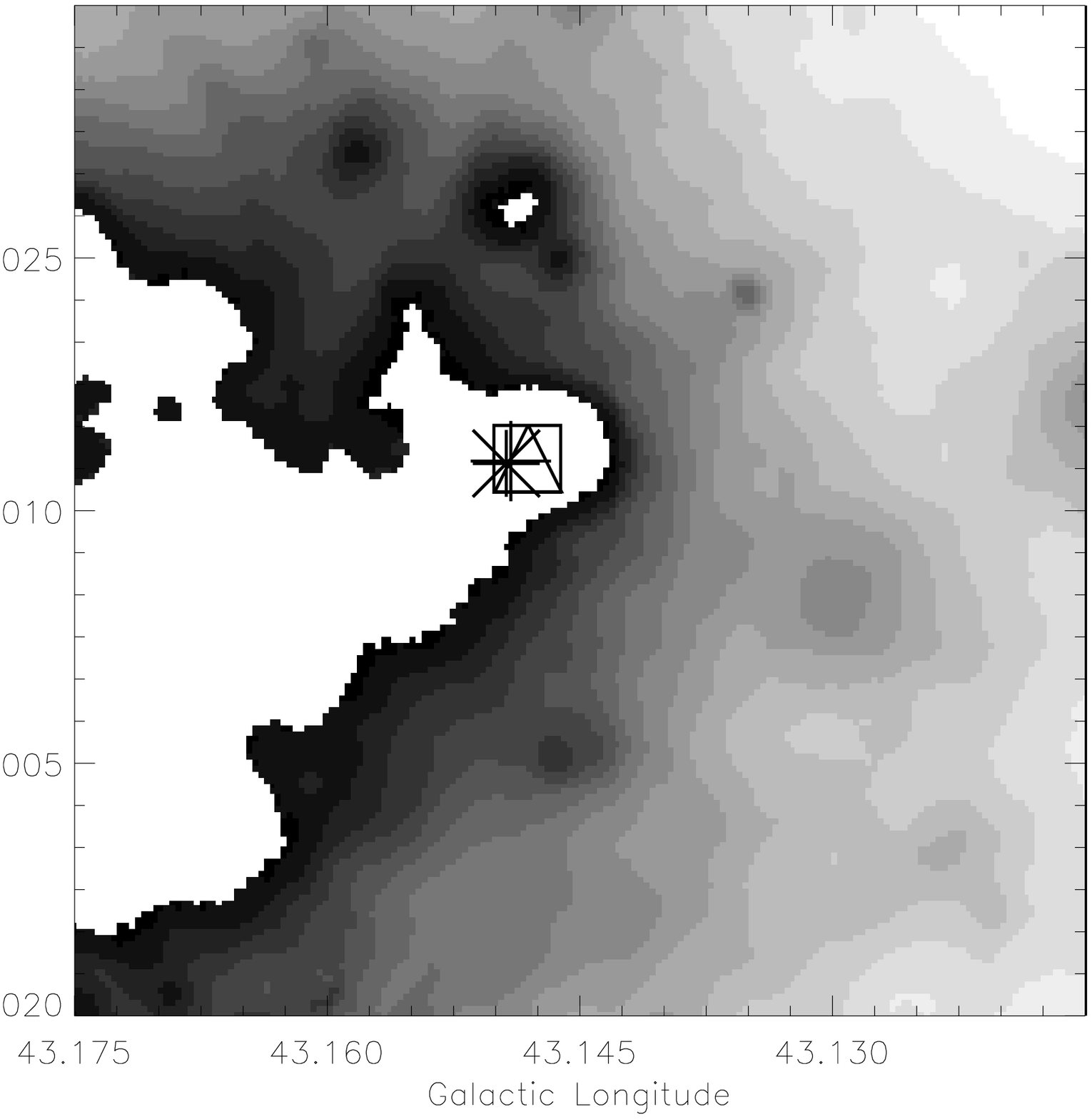}
\\
\includegraphics[bb=0 260 550 842,angle=90,width=6cm]{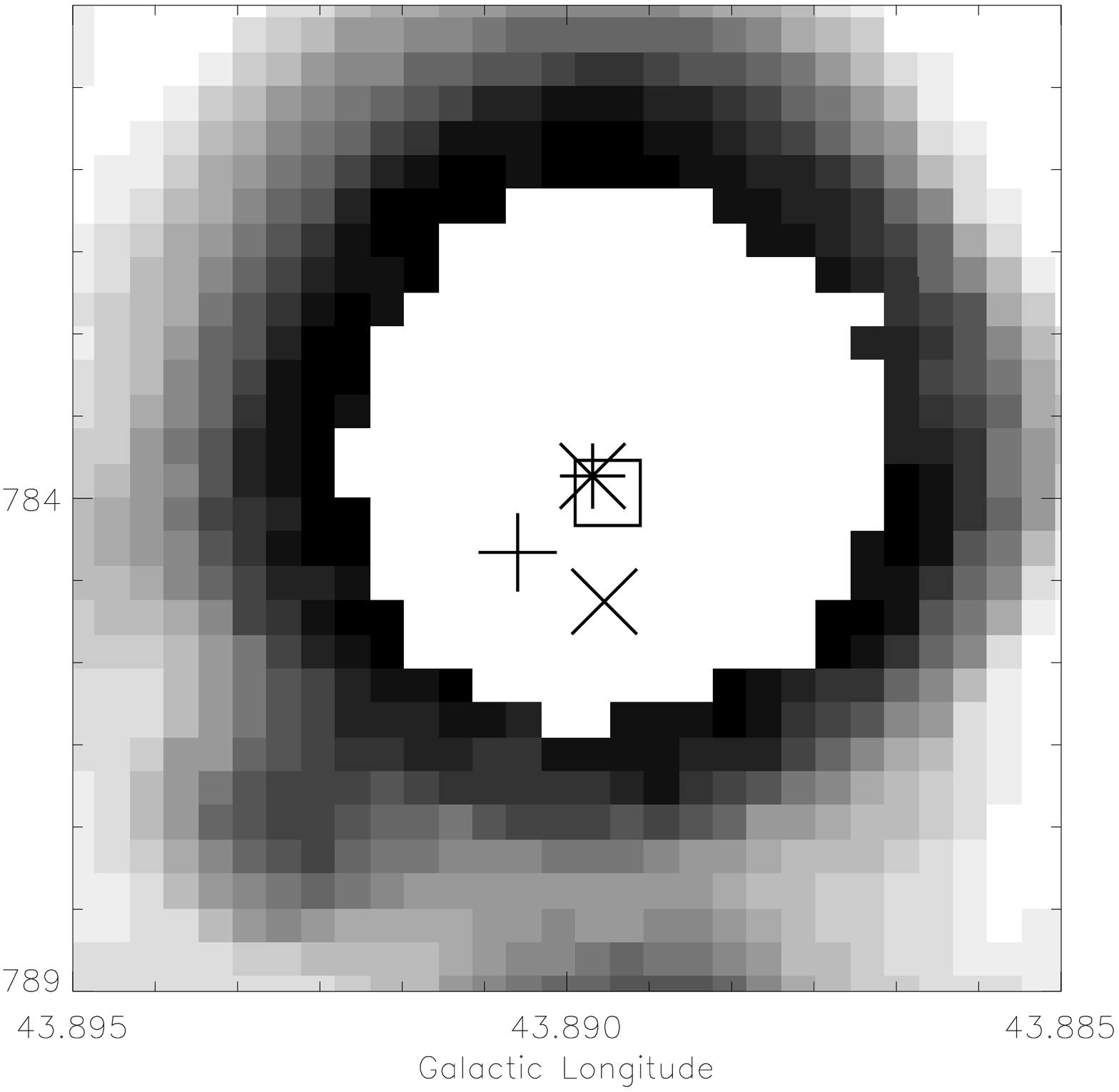}
\includegraphics[bb=0 260 550 842,angle=90,width=6cm]{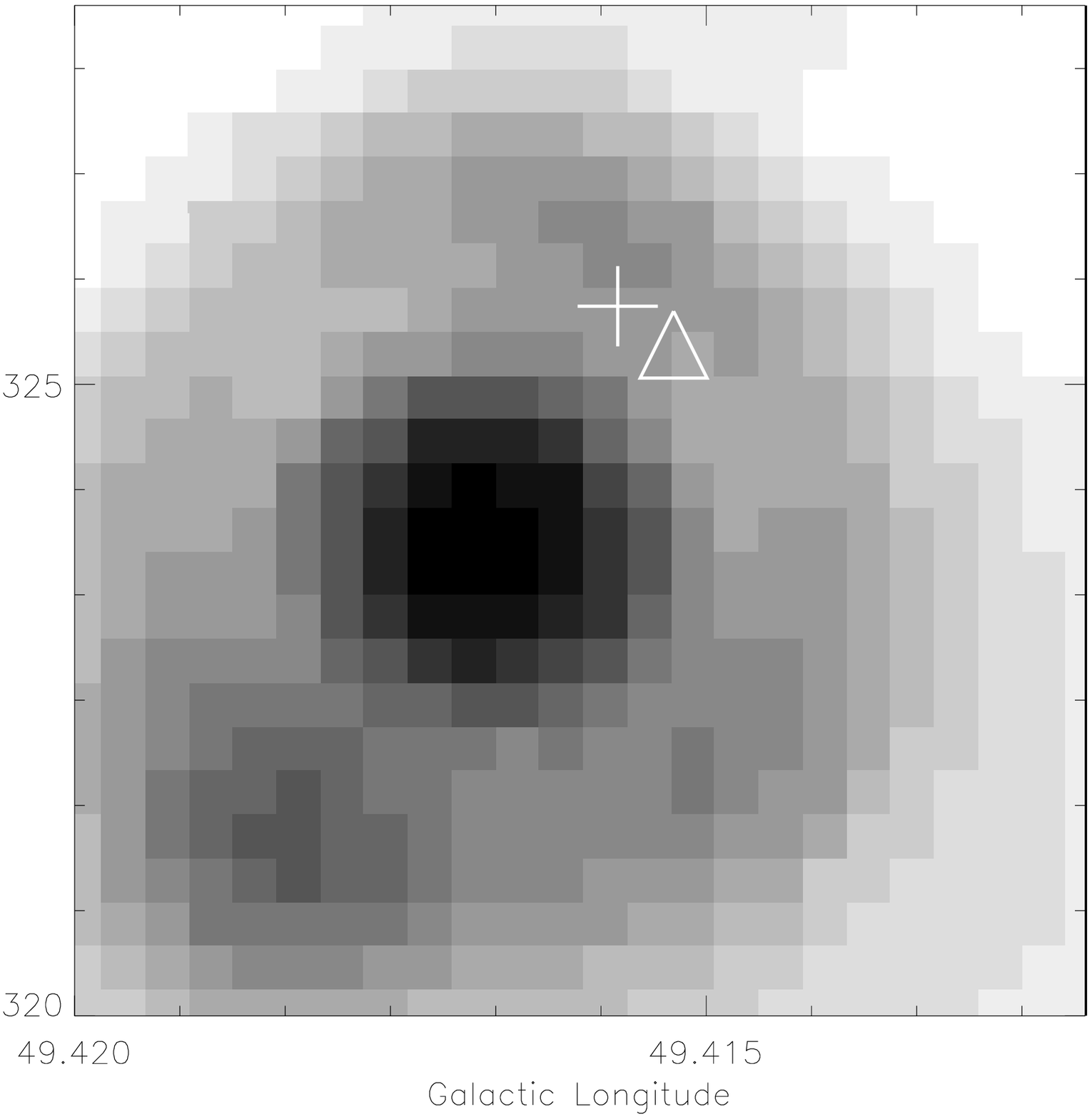}
\includegraphics[bb=0 260 550 842,angle=90,width=6cm]{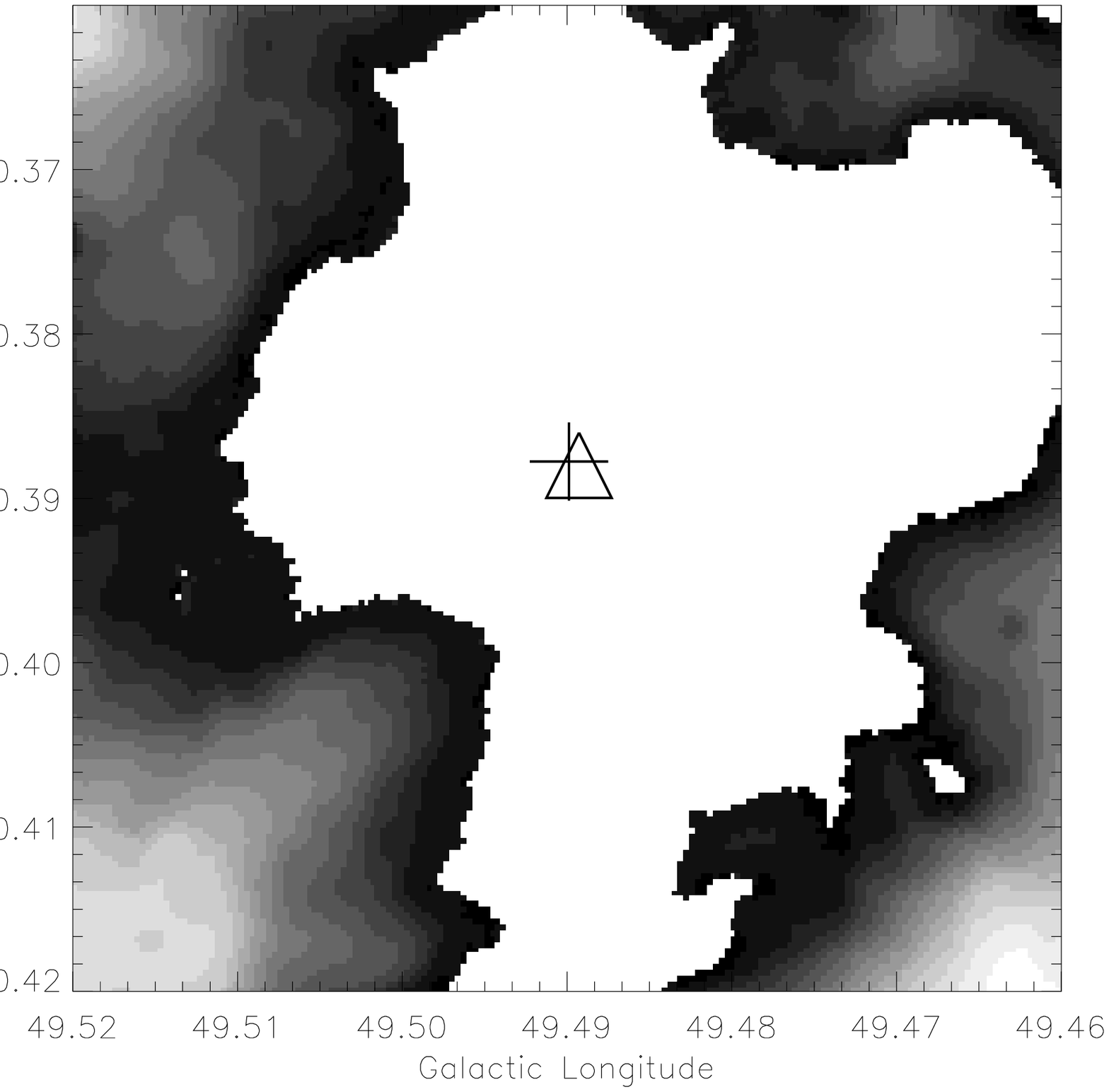}
\end{tabular}
\caption{Gray scale is the MIPSGAL 24 $\mu$m emission, the pluses
are the positions of 6.7-GHz methanol masers, the triangles are the
positions of GLIMPSE point sources, the squares are the positions of
2MASS point sources, the stars are the positions of MSX point
sources, and the crosses are the positions of $IRAS$ point sources.
Moving from the top left to the bottom right, the sources are
G8.8316-0.0281, G8.8722-0.4928, G14.1014+0.0869, G23.0099-0.4107,
G23.2068-0.3777, G24.1480-0.0092, G24.3287+0.1440, G27.2198+0.2604,
G27.3652-0.1659, G29.8630-0.0442, G30.1987-0.1687, G30.2251-0.1796,
G30.8987+0.1616, 18556+0136, G43.15+0.02, 19120+0917, 19186+1440,
and W51e2.}
\end{figure*}

\begin{longtable}{lllrllllllll}
\caption{Parameters of 592 6.7-GHz methanol masers. The first column
lists the source name. The next two columns give their J2000
equatorial coordinates. Column. (4) shows the radial velocity of
peak emission from molecular lines, such as CS, CO and NH$_{3}$,
etc., if available, otherwise from the maser peak. Columns. (5) -
(7) present the Galactic center distance, far and near kinematic
distances calculated using the Galactic rotation model of Wouterloot
\& Brand (1989), assuming $R_{0}$ = 8.5 kpc and $\Theta_{0}$ = 220
km s$^{-1}$. Columns. (8) - (10) are the same as Cols. (5) - (7),
but
revised using new $R_{0}$ and $\Theta_{0}$ from Reid et al. (2009).}\\
\hline \hline
 & & & & \multicolumn{3}{c}{$\Theta_{0}$ = 220 km~s$^{-1}$} & & \multicolumn{3}{c}{$\Theta_{0}$ = 254 km~s$^{-1}$} &  \\
\cline{5-7} \cline{9-11}
Source  & R.A.(2000) &  DEC(2000) &$\rm V_{LSR}$ & R & $\rm d_{far}$ & $\rm d_{near}$ & & R & $\rm d_{far}$ & $\rm d_{near}$ & ref \\
 Name   & $\mathrm{(^h\;\;\;^m\;\;\;^s)}$ & ($\degr\;\;\;\arcmin\;\;\;\arcsec$) & \scriptsize (km s$^{-1}$) & \scriptsize (kpc)
& \scriptsize (kpc) & \scriptsize (kpc) & & \scriptsize (kpc) & \scriptsize (kpc) & \scriptsize (kpc) &  \\
\hline
\endfirsthead
\caption{continued.}\\
\hline\hline
 & & & & \multicolumn{3}{c}{$\Theta_{0}$ = 220 km~s$^{-1}$} & & \multicolumn{3}{c}{$\Theta_{0}$ = 254 km~s$^{-1}$} & \\
\cline{5-7} \cline{9-11}
Source  & R.A.(2000) &  DEC(2000) &$\rm V_{LSR}$ & R & $\rm d_{far}$ & $\rm d_{near}$ & & R & $\rm d_{far}$ & $\rm d_{near}$ & ref\\
 Name   & $\mathrm{(^h\;\;\;^m\;\;\;^s)}$ & ($\degr\;\;\;\arcmin\;\;\;\arcsec$) & \scriptsize (km s$^{-1}$) & \scriptsize (kpc)
& \scriptsize (kpc) & \scriptsize (kpc) & & \scriptsize (kpc) & \scriptsize (kpc) & \scriptsize (kpc) & \\
\hline
\endhead
\hline

0.21$-$0.00    & 17\ 46\ 07.66   & $-$28\ 45\ 20.0     &     37.5$^{19}$  &   0.2  &   8.7  &  8.3  &  &   0.2  &   8.6  &  8.2  & 6,43,44   \\
0.31$-$0.20    & 17\ 47\ 09.12   & $-$28\ 46\ 16.0     &     18.7$^{3}$   &   0.5  &   9.0  &  8.0  &  &   0.5  &   8.9  &  7.9  & 6,43,44   \\
0.37$+$0.04    & 17\ 46\ 21.42   & $-$28\ 35\ 39.5     &     84.4$^{3}$   &   0.1  &   8.6  &  8.4  &  &   0.1  &   8.5  &  8.3  & 4,6,43    \\
0.39$-$0.03    & 17\ 46\ 41.12   & $-$28\ 37\ 05.5     &     28.7$^{27}$  &   0.4  &   8.9  &  8.1  &  &   0.4  &   8.8  &  8.0  & 6         \\
0.49$+$0.18    & 17\ 46\ 04.00   & $-$28\ 24\ 51.5     &   $-$3.8$^{3}$   &  25.0  &  33.5  &       &  &  25.0  &  33.4  &       & 6,43,44   \\
0.53$+$0.18    & 17\ 46\ 09.8    & $-$28\ 23\ 29       &   $-$3.8$^{3}$   &  25.0  &  33.5  &       &  &  25.0  &  33.4  &       & 32,43,33  \\
0.54$-$0.85    & 17\ 50\ 14.53   & $-$28\ 54\ 31.0     &     16.6$^{30}$  &   0.9  &   9.4  &  7.6  &  &   0.9  &   9.3  &  7.5  & 20,4,43,44  \\
0.64$-$0.04    & 17\ 47\ 18.65   & $-$28\ 24\ 25.0     &     60.5$^{30}$  &   0.3  &   8.8  &  8.2  &  &   0.3  &   8.7  &  8.1  & 20,4,6,43,44 33 \\
\hline

\end{longtable}

References: \\

1 \ Anglada et al. 1996. \ 2 \ Beuther \& Sridharan 2007. \ 3 \
Bronfman et al. 1996. \ 4 \ Caswell et al. 1995. \ 5 \ Caswell et
al. 1996a. 6 \ Caswell et al. 1996b. \ 7 \ Churchwell et al. 1990. \
8 \ Ellingsen et al. 1996. \ 9 \ Ellingsen et al. 2007. \ 10 \ Felli
et al. 1992. 11 \ Fontani et al. 2006. \ 12 \ Garay et al. 1993.  \
13 \ Gaylard \& Macleod 1993. \ 14 \ Juvela et al. 1996. \ 15 \ Kim
\& Koo 2003. 16 \ Larionov et al. 1999. \ 17 \ Macleod \& Gaylard
1992. \ 18 \ Macleod et al. 1993. \ 19 \ Mauersberger et al. 1986. \
20 \ Menten 1991. 21 \ Minier et al. 1998. \ 22 \ Minier et al.
2000. \ 23 \ Minier et al. 2001. \ 24 \ Minier et al. 2003. \ 25 \
Molinari et al. 1996. 26 \ Pandian et al. 2007. \ 27 \ Pestalozzi et
al. 2005. \ 28 \ Pillai et al. 2006. \ 29 \ Pirogov et al. 2006. \
30 \ Plume et al. 1992. 31 \ Russeil 2003. \ 32 \ Schutte et al.
1993. \ 33 \ Slysh et al. 1999. \ 34 \ Solomon et al. 1987. \ 35 \
Szymczak et al. 2000. 36 \ Szymczak et al. 2002. \ 37 \ Szymczak et
al. 2008. \ 38 \ Teyssier et al. 2002. \ 39 \ van der Walt et al.
1995. \ 40 \ van der Walt et al. 1996. 41 \ van der Walt et al.
2007. \ 42 \ Vilas-Boas et al. 2000. \ 43 \ Walsh et al. 1997. \ 44
\ Walsh et al. 1998. \ 45 \ Wouterloot et al. 1989. 46 \ Wouterloot
et al. 1993. \ 47 \ Wu et al. 2006. \ 48 \ Xu et al. 2008. \ 49 \
Zhang et al. 2005. \ 50 \ Zinchenko et al. 1995. 51 \ MacLeod et al.
1998. \ 52 \ this paper.

\end{document}